\documentstyle[subfigure,12pt,amstex,floatfig,epsf,epsfig,a4]{article}
\epsfverbosetrue
\newcommand{\longpage}{\enlargethispage{\baselineskip}}
\begin{document}
\setcounter{tocdepth}{2}
\thispagestyle{empty}
\date{~}
\title{\vspace*{ -1cm}
\centerline{\large EUROPEAN ORGANIZATION FOR NUCLEAR RESEARCH (CERN)}
\begin{flushright}{\normalsize
\mbox{~}\\CERN-EP/2001-027\\[-.3cm]March 2, 2001
}
\end{flushright} 
 \vspace{3cm}
{\bf Measurement of the Tau Polarisation at~LEP}}
\vspace{1cm}
\author{The ALEPH collaboration\thanks{See next pages for the list of authors.}}
\maketitle
\vspace{2cm}
\begin{abstract}
The polarisation of  $\tau$'s produced in  Z decay is measured using
160 pb$^{-1}$ of data accumulated at LEP  by the ALEPH detector
between 1990 and 1995.
The variation of the polarisation 
with polar angle yields 
the two parameters ${\cal A}_e = 0.1504 \pm 0.0068 $  and
${\cal A}_{\tau} = 0.1451 \pm 0.0059$ which are consistent with the
hypothesis of  $e$-$\tau$  universality.
Assuming  universality, the value ${\cal A}_{e\mbox{-}\tau} = 0.1474 \pm
0.0045$ is obtained
from which the effective weak mixing angle 
 $\sin^2 {\theta_{\mathrm{W}}^{\mathrm{eff}}} =0.23147 \pm 0.00057 $ is derived.\\[3cm]

\end{abstract}
\centerline{\sl To be submitted to The European Physical Journal C}
\thispagestyle{empty}
\newpage
\pagestyle{empty}
\newpage
\small
%
\newlength{\saveparskip}
\newlength{\savetextheight}
\newlength{\savetopmargin}
\newlength{\savetextwidth}
\newlength{\saveoddsidemargin}
\newlength{\savetopsep}
\setlength{\saveparskip}{\parskip}
\setlength{\savetextheight}{\textheight}
\setlength{\savetopmargin}{\topmargin}
\setlength{\savetextwidth}{\textwidth}
\setlength{\saveoddsidemargin}{\oddsidemargin}
\setlength{\savetopsep}{\topsep}
%
%
\setlength{\parskip}{0.0cm}
\setlength{\textheight}{25.0cm}
\setlength{\topmargin}{-1.5cm}
\setlength{\textwidth}{16 cm}
\setlength{\oddsidemargin}{-0.0cm}
\setlength{\topsep}{1mm}
\pretolerance=10000
\centerline{\large\bf The ALEPH Collaboration}
\footnotesize
\vspace{0.5cm}
{\raggedbottom
\begin{sloppypar}
\samepage\noindent
A.~Heister,
S.~Schael
\nopagebreak
\begin{center}
\parbox{15.5cm}{\sl\samepage
Physikalisches Institut das RWTH-Aachen, D-52056 Aachen, Germany}
\end{center}\end{sloppypar}
\vspace{2mm}
\begin{sloppypar}
\noindent
R.~Barate,
I.~De~Bonis,
D.~Decamp,
P.~Ghez,
C.~Goy,
\mbox{J.-P.~Lees},
E.~Merle,
\mbox{M.-N.~Minard},
B.~Pietrzyk
\nopagebreak
\begin{center}
\parbox{15.5cm}{\sl\samepage
Laboratoire de Physique des Particules (LAPP), IN$^{2}$P$^{3}$-CNRS,
F-74019 Annecy-le-Vieux Cedex, France}
\end{center}\end{sloppypar}
\vspace{2mm}
\begin{sloppypar}
\noindent
\mbox{R.~Alemany},
\mbox{S.~Bravo},
\mbox{M.P.~Casado},
\mbox{M.~Chmeissani},
\mbox{J.M.~Crespo},
\mbox{E.~Fernandez},\\
\mbox{M.~Fernandez-Bosman},
Ll.~Garrido,$^{15}$
E.~Graug\'{e}s,
M.~Martinez,
G.~Merino,
R.~Miquel,
Ll.M.~Mir,
A.~Pacheco,
H.~Ruiz
\nopagebreak
\begin{center}
\parbox{15.5cm}{\sl\samepage
Institut de F\'{i}sica d'Altes Energies, Universitat Aut\`{o}noma
de Barcelona, E-08193 Bellaterra (Barcelona), Spain$^{7}$}
\end{center}\end{sloppypar}
\vspace{2mm}
\begin{sloppypar}
\noindent
A.~Colaleo,
D.~Creanza,
M.~de~Palma,
G.~Iaselli,
G.~Maggi,
M.~Maggi,$^{1}$
S.~Nuzzo,
A.~Ranieri,
G.~Raso,$^{23}$
F.~Ruggieri,
G.~Selvaggi,
L.~Silvestris,
P.~Tempesta,
A.~Tricomi,$^{3}$
G.~Zito
\nopagebreak
\begin{center}
\parbox{15.5cm}{\sl\samepage
Dipartimento di Fisica, INFN Sezione di Bari, I-70126
Bari, Italy}
\end{center}\end{sloppypar}
\vspace{2mm}
\begin{sloppypar}
\noindent
X.~Huang,
J.~Lin,
Q. Ouyang,
T.~Wang,
Y.~Xie,
R.~Xu,
S.~Xue,
J.~Zhang,
L.~Zhang,
W.~Zhao
\nopagebreak
\begin{center}
\parbox{15.5cm}{\sl\samepage
Institute of High Energy Physics, Academia Sinica, Beijing, The People's
Republic of China$^{8}$}
\end{center}\end{sloppypar}
\vspace{2mm}
\begin{sloppypar}
\noindent
D.~Abbaneo,
P.~Azzurri,
G.~Boix,$^{6}$
O.~Buchm\"uller,
M.~Cattaneo,
F.~Cerutti,
B.~Clerbaux,
G.~Dissertori,
H.~Drevermann,
R.W.~Forty,
M.~Frank,
T.C.~Greening,
J.B.~Hansen,
J.~Harvey,
P.~Janot,
B.~Jost,
M.~Kado,
P.~Mato,
A.~Moutoussi,
F.~Ranjard,
L.~Rolandi,
D.~Schlatter,
M.~Schmitt,$^{27}$
O.~Schneider,$^{2}$
P.~Spagnolo,
W.~Tejessy,
F.~Teubert,
E.~Tournefier,$^{25}$
J.~Ward,
A.E.~Wright
\nopagebreak
\begin{center}
\parbox{15.5cm}{\sl\samepage
European Laboratory for Particle Physics (CERN), CH-1211 Geneva 23,
Switzerland}
\end{center}\end{sloppypar}
\vspace{2mm}
\begin{sloppypar}
\noindent
Z.~Ajaltouni,
F.~Badaud,
A.~Falvard,$^{22}$
P.~Gay,
P.~Henrard,
J.~Jousset,
B.~Michel,
S.~Monteil,
\mbox{J-C.~Montret},
D.~Pallin,
P.~Perret,
F.~Podlyski
\nopagebreak
\begin{center}
\parbox{15.5cm}{\sl\samepage
Laboratoire de Physique Corpusculaire, Universit\'e Blaise Pascal,
IN$^{2}$P$^{3}$-CNRS, Clermont-Ferrand, F-63177 Aubi\`{e}re, France}
\end{center}\end{sloppypar}
\vspace{2mm}
\begin{sloppypar}
\noindent
J.D.~Hansen,
J.R.~Hansen,
P.H.~Hansen,
B.S.~Nilsson,
A.~W\"a\"an\"anen
\nopagebreak
\begin{center}
\parbox{15.5cm}{\sl\samepage
Niels Bohr Institute, DK-2100 Copenhagen, Denmark$^{9}$}
\end{center}\end{sloppypar}
\vspace{2mm}
\begin{sloppypar}
\noindent
G.~Daskalakis,
A.~Kyriakis,
C.~Markou,
E.~Simopoulou,
A.~Vayaki
\nopagebreak
\begin{center}
\parbox{15.5cm}{\sl\samepage
Nuclear Research Center Demokritos (NRCD), GR-15310 Attiki, Greece}
\end{center}\end{sloppypar}
\vspace{2mm}
\begin{sloppypar}
\noindent
A.~Blondel,$^{12}$
G.~Bonneaud,
\mbox{J.-C.~Brient},
A.~Roug\'{e},
M.~Rumpf,
M.~Swynghedauw,
M.~Verderi,
\linebreak
H.~Videau
\nopagebreak
\begin{center}
\parbox{15.5cm}{\sl\samepage
Laboratoire de Physique Nucl\'eaire et des Hautes Energies, Ecole
Polytechnique, IN$^{2}$P$^{3}$-CNRS, \mbox{F-91128} Palaiseau Cedex, France}
\end{center}\end{sloppypar}
\vspace{2mm}
\begin{sloppypar}
\noindent
E.~Focardi,
G.~Parrini,
K.~Zachariadou
\nopagebreak
\begin{center}
\parbox{15.5cm}{\sl\samepage
Dipartimento di Fisica, Universit\`a di Firenze, INFN Sezione di Firenze,
I-50125 Firenze, Italy}
\end{center}\end{sloppypar}
\vspace{2mm}
\begin{sloppypar}
\noindent
A.~Antonelli,
M.~Antonelli,
G.~Bencivenni,
G.~Bologna,$^{4}$
F.~Bossi,
P.~Campana,
G.~Capon,
V.~Chiarella,
P.~Laurelli,
G.~Mannocchi,$^{5}$
F.~Murtas,
G.P.~Murtas,
L.~Passalacqua,
\mbox{M.~Pepe-Altarelli}$^{24}$
\nopagebreak
\begin{center}
\parbox{15.5cm}{\sl\samepage
Laboratori Nazionali dell'INFN (LNF-INFN), I-00044 Frascati, Italy}
\end{center}\end{sloppypar}
\vspace{2mm}
\begin{sloppypar}
\noindent
A.W. Halley,
J.G.~Lynch,
P.~Negus,
V.~O'Shea,
C.~Raine,
A.S.~Thompson
\nopagebreak
\begin{center}
\parbox{15.5cm}{\sl\samepage
Department of Physics and Astronomy, University of Glasgow, Glasgow G12
8QQ,United Kingdom$^{10}$}
\end{center}\end{sloppypar}
\vspace{2mm}
\begin{sloppypar}
\noindent
S.~Wasserbaech
\nopagebreak
\begin{center}
\parbox{15.5cm}{\sl\samepage
Department of Physics, Haverford College, Haverford, PA 19041-1392, U.S.A.}
\end{center}\end{sloppypar}
\vspace{2mm}
\pagebreak
\begin{sloppypar}
\noindent
R.~Cavanaugh,
S.~Dhamotharan,
C.~Geweniger,
P.~Hanke,
G.~Hansper,
V.~Hepp,
E.E.~Kluge,
A.~Putzer,
J.~Sommer,
K.~Tittel,
S.~Werner,$^{19}$
M.~Wunsch$^{19}$
\nopagebreak
\begin{center}
\parbox{15.5cm}{\sl\samepage
Kirchhoff-Institut f\"ur Physik, Universit\"at Heidelberg, D-69120
Heidelberg, Germany$^{16}$}
\end{center}\end{sloppypar}
\vspace{2mm}
\begin{sloppypar}
\noindent
R.~Beuselinck,
D.M.~Binnie,
W.~Cameron,
P.J.~Dornan,
M.~Girone,$^{1}$
N.~Marinelli,
J.K.~Sedgbeer,
J.C.~Thompson$^{14}$
\nopagebreak
\begin{center}
\parbox{15.5cm}{\sl\samepage
Department of Physics, Imperial College, London SW7 2BZ,
United Kingdom$^{10}$}
\end{center}\end{sloppypar}
\vspace{2mm}
\begin{sloppypar}
\noindent
V.M.~Ghete,
P.~Girtler,
E.~Kneringer,
D.~Kuhn,
G.~Rudolph
\nopagebreak
\begin{center}
\parbox{15.5cm}{\sl\samepage
Institut f\"ur Experimentalphysik, Universit\"at Innsbruck, A-6020
Innsbruck, Austria$^{18}$}
\end{center}\end{sloppypar}
\vspace{2mm}
\begin{sloppypar}
\noindent
E.~Bouhova-Thacker,
C.K.~Bowdery,
A.J.~Finch,
F.~Foster,
G.~Hughes,
R.W.L.~Jones,$^{1}$
M.R.~Pearson,
N.A.~Robertson
\nopagebreak
\begin{center}
\parbox{15.5cm}{\sl\samepage
Department of Physics, University of Lancaster, Lancaster LA1 4YB,
United Kingdom$^{10}$}
\end{center}\end{sloppypar}
\vspace{2mm}
\begin{sloppypar}
\noindent
I.~Giehl,
K.~Jakobs,
K.~Kleinknecht,
G.~Quast,
B.~Renk,
E.~Rohne,
\mbox{H.-G.~Sander},
H.~Wachsmuth,
C.~Zeitnitz
\nopagebreak
\begin{center}
\parbox{15.5cm}{\sl\samepage
Institut f\"ur Physik, Universit\"at Mainz, D-55099 Mainz, Germany$^{16}$}
\end{center}\end{sloppypar}
\vspace{2mm}
\begin{sloppypar}
\noindent
A.~Bonissent,
J.~Carr,
P.~Coyle,
O.~Leroy,
P.~Payre,
D.~Rousseau,
M.~Talby
\nopagebreak
\begin{center}
\parbox{15.5cm}{\sl\samepage
Centre de Physique des Particules, Universit\'e de la M\'editerran\'ee,
IN$^{2}$P$^{3}$-CNRS, F-13288 Marseille, France}
\end{center}\end{sloppypar}
\vspace{2mm}
\begin{sloppypar}
\noindent
M.~Aleppo,
F.~Ragusa
\nopagebreak
\begin{center}
\parbox{15.5cm}{\sl\samepage
Dipartimento di Fisica, Universit\`a di Milano e INFN Sezione di Milano,
I-20133 Milano, Italy}
\end{center}\end{sloppypar}
\vspace{2mm}
\begin{sloppypar}
\noindent
A.~David,
H.~Dietl,
G.~Ganis,$^{26}$
K.~H\"uttmann,
G.~L\"utjens,
C.~Mannert,
W.~M\"anner,
\mbox{H.-G.~Moser},
R.~Settles,$^{1}$
H.~Stenzel,
W.~Wiedenmann,
G.~Wolf
\nopagebreak
\begin{center}
\parbox{15.5cm}{\sl\samepage
Max-Planck-Institut f\"ur Physik, Werner-Heisenberg-Institut,
D-80805 M\"unchen, Germany\footnotemark[16]}
\end{center}\end{sloppypar}
\vspace{2mm}
\begin{sloppypar}
\noindent
J.~Boucrot,$^{1}$
O.~Callot,
S.~Chen,
M.~Davier,
L.~Duflot,
\mbox{J.-F.~Grivaz},
Ph.~Heusse,
A.~Jacholkowska,$^{22}$
J.~Lefran\c{c}ois,
I.~Nikolic,
\mbox{J.-J.~Veillet},
I.~Videau,
C.~Yuan
\nopagebreak
\begin{center}
\parbox{15.5cm}{\sl\samepage
Laboratoire de l'Acc\'el\'erateur Lin\'eaire, Universit\'e de Paris-Sud,
IN$^{2}$P$^{3}$-CNRS, F-91898 Orsay Cedex, France}
\end{center}\end{sloppypar}
\vspace{2mm}
\begin{sloppypar}
\noindent
G.~Bagliesi,
T.~Boccali,
G.~Calderini,
V.~Ciulli,
L.~Fo\`{a},
A.~Giassi,
F.~Ligabue,
A.~Messineo,
F.~Palla,
G.~Sanguinetti,
A.~Sciab\`a,
G.~Sguazzoni,
R.~Tenchini,$^{1}$
A.~Venturi,
P.G.~Verdini
\samepage
\begin{center}
\parbox{15.5cm}{\sl\samepage
Dipartimento di Fisica dell'Universit\`a, INFN Sezione di Pisa,
e Scuola Normale Superiore, I-56010 Pisa, Italy}
\end{center}\end{sloppypar}
\vspace{2mm}
\begin{sloppypar}
\noindent
\mbox{G.A.~Blair},
\mbox{G.~Cowan},
\mbox{M.G.~Green},
\mbox{T.~Medcalf},
\mbox{J.A.~Strong},
\mbox{P.~Teixeira-Dias},\\
\mbox{J.H.~von~Wimmersperg-Toeller}
\nopagebreak
\begin{center}
\parbox{15.5cm}{\sl\samepage
Department of Physics, Royal Holloway \& Bedford New College,
University of London, Egham, Surrey TW20 OEX, United Kingdom$^{10}$}
\end{center}\end{sloppypar}
\vspace{2mm}
\begin{sloppypar}
\noindent
R.W.~Clifft,
T.R.~Edgecock,
P.R.~Norton,
I.R.~Tomalin
\nopagebreak
\begin{center}
\parbox{15.5cm}{\sl\samepage
Particle Physics Dept., Rutherford Appleton Laboratory,
Chilton, Didcot, Oxon OX11 OQX, United Kingdom$^{10}$}
\end{center}\end{sloppypar}
\vspace{2mm}
\begin{sloppypar}
\noindent
\mbox{B.~Bloch-Devaux},$^{1}$
P.~Colas,
S.~Emery,
W.~Kozanecki,
E.~Lan\c{c}on,
\mbox{M.-C.~Lemaire},
E.~Locci,
P.~Perez,
J.~Rander,
\mbox{J.-F.~Renardy},
A.~Roussarie,
\mbox{J.-P.~Schuller},
J.~Schwindling,
A.~Trabelsi,$^{21}$
B.~Vallage
\nopagebreak
\begin{center}
\parbox{15.5cm}{\sl\samepage
CEA, DAPNIA/Service de Physique des Particules,
CE-Saclay, F-91191 Gif-sur-Yvette Cedex, France$^{17}$}
\end{center}\end{sloppypar}
\vspace{2mm}
\begin{sloppypar}
\noindent
N.~Konstantinidis,
A.M.~Litke,
G.~Taylor
\nopagebreak
\begin{center}
\parbox{15.5cm}{\sl\samepage
Institute for Particle Physics, University of California at
Santa Cruz, Santa Cruz, CA 95064, USA$^{13}$}
\end{center}\end{sloppypar}
\vspace{2mm}
\begin{sloppypar}
\noindent
C.N.~Booth,
S.~Cartwright,
F.~Combley,
M.~Lehto,
L.F.~Thompson
\nopagebreak
\begin{center}
\parbox{15.5cm}{\sl\samepage
Department of Physics, University of Sheffield, Sheffield S3 7RH,
United Kingdom$^{10}$}
\end{center}\end{sloppypar}
\vspace{2mm}
\begin{sloppypar}
\noindent
K.~Affholderbach,
A.~B\"ohrer,
S.~Brandt,
C.~Grupen,
A.~Misiejuk,
A.~Ngac,
G.~Prange,
U.~Sieler
\nopagebreak
\begin{center}
\parbox{15.5cm}{\sl\samepage
Fachbereich Physik, Universit\"at Siegen, D-57068 Siegen,
 Germany$^{16}$}
\end{center}\end{sloppypar}
\vspace{2mm}
\begin{sloppypar}
\noindent
G.~Giannini
\nopagebreak
\begin{center}
\parbox{15.5cm}{\sl\samepage
Dipartimento di Fisica, Universit\`a di Trieste e INFN Sezione di Trieste,
I-34127 Trieste, Italy}
\end{center}\end{sloppypar}
\vspace{2mm}
\begin{sloppypar}
\noindent
J.~Rothberg
\nopagebreak
\begin{center}
\parbox{15.5cm}{\sl\samepage
Experimental Elementary Particle Physics, University of Washington, Seattle, 
WA 98195 U.S.A.}
\end{center}\end{sloppypar}
\vspace{2mm}
\begin{sloppypar}
\noindent
S.R.~Armstrong,
K.~Cranmer,
P.~Elmer,
D.P.S.~Ferguson,
Y.~Gao,$^{20}$
S.~Gonz\'{a}lez,
O.J.~Hayes,
H.~Hu,
S.~Jin,
J.~Kile,
P.A.~McNamara III,
J.~Nielsen,
W.~Orejudos,
Y.B.~Pan,
Y.~Saadi,
I.J.~Scott,
J.~Walsh,
Sau~Lan~Wu,
X.~Wu,
G.~Zobernig
\nopagebreak
\begin{center}
\parbox{15.5cm}{\sl\samepage
Department of Physics, University of Wisconsin, Madison, WI 53706,
USA$^{11}$}
\end{center}\end{sloppypar}
}
\footnotetext[1]{Also at CERN, 1211 Geneva 23, Switzerland.}
\footnotetext[2]{Now at Universit\'e de Lausanne, 1015 Lausanne, Switzerland.}
\footnotetext[3]{Also at Dipartimento di Fisica di Catania and INFN Sezione di
 Catania, 95129 Catania, Italy.}
\footnotetext[4]{Deceased.}
\footnotetext[5]{Also Istituto di Cosmo-Geofisica del C.N.R., Torino,
Italy.}
\footnotetext[6]{Supported by the Commission of the European Communities,
contract ERBFMBICT982894.}
\footnotetext[7]{Supported by CICYT, Spain.}
\footnotetext[8]{Supported by the National Science Foundation of China.}
\footnotetext[9]{Supported by the Danish Natural Science Research Council.}
\footnotetext[10]{Supported by the UK Particle Physics and Astronomy Research
Council.}
\footnotetext[11]{Supported by the US Department of Energy, grant
DE-FG0295-ER40896.}
\footnotetext[12]{Now at Departement de Physique Corpusculaire, Universit\'e de
Gen\`eve, 1211 Gen\`eve 4, Switzerland.}
\footnotetext[13]{Supported by the US Department of Energy,
grant DE-FG03-92ER40689.}
\footnotetext[14]{Also at Rutherford Appleton Laboratory, Chilton, Didcot, UK.}
\footnotetext[15]{Permanent address: Universitat de Barcelona, 08208 Barcelona,
Spain.}
\footnotetext[16]{Supported by the Bundesministerium f\"ur Bildung,
Wissenschaft, Forschung und Technologie, Germany.}
\footnotetext[17]{Supported by the Direction des Sciences de la
Mati\`ere, C.E.A.}
\footnotetext[18]{Supported by the Austrian Ministry for Science and Transport.}
\footnotetext[19]{Now at SAP AG, 69185 Walldorf, Germany.}
\footnotetext[20]{Also at Department of Physics, Tsinghua University, Beijing, The People's Republic of China.}
\footnotetext[21]{Now at D\'epartement de Physique, Facult\'e des Sciences de Tunis, 1060 Le Belv\'ed\`ere, Tunisia.}
\footnotetext[22]{Now at Groupe d' Astroparticules de Montpellier, Universit\'e de Montpellier II, 34095 Montpellier, France.}
\footnotetext[23]{Also at Dipartimento di Fisica e Tecnologie Relative, Universit\`a di Palermo, Palermo, Italy.}
\footnotetext[24]{Now at CERN, 1211 Geneva 23, Switzerland.}
\footnotetext[25]{Now at ISN, Institut des Sciences Nucl\'eaires, 53 Av. des Martyrs, 38026 Grenoble, France.} 
\footnotetext[26]{Now at INFN Sezione di Roma II, Dipartimento di Fisica, Universit\'a di Roma Tor Vergata, 00133 Roma, Italy.} 
\footnotetext[27]{Now at Harvard University, Cambridge, MA 02138, U.S.A.}
%
\setlength{\parskip}{\saveparskip}
\setlength{\textheight}{\savetextheight}
\setlength{\topmargin}{\savetopmargin}
\setlength{\textwidth}{\savetextwidth}
\setlength{\oddsidemargin}{\saveoddsidemargin}
\setlength{\topsep}{\savetopsep}
\normalsize
\newpage
\pagestyle{plain}
\setcounter{page}{1}
\newpage

\section{Introduction}
Owing to  parity violation in Z production and decay, the $\tau$ leptons
produced in the $e^+e^-\to\mbox{Z}\to\tau^+\tau^-$ reaction  are polarised.
Except for very small ${\cal O}(m_\tau^2/m_Z^2)$ corrections, the
$\tau^+$ and $\tau^-$ have opposite helicities, hence opposite
polarisations.
 In the present paper,  
$P_\tau$ is defined as the $\tau^-$ polarisation. 

The $P_\tau$ dependence   on the angle $\theta$ formed by the $\tau^-$ direction
and the $e^-$ beam is given, in the improved Born approximation at the
Z peak, by 
\begin{equation}\label{eq:p-theta} 
P_\tau(\cos\theta)=-\frac{{\cal A}_\tau(1+\cos^2\theta)+{\cal
    A}_e(2\cos\theta)}{(1+\cos^2\theta)+\frac{4}{3}{\cal A}_{fb}(2\cos\theta)},
\end{equation}
where ${\cal A}_{fb}$ is the forward-backward charge asymmetry of $\tau$
production and
\begin{equation}
{\cal A}_l\equiv 2g_V^lg_A^l/[(g_V^l)^2+(g_A^l)^2].
\end{equation}
In the standard model, the vector ($g_V^l$) and axial vector ($g_A^l$)
couplings of the Z to lepton $l$ are independent of the lepton flavour
and related to the effective weak mixing angle by
\begin{equation}
g^l_V/g^l_A=1-4\sin^2\theta^{eff}_W.
\end{equation}
Therefore, the measurement of $P_\tau(\cos\theta)$ allows a test of the $e$-$\tau$
universality prediction ${\cal A}_\tau={\cal A}_e$ and, assuming
universality, gives a determination of the weak mixing angle.

The analyses presented here use the complete set of data (160 pb$^{-1}$) accumulated
by the ALEPH detector at LEP~I from 1990 to 1995. 
They supersede the previously published results~[1-3]
 
The $\tau$ polarisation is extracted
from five decay channels which  amount  to a total branching ratio of
about 90\%, namely $\tau\to e\nu\bar\nu$, $\tau\to
\mu\nu\bar\nu$, $\tau\to\pi\nu$, $\tau\to\rho\nu$, $\tau \to a_1\nu$.
 The channels $\tau\to K\nu$ and $\tau\to K\pi^0\nu$ 
are included in
$\tau\to\pi\nu$  and $\tau\to\rho\nu$ respectively, and the $a_1$ decay is observed 
as both $a_1\to\pi^+\pi^-\pi^\pm$ ($3h$) and 
$a_1\to\pi^0\pi^0\pi^\pm$ ($h2\pi^0$).

The methods employed for the $P_\tau$ measurement in each channel
are described in the following section.  At each energy, a  fit of 
Eq.~\ref{eq:p-theta}
allows to determine uncorrected parameters  ${A}_\tau$ and ${ A}_e$. 
Final corrections are needed to get the effective couplings at the Z,
whose combinations $2g_V^lg_A^l/[(g_V^l)^2+(g_A^l)^2]$ are written ${\cal A}_l$ to distinguish them from
the uncorrected $A_l$. They  are
computed with the help of the ZFITTER program~\cite{zfit}.

\section{Methods for \boldmath$\tau$ polarisation measurement}
\label{sec:method}
\subsection{Information from a polarised {\boldmath$\tau$} decay}
The  general method to measure the $\tau$ polarisation in an optimal
way~\cite{optimal} takes advantage of the linear dependence on the polarisation of the
distribution of decay products.

In each decay channel, the decay products are described by a set of
$n$ observables {\boldmath$\xi$} and their distribution reads
\begin{equation}\label{eq:distxi}
W(\mbox{\boldmath$\xi$})=f(\mbox{\boldmath$\xi$})+P_\tau g(\mbox{\boldmath$\xi$}),
\end{equation} 
with the normalization and positivity conditions 
$\int\! f(\mbox{\boldmath$\xi$})d^n\mbox{\boldmath$\xi$}=1$, $\int\!
g(\mbox{\boldmath$\xi$})d^n\mbox{\boldmath$\xi$}=0$, $f\ge 0$ and $|g|\le f$.

With the optimal variable for the channel defined by
\begin{equation}
\omega=\frac{g(\mbox{\boldmath$\xi$})}{f(\mbox{\boldmath$\xi$})},
\end{equation}
the distribution in Eq.~\ref{eq:distxi}
can be cast,  with no loss of information, in the general reduced form
\begin{equation}\label{eq:distomega}
\hat{W}(\omega)=\hat{f}(\omega)\left[1+P_\tau\,\omega\right]=
\frac{1}{2}\left[(1+P_\tau)\hat{W}^+(\omega)+(1-P_\tau)\hat{W}^-(\omega)\right],
\end{equation}
where $W^+$ and $W^-$ are the distributions for positive and negative helicity respectively.

\begin{table}[b]
\caption[sens]{\label{table:sens}\protect\small Ideal sensitivities for the polarisation measurement in the $\tau$ 
decay channels without and  with the $\tau$ direction $\vec\tau$~\protect\cite{optimal}. 
The sensitivity is defined as $S=1/\Delta\sqrt{N}$, where $\Delta$ is 
the statistical error  expected for a sample of $N$ events.
}\vspace{.5cm}
\centerline{
\begin{tabular}{lcc}
\hline
Channel&\multicolumn{2}{c}{Sensitivity}\\
&without $\vec\tau$&with $\vec\tau$\\\hline
$\pi\nu$&0.58&0.58\\
$\rho\nu$&0.49&0.58\\
$a_1\nu$&0.45&0.58\\
$l\nu\bar\nu$&0.22&$-$\\\hline
\end{tabular}
}
\end{table}

For events in which both $\tau$'s decay into hadrons, it is possible to
get information on the line of flight of the two $\tau$'s~\cite{3dip}. It is 
 shown in Ref.~\cite{optimal} that, in an ideal measurement, using the knowledge
of the $\tau$ direction to construct the $\omega$ variable gives for
$\rho$ and $a_1$ decays the same sensitivity as that for $\pi$ decay (Table~\ref{table:sens}).

The polarisation is obtained by means of a fit of 
  $\hat{W}^+(\omega)$ and $\hat{W}^-(\omega)$ functions
 to the 
experimental $\hat W(\omega)$ distribution.
The $W^\pm$ functions are constructed from Monte Carlo events and normalized 
to the acceptances.
Non-tau background is taken into account  in the fit.

The validity of the standard model for the description of $\tau$
decays has been  checked elsewhere~\cite{michelparam} by the measurement
of the $\tau$ decay parameters.

In  the standard model, which is assumed here, the decay distributions for the 
 $\tau\to l\nu\bar\nu$,  $\tau\to\pi\nu$, and  $\tau\to\rho\nu$
 channels are completely determined by Lorentz invariance and there is
 no hadronic decay model dependence in the definition of $\omega$. Furthermore,
 because the $\tau$ polarisation always appears  through the product
 $\chi P$, where $\chi$ is the handedness of the neutrino~\cite{polarb},
 the definition of the $\omega$ variable is the
 same for $\tau^+$ and $\tau^-$, as are its distributions when
 expressed in terms of the $\tau^-$ polarisation $P_\tau=P_{\tau^-}=-P_{\tau^+}$.

The situation is more intricate in the case of the $a_1$ decay. There,
in addition to the decay angles, the two-pion masses are required to
describe the hadronic system. Their dependence is embodied in four
``structure functions'', $W_A$, $W_C$, $W_D$, and $W_E$~\cite{kuhn},
whose computation requires a model of the $a_1$ decay. The model~\cite{kuhn} that
is  used in the present study assumes a sequential $a_1\to\rho\pi$ decay.
The sign of the $W_E$ function depends on the charge of the $\tau$
so that the expression of the $\omega$ variable is not exactly the
same for $\tau^+$ and $\tau^-$.

For the $\tau\to l\nu\bar\nu$ and $\tau\to\pi\nu$ decay modes, the
optimal variable is the ratio $x=E/E_{\mathrm{beam}}$ 
of the energy of the charged particle ($l$ or
$\pi$) to the beam energy.

For the other channels, the $\tau$ decay is described by one angle and
the mass of the hadronic system.  The subsequent decay of the hadron
is described by two angles in the case of the $\rho$, and by two effective
masses and three Euler angles in the case of the $a_1$.
If the $\tau$ direction is not available, one of the angles used to
describe the hadron decay is not measurable  and the analytical form
of the decay distributions is made more complex by the Wigner rotation
between the laboratory and $\tau$ rest frame helicity axes.
The expressions used for the construction of the $\omega$
variables can be found in Refs.~\cite{polarb} and~\cite{kuhn}.

Although all the information relevant to the polarisation measurement
is contained in the sole $\omega$ variable, the distributions of the
other parameters are useful when checking the quality of the
simulations and the understanding of energy calibrations. For example,
the two angles used to describe the $\tau\to\rho\nu$ decay  when the
$\tau$ direction is not available are~\cite{polarb}
$\cos\psi_\tau\propto (2(E_{\pi^{\pm}}+E_{\pi^0})/E_{\mathrm{beam}}-1)$ and
$\cos\psi_\rho\propto (E_{\pi^{\pm}}-E_{\pi^0})/
(E_{\pi^{\pm}}+E_{\pi^0})$. The comparison of their distributions
in data and Monte Carlo is   the best test of the quality of the measurement. 
\subsection{The analyses}

Two  complementary analyses  have been performed, which emphasize
differently the sensitivity of the estimators and the reduction of the
systematic uncertainties.

In the first one, 
the information from the opposite 
hemisphere of the event is used  for background rejection
 when studying a $\tau$ decay.
This approach, which was followed in previous analyses~[1-3], 
 is called hereafter the ``single $\tau$ method''. In this method, the information exploited
 to construct the $\omega$ variable comes  from only one of the
two hemispheres defined by the plane perpendicular to the thrust axis.

As mentioned in the previous section,
for events in which both $\tau$'s decay into hadrons it is possible to
get information on the line of flight of the two $\tau$'s and
construct an $\omega$ variable giving an improved sensitivity for the
$\tau\to\rho\nu$ and $\tau\to a_1\nu$ decays.
This more global approach is followed in the second analysis and is
called hereafter the ``$\tau$ direction
method''.    

These two  methods use the same set of data and many
common standard ALEPH tools. 
The two analyses have nevertheless  been performed independently and thus
are  described in detail in  Sections~\ref{section:singletau} 
and~\ref{sec:taudir}.
The combination of their results is presented in Section~\ref{section:final}.
\section{Detector and data sets}
\subsection{The ALEPH detector}\label{sec:calib}
The ALEPH detector and its performance are described elsewhere~\cite{detector,perfo}. 
The study of $\tau$ polarisation
requires a very special care
in the alignment and calibration procedures to master 
the differences between real and simulated data at the required level.

The charged track measurement rests on three elements, the Vertex Detector\break (VDET),
the Inner Tracking Chamber (ITC), and the Time Projection Chamber (TPC). 

The VDET,
a double-sided silicon strip detector, provides precise measurements 
close to the interaction point and, therefore,
plays an important role in determining  the $\tau$
direction.

The ITC is a drift chamber with wires parallel to the beam and a  short drift time.
Besides providing a trigger, it is  an efficient 
tool against cosmic rays.
Because its complete drift time is about 200~ns, to be compared to the
 11~$\mu$s interval between bunch collisions,  its sensitive time window is small.
Therefore, there are few if any  ITC hits in most   triggers on cosmic rays. 
If present, they are not reconstructed with
the correct timing and the reconstructed track is distorted leading to
erroneous distances of approach to the interaction point.

The TPC is a very large volume drift chamber.
Together with the VDET and ITC, it brings a resolution on  the track momentum which
reaches $ 6\times 10^{-4}p$  ($p$ in GeV), but to get this accuracy a precise 
 calibration
is needed. For these analyses, the distortions from the TPC are derived 
from the study of nonradiative $\mu^+\mu^-$ events in $\phi$ and $\cos \theta$ bins, where $\phi$
and $\theta$ are the azimuthal and polar angles of the $\mu^-$ with respect to the $e^-$ beam.
The  departures from the corresponding observations on Monte Carlo are
corrected for, considering that the distortions  on the sum of 
positive and negative track momenta are due to field effects and that
the distortions  on the difference are due to sagitta measurement
effects, i.e. alignment effects. 

Apart from momentum and angle measurements, the tracking system is used to measure
the impact parameter relative to the beam 
axis, denoted $d_0$.

The tracking detectors are surrounded by  the electromagnetic calorimeter (ECAL).
All these elements are installed inside a superconducting coil which provides a
1.5 ~T axial field. The return yoke of the magnetic field is instrumented  to form 
a hadron calorimeter (HCAL).

The ECAL is a lead/proportional wire chamber sampling device of 22 radiation 
length thickness. The insensitive regions between its modules (cracks) 
represent 2\% of the barrel and 6\% 
of the endcap areas.
 The anode wire signal is read plane by plane.
 The cathodes are divided in pads, making 74,000  towers pointing 
to the interaction point. These towers are read out in three sections
in depth (``storeys''or ``stacks'').    
The use of 
 the energies measured with the pad towers is mandatory whenever  a precise
direction 
is needed  but, in the case of $\tau$
 events, the energy collected  on the wires
may be  advantageous. It does not suffer from disconnected channels 
ensuring that the total energy left in each module is properly known.
This comes from the fact that the lack  of a
plane is, to first order, corrected for by the energy calibration.
It has a very low level of noise (about 10~MeV for a complete module).
The energy resolution therefore does not suffer  from threshold effects and
is slightly better than the resolution of the pad signal.
Another interesting feature is that the 45 plane signals
 provide a complete depth profile of the energy deposit. 

The calibration of the ECAL was performed independently in the two
analyses presented below but the salient features are common and
the technique is the same for pad and wire signals.
First a correction map is applied which takes care of local gain variations
related to mechanical distortions, then gain dependencies as a function
of $\phi$ and $\theta$ are applied to each module, finally the modules are 
inter-calibrated and normalized to the response observed in Monte Carlo.
A problem in this procedure comes from the nonlinearity with energy
of the calorimeter response which reaches 4\% at 50~GeV.
Bhabha events provide a calibration at the beam energy but
not a measurement of the nonlinearity which
can be parametrized in first order by a constant $\alpha$ in the expression
giving the true energy as a function of the measured one, 
$E_t = E_m (1 + \alpha E_m)$.
To extract $\alpha$, clean electrons are selected and the $E/p$ ratio is compared,
as a function of energy, between real data and  Monte Carlo where the saturation effects
have not been simulated. This is done independently for barrel and endcaps.

The HCAL is made of 23 layers of streamer tubes inserted between 
5~cm thick iron plates. The individual tube hits are recorded
digitally and the energy  is read out in 4,788 projective towers.
This calorimeter is used primarily for discriminating between pions
and muons.
 This does not require a specific energy calibration. 
For the current purpose, use is made 
essentially of the digital pattern provided by the firing of the 
streamer tubes. 
In particular, the pattern recognition 
described in Ref.~\cite{polarb} is used to get variables describing
the shape of the shower.

The HCAL is surrounded by two double layers of streamer tubes, the muon chambers, 
providing additional information for muon identification.

The luminosity calorimeters are not utilized in the 
analyses presented here.
\subsection{The data and Monte Carlo samples}
 The data used in these analyses were accumulated with the ALEPH detector
during the years 1990 to 1995 at centre-of-mass energies close to the Z mass.
This set, often referred to as the LEP~I data set,
corresponds to about 160~pb$^{-1}$ of  data.
 A detailed description of the 
 characteristics of each period can be found in Ref.~\cite{bib:aleph_ew}. 
For the present study no requirement on the
luminosity measurement quality is made.

Monte Carlo simulation samples were generated for the $\tau^{+}\tau^{-}$ events
as well as for the different backgrounds through the complete 
detector simulation chain. A sample of 
$1.9\times  10^{6}$ $\tau^+\tau^-$ events was generated using KORALZ~\cite{koral}. 
For each year and energy of data taking a sample of about six times
the real data sample was simulated with the same energy and the
effective detector geometry of the year.

 A $\mu^{+}\mu^{-}$  sample of about 4$\times10^{5}$ events was simulated using 
also KORALZ  for different years and energies.
 About 5$\times 10^{5}$ large angle Bhabha events were produced using 
 UNIBAB~\cite{unibab}, and some with BABAMC~\cite{babamc}, for 
different energies.
 For the two-photon  interaction processes,  
 the simulation used the generator PHOT02~\cite{phot02}. 
 Finally the four-fermion processes $\ell^{+}\ell^{-}{\rm V}$,
defined \cite{bib:aleph_ew} as a pair of leptons
 associated with a two particle system (V) of  lower mass, 
 were generated using FERMISV~\cite{bib:4ferm}.
\section{Single {\boldmath$\tau$} analysis}
\label{section:singletau}
\subsection{Introduction}
The aim of this analysis is to reduce 
the systematic uncertainties, even at the price of some
inefficiency. Data have been processed separately  for each year 
of LEP operation in order to take
 as much care as possible of  time-variations in the behaviour of the detector.

Particle and channel identification, as well as  non-$\tau$ background rejection, 
are achieved by means of  likelihood estimators, in order to optimize the efficiencies.
These estimators are constructed from Monte Carlo distributions of the relevant variables, 
but their efficiencies are directly measured on
selected samples of data, which are used also to 
study the properties of the remaining backgrounds.

Specific procedures have been designed for the $\gamma \gamma$ and cosmic ray
background rejection. 
They are
described in detail in the following.

To cross-check  the photon identification and reconstruction, an inclusive  one-prong hadronic analysis,
which infers the polarisation from the charged pion momentum distribution, has been
devised, in addition to the standard study of the previously listed channels.

\subsection{Tools}
The general tools applied in the analysis are presented here.
Additional information on their performances is given in  
Section~\ref{section:single_syst} dedicated to the study of the systematic uncertainties.
\subsubsection{Charged particle identification}\label{sec:chargid}
The charged particle identification procedure (PID)
 described in Ref.~\cite{bib:aleph_ew} is used
for the non-$\tau$ background rejection and for the
 decay channel selection.

Charged particles  are classified as 
 electron, muon, or  hadron (pions and  kaons are
 treated as a single class thereafter referred to as ``pions'').
 Since there is  negligible probability of  mistaking  electron and muon
($\sim\,$2.5$\times$10$^{-4}$),
 the PID is based on  two  likelihood  estimators which give
 the relative probability for a track 
   to be an electron versus a pion ($e/\pi$),
and  to be a muon versus a pion ($\mu/\pi$), respectively.
 The $\mu/\pi$ PID is restricted to
 charged tracks with momentum above 1.3 GeV/$c$.
 An identification efficiency
 of about 99\% with a rate of misidentification of
 about 1\% is achieved for the $\mu$'s in the previously defined region
and for the electrons inside a fiducial region which excludes about twice 
the size of the physical ECAL cracks.
A study of selected data samples \cite{bib:aleph_ew} has shown that
the efficiencies are reproduced by the Monte Carlo at the two per mil
level. The small resultant corrections are applied in the
polarisation measurements.
\subsubsection{Photon reconstruction}
The reconstruction of photons from clusters of cells in the
electromagnetic calorimeter is performed with the ALEPH standard 
 algorithm \cite{perfo} which searches for local energy maxima in the 
 ECAL pad clusters. Fluctuations of  a    shower 
 can generate 
``fake  photons''~\cite{hbr} which are artefacts of the clustering algorithm or
true photons produced by  secondary  interactions in the ECAL.

\begin{figure}[ht]
\centerline{
\epsfig{file=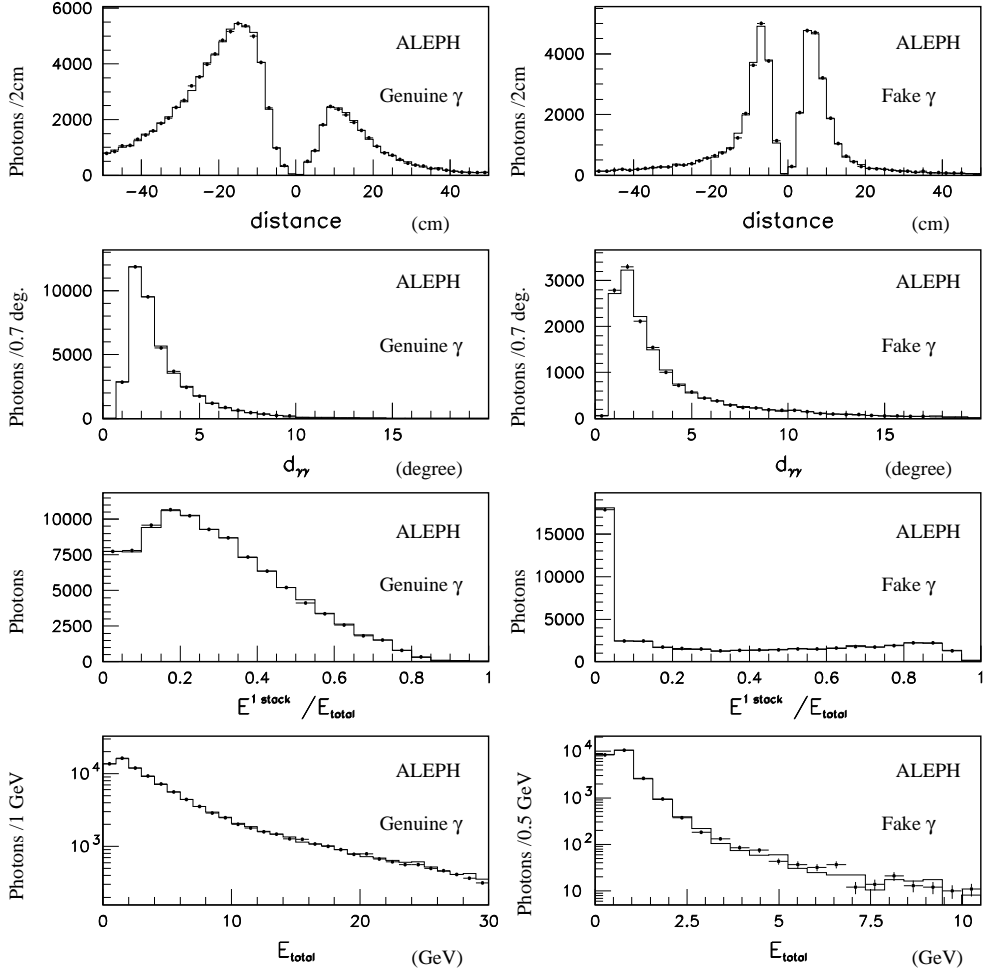,width=13.65cm
}}\mbox{~}\\[-1.5cm]
\caption[.]
{\protect\small Distributions of the variables used to define estimators for genuine and fake
photons. Data are represented by points while the histograms are the
predictions of the simulation. Discriminant variables are described in the text.
In each plot the distributions for data and Monte Carlo are normalized to 
the same number of photons. The distance  $d_{\gamma\gamma}$ \cite{hbr} is used in the 
second analysis only.}
\label{ref_photon}
\end{figure}

In order to reduce the number of such fake photons,
  an estimator of fakeness ({\it P$_{fak}$}) is built from the relevant parameters~\cite{hbr}:
the oriented distance to the closest charged track, the  fraction of the
  shower energy
deposited in the first ECAL stack, and the total ECAL shower energy.
The sign of the distance is computed depending of the position of the shower with respect
to the track bending in the $r-\phi$ projection. 
The reference distributions  are taken from Monte Carlo, with the
  ECAL barrel and  end-cap being handled independently. 
 A cut on the fakeness likelihood estimator  provides
 a partition of the photon candidate sample
 in two subsamples: the ``fake photon'' sample ({\it P$_{fak}$} $>$0.75), enriched in 
 fake photons and the ``genuine photon'' sample ({\it P$_{fak}$} $<$0.75), enriched in genuine
photons.
 The distributions of the variables used
 to construct the {\it P$_{fak}$} estimator  are shown  in 
Fig.~\ref{ref_photon}
 for each of the two subsamples of photon
 candidates.
 A good agreement between 
 data and $\tau^{+}\tau^{-}$ Monte Carlo can be observed.
\subsubsection{Non-{\boldmath$\tau$} background rejection}\label{sec:s-nontaubck}
Non-$\tau$ background in the selected samples has four main sources:
 Bhabha events, $Z\to\mu^+\mu^-$ decays,
 $Z\to q \bar{q}$ decays,  and $\gamma\gamma$ processes.
As stated before, 
likelihood estimators exploiting  the information of  a single hemisphere are used for the 
suppression of the backgrounds that arise from the first three categories.

 The Bhabha estimator (${\cal{E}}^{e^{+}e^{-}}$)  provides the likelihood
 for one hemisphere to be part of a Bhabha event
 rather than of a $\tau^+\tau^-$. It
  is  based on three variables: the  $e/\pi$ probability of the track,
 its momentum, and the ECAL wire energy.
 Specific reference distributions are used for tracks pointing to an ECAL crack.

 The $Z\to\mu^+\mu^-$ estimator (${\cal{E}}^{\mu^{+}\mu^{-}}$) is similar to the previous one
 with the  $\mu/\pi$ track probability  replacing
 the  $e/\pi$ probability.

  A third  estimator (${\cal{E}}^{q \bar{q}}$) tags a Z hadronic decay in one hemisphere.
 It makes use of  the opening angle between the
 charged tracks, the number of charged tracks, the number of
 objects reconstructed from the pattern of the energy deposited in the 
 calorimeters, and  the mass and  energy of the jet.
  It is very similar to the estimator
 used for the leptonic cross section measurement and is
  described in Refs.~\cite{bib:aleph_ew} and \cite{bib:etaomeg}.
 Different reference distributions are taken for each ECAL
 region (barrel, endcaps, and their overlap).

 The $\gamma \gamma$ background is rejected by cuts on  the three
 pertinent variables: visible
 energy,  missing transverse momentum, and  acollinearity of the
 two jets. The cut values  are optimized for each final state.
 
  Two less important backgrounds can also contribute:
  $\ell^{+}\ell^{-}{\rm V}$ and cosmic ray events.
The $\ell^{+}\ell^{-}{\rm V}$ events
are eliminated  by the same cuts that reject
 Bhabha, $Z\to\mu^+\mu^-$, and $Z\to q \bar{q}$ events.

 Cosmic rays can contaminate the
  $ \tau \to h  \nu  $ and 
 $\tau \to \mu 
 \nu\bar\nu $ channels.
   They appear as
  events with  two collinear tracks of the same momentum coming from the vertex region
and a low energy deposit in the electromagnetic calorimeter.
A first set of cuts is applied to select  them: ECAL energy
below 2.5 GeV, $|p_1|-|p_2|<5.0\,\hbox{GeV/$c$}$,
$\min(|d_0^1|,|d_0^2|)>0.1\,\hbox{cm}$, and $|d_0^1+d_0^2|<0.5\,\hbox{cm}$,
where $p_i$ and $d_0^i$ are respectively the momentum and signed impact parameter of the
track $i$. The short gating time of the ITC is then used \cite{lbr}.
 On the sample of cosmic ray candidates selected by the preceding cuts,
 the distribution of the  number of ITC hits is measured.
 It is found that 68\% of the events have
 no hits  and 88\% less than eight.
 Since  the number of ITC hits is related only to the arrival time of the
 particles,
 the comparison of its distributions in the two samples, cosmic ray
 candidates and $\tau$ decays,
is used first to estimate quantitatively the residual background in the $\tau$ samples. Finally, 
the events in the previously defined sample are rejected, along with events satisfying 
the same criteria on ECAL energy and momenta, and $\hbox{N}_{ITC}<8$.
\subsection{Channel selection}
The ALEPH standard leptonic pre-selection
described in Ref.~\cite{bib:aleph_ew}, which essentially requires a 
multiplicity of charged tracks between two and eight,
 is used first.
The different channels are then selected in the angular acceptance  
($|\cos\theta|<0.9$) with the help of
 the charged particle  and  photon identification.
\subsubsection{Leptonic decays}
\begin{figure}[ht]
    \mbox{\epsfig{file=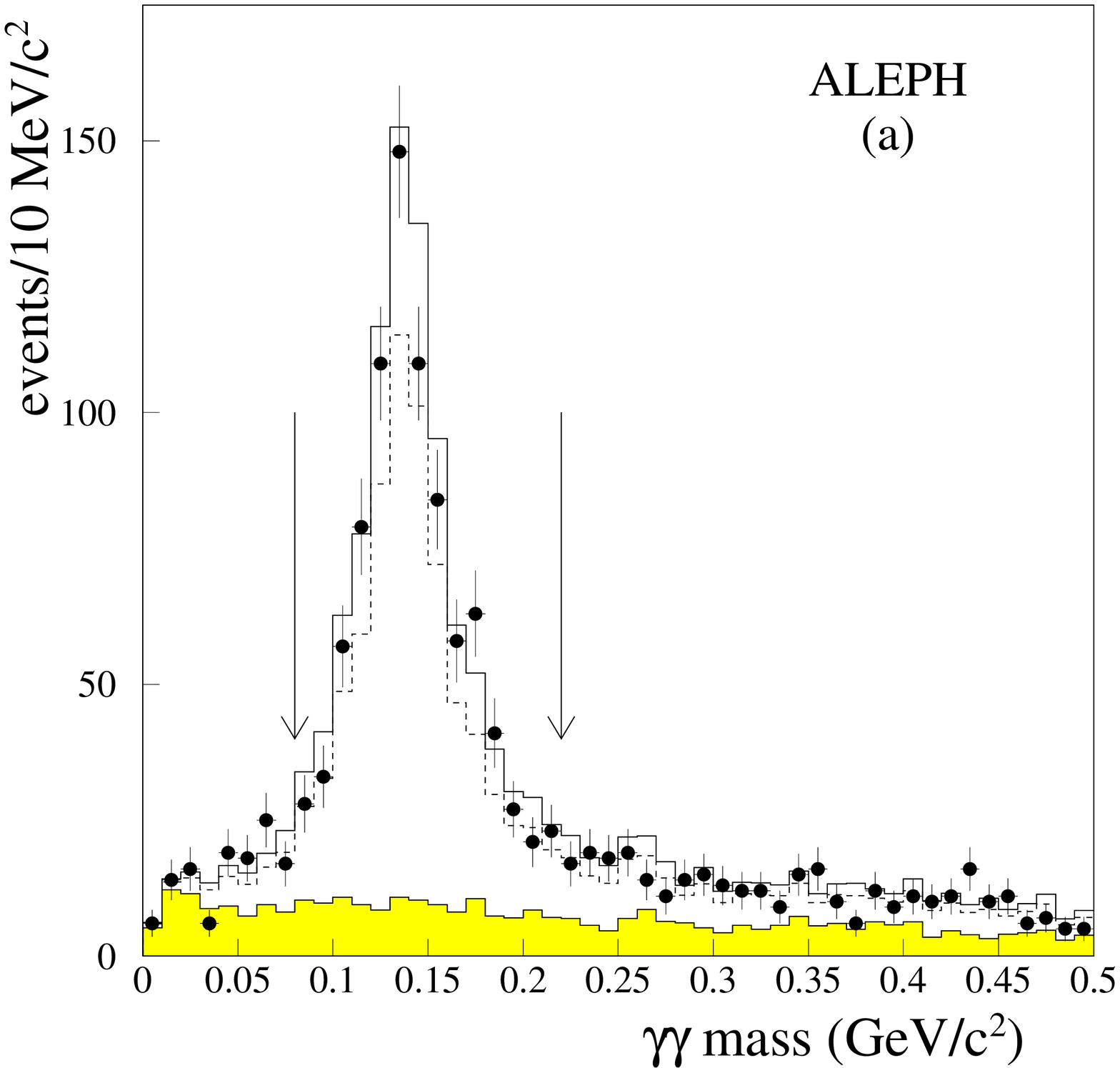,width=7.75cm}}\mbox{
      \epsfig{file=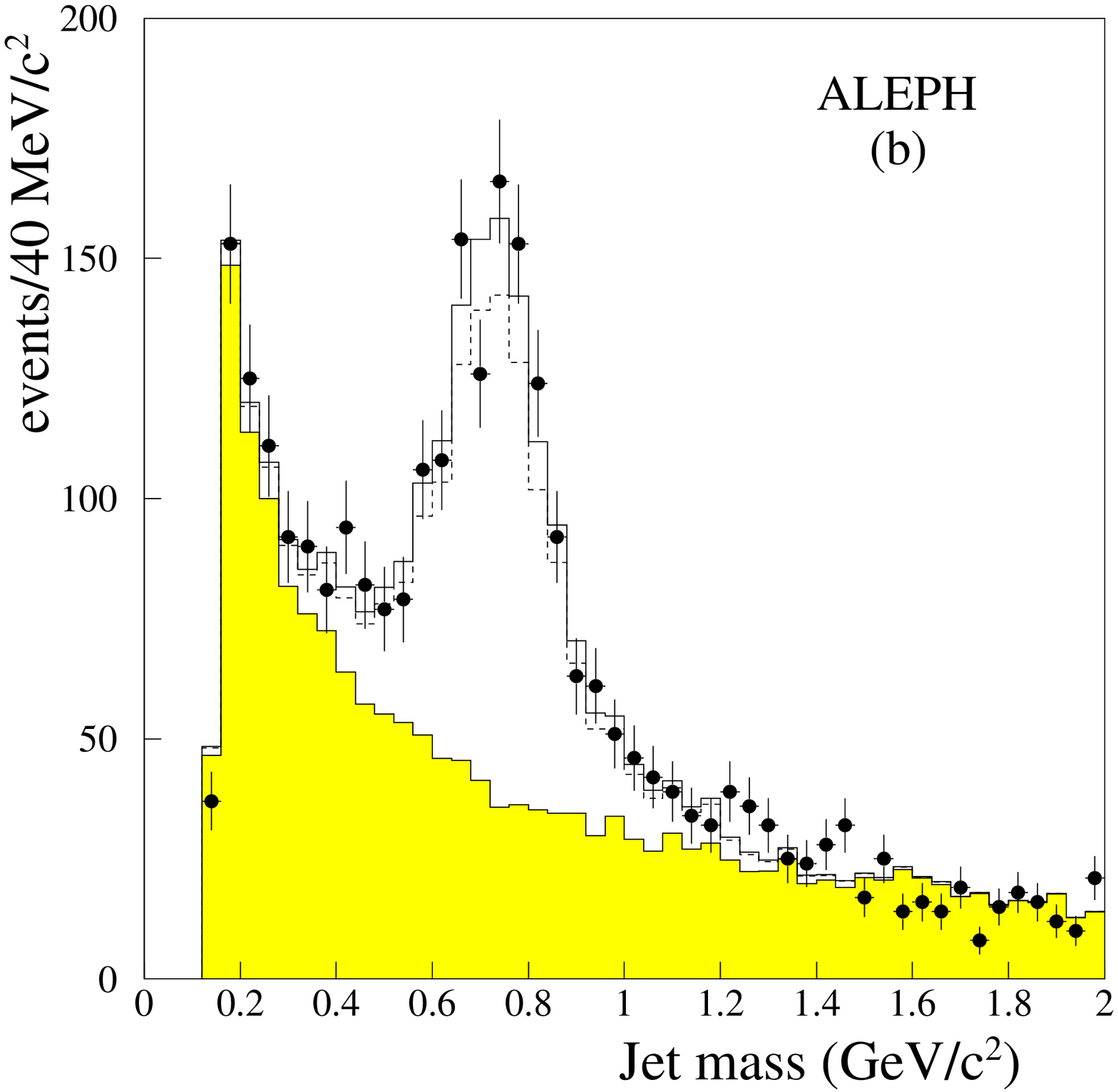,width=7.75cm}}
    \caption{\label{fig:ep_hadveto}\protect\small (a) The $\gamma \gamma$ mass distribution  
      for the ``electron'' hemispheres with a jet  mass larger than 0.5 GeV/$c^2$.
      The shaded area is 
      the contribution of $\tau\to e\nu\bar\nu$ taken from the $\tau^{+}\tau^{-}$ 
      Monte Carlo.
  (b)     The ``$\pi^{\pm}$''($\gamma$(s)) mass distribution  in the muon sample
      before the hadronic veto.
      The shaded area shows 
      the contribution 
      of $\tau\to \mu \nu \bar\nu $ taken from the $\tau^{+}\tau^{-}$ 
      Monte Carlo.  The black dots are the data, the solid (dashed) histogram 
      the $\tau^{+}\tau^{-}$ Monte Carlo prediction   taking (not taking) into account the
      correction to the  hadron misidentification rate described in Section~\ref{sec:chargid}.}
\end{figure}

The leptonic $\tau$ decay sample is selected from hemispheres
containing  only one charged track coming from the interaction region
and identified as electron or $\mu$. In the case of electron, it is required that the
charged track does not point to an ECAL crack.  
To reduce the hadronic $\tau$ decay background
possibly arising  from  particle misidentification, 
it is required  that there is no genuine $\gamma$
pair with an  invariant mass  compatible with the $\pi^0$ mass,
and that the jet (hemisphere) mass is  smaller than 0.5 GeV/$c^2$.
Since  bremsstrahlung is only significant  for  electrons,
muon hemispheres with one or more photons are also rejected
when the hemisphere mass lies between 0.5 GeV/$c^2$ and 1.6 GeV/$c^2$.
This ``hadronic veto'' is illustrated by Fig.~\ref{fig:ep_hadveto}
which shows the $ \gamma \gamma$ mass in electron hemispheres,
and the jet mass in  muon hemispheres
containing at least one photon.

\begin{figure}[ht]
\centerline{    \epsfig{file=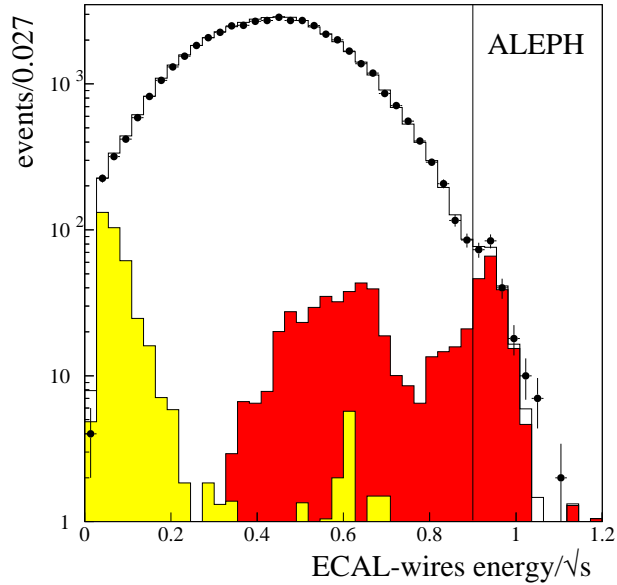,width=8.5cm} }
    \caption{\label{fig:ep_ecalwires}\protect\small Distribution of the total ECAL wire
      energy, normalized to the centre-of-mass energy,
      in the electron channel.
      The black dots represent the data, the line histogram
      the sum of all contributions, $\tau^{+}\tau^{-}$, $\gamma\gamma \to e^+e^-$, 
      $\gamma\gamma \to \tau^+\tau^-$, and Bhabha events,
      given by Monte Carlo.
      Grey histograms, represent the backgrounds: dark grey for Bhabha events and light grey for
$\gamma\gamma$ interactions.
      The vertical line displays the cut value. }
\end{figure}

The non-$\tau$ background is rejected by cuts on the above described
${\cal{E}}^{e^{+}e^{-}}$ and ${\cal{E}}^{\mu^{+}\mu^{-}}$ estimators.
To further  suppress the Bhabha event contamination in the electron channel,
the total ECAL wire energy is required  to be smaller than 0.9 $\sqrt{s}$.
The ECAL wire energy distribution,
before this last cut, is shown in Fig.~\ref{fig:ep_ecalwires}. The  agreement observed
between the data and the sum of signal and background Monte Carlo
shows the good understanding of the non-$\tau$ background contamination
in the electron channel. 
The two-photon background is rejected by the cuts defined in the previous section.

\subsubsection{\boldmath $\tau \to {h}\nu$}
For the selection of  hadronic decays without a  neutral hadron, the presence of
only one charged track
coming from the vertex region,
with a momentum larger than 1.3~GeV/$c$ (limit of the validity of $\mu$-PID), 
is required. It is also required that
this  track is not  identified  as an electron or a
muon by the PID. Two further cuts are imposed on the same hemisphere:
no  genuine reconstructed photon
and  a calorimetric energy deposit  consistent
with the momentum of the charged hadron. This last cut reduces the $K^0_L\pi$ background.

The rejection of the hadronic Z decays and other non-$\tau$ backgrounds
is performed
by cuts   on the values of the previously defined likelihood estimators
computed on the  opposite hemisphere.
The two-photon background is rejected by the  cuts described in Section~\ref{sec:s-nontaubck}. 

\subsubsection{\boldmath  $\tau \to  \rho\nu$}\label{sec:EP_sel_rho}
This selection is similar to the $h\nu$ channel selection  as far as the characteristics
of the charged track and the non-$\tau$ background rejection are
concerned.
The requirements on the photons in the charged track hemisphere are
opposite. The presence of  one or two genuine photons is required.
When a single photon is reconstructed, its energy must be greater than 0.5\,GeV and a cut on the 
$\pi \gamma$ mass (0.4\,GeV/$c^2<m_{\pi\gamma}<\,$1.6\,GeV/$c^2$)
is imposed.
When two photons are reconstructed, no cut on the $\pi 2 \gamma$ mass is applied 
but  the  $\gamma \gamma$ mass  has to be compatible with a $\pi^0$ mass
(80\,MeV/$c^2<m_{\gamma\gamma}<$\,250\,MeV/$c^2$).

\begin{figure}[h]
    \centerline{\epsfig{file=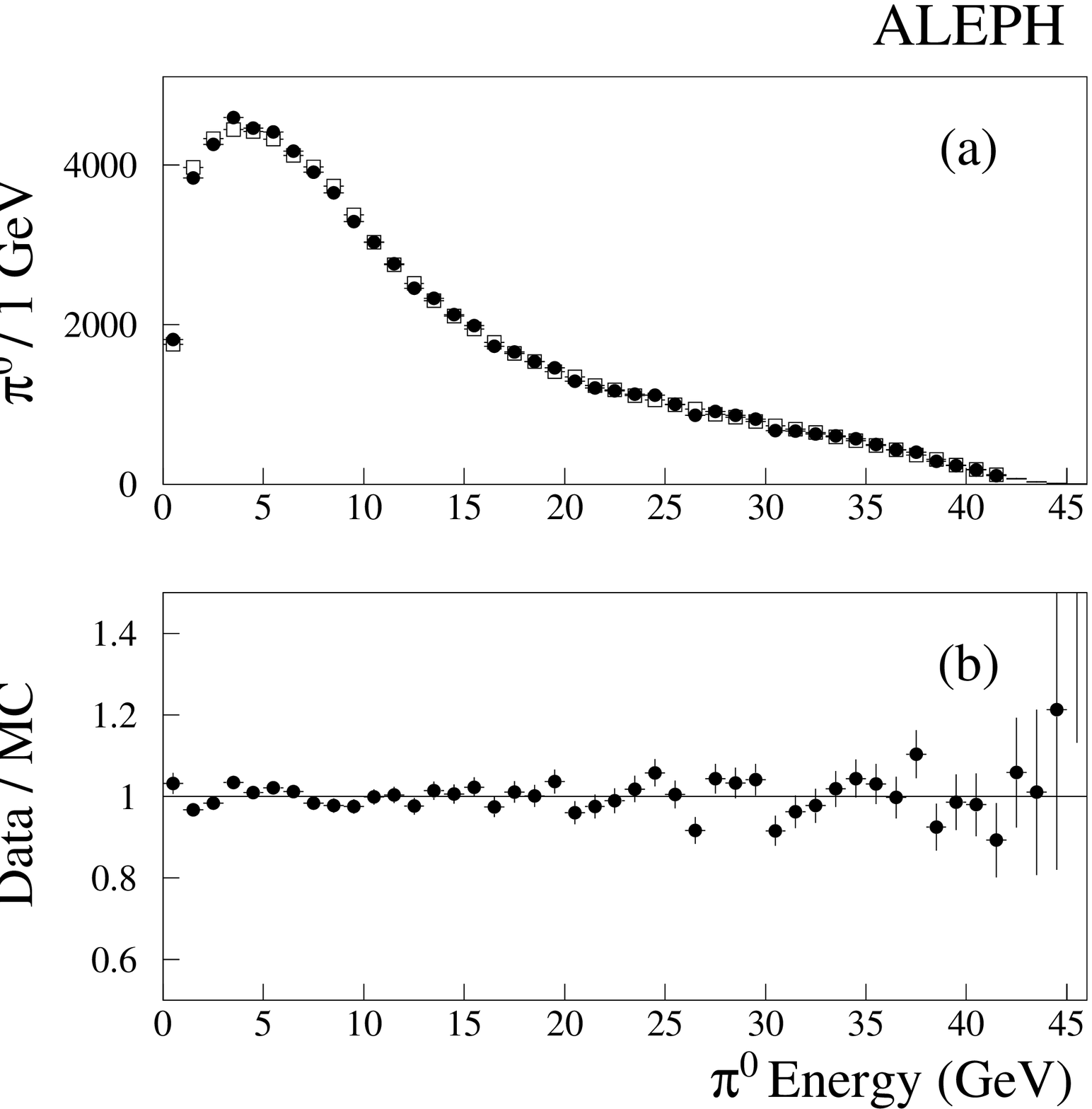,width=7.75cm}
      \epsfig{file=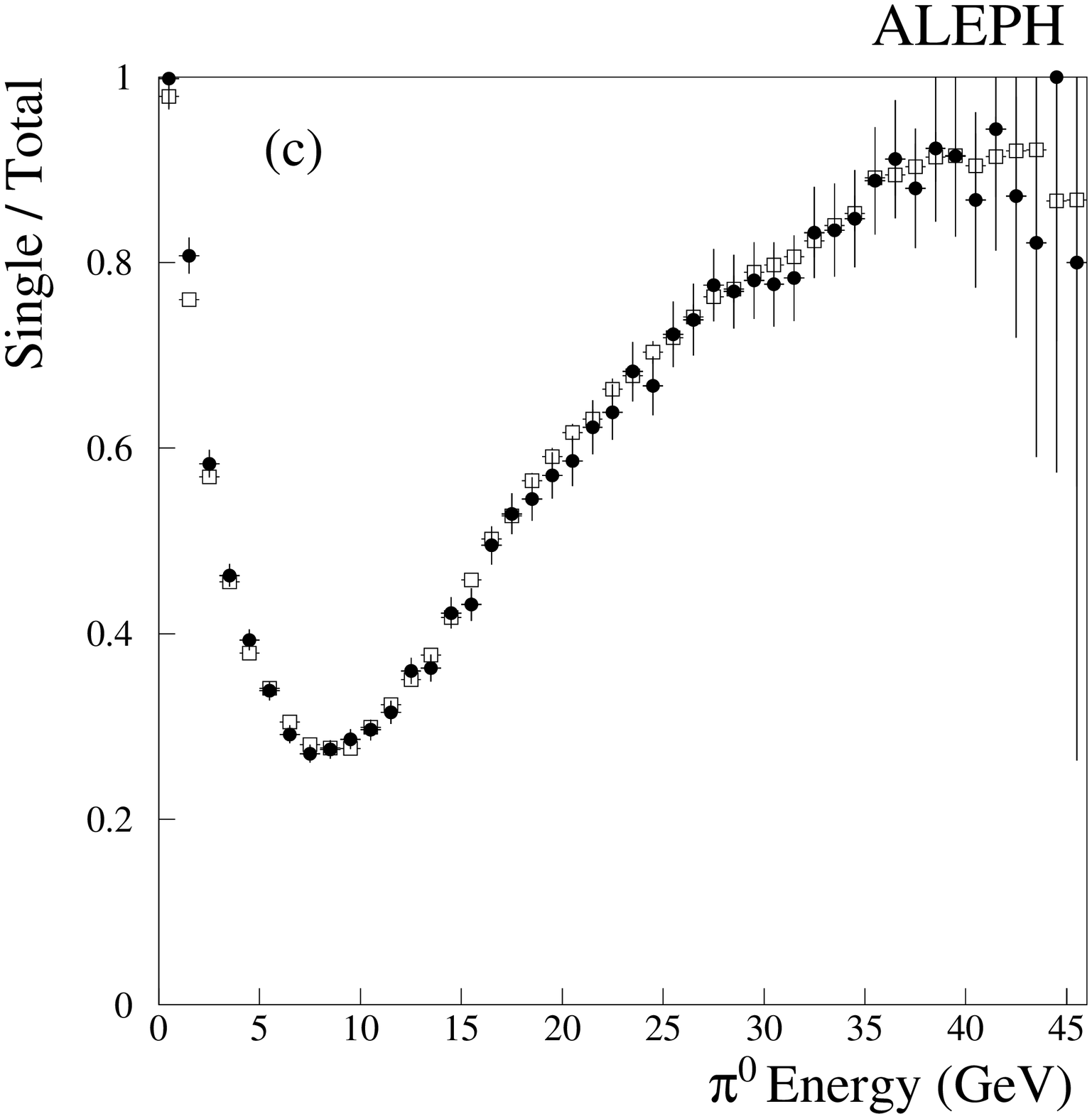,width=7.75cm}}
    \caption{\protect\small (a) Distribution of the $\pi^0$ energy    data (black dots)
      and $\tau^{+}\tau^{-}$ Monte Carlo (empty squares).
      (b) Data over Monte Carlo ratio. 
      (c) Single photon fraction in the  $\pi^0$ sample
      as a function of the $\pi^0$ energy.\label{fig:epi0_rho} }
\end{figure}

\begin{figure}[htp]
    \centerline{\epsfig{file=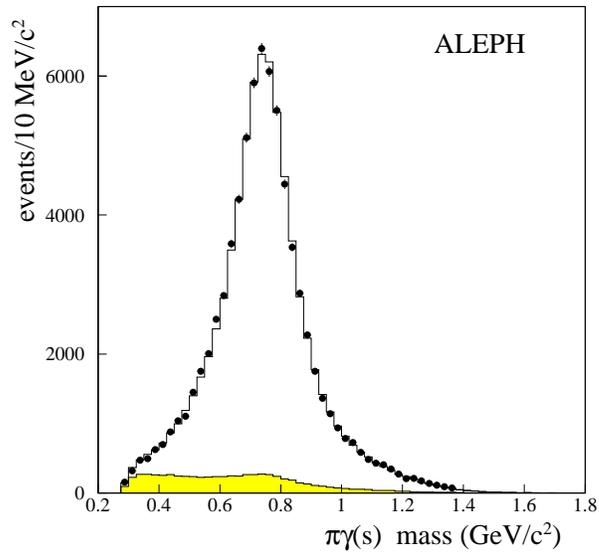,width=8cm}}
    \caption{\protect\small Distribution of the $\pi^{\pm} \gamma$(s) mass for data (black dots)
      and $\tau^{+}\tau^{-}$ Monte Carlo (histogram), in the $\tau\to\rho\nu$ sample. 
      The shaded histogram shows the contribution of $\tau$ background.}
    \label{fig:jetmas_rho}
  \end{figure}

The quality of the simulation
is illustrated in
Fig.~\ref{fig:epi0_rho}, which shows the $\pi^0$ (one and two $\gamma$(s))
energy distribution
in the selected $\rho$ sample along with the single photon
fraction, and in Fig.~\ref{fig:jetmas_rho}, where
the  jet mass in the one $\gamma$ sample (before the above described cut on this mass) is 
displayed.

\subsubsection{\boldmath $\tau \to  {a}_{1}\nu$,
  $a_{1} \to h 2\pi^0$}
Here also,  the selection criteria  on the charged track 
and the non-$\tau$ background rejection are 
the same as in the  $h\nu$ case.
The  additional requirements on the photons in the selected
hemisphere are as follows.
The number of genuine reconstructed photons, $n_\gamma$, ranges from  two to four.
For $n_\gamma=2$,
the  $\gamma \gamma$ mass has to be larger than the $\pi^0$ mass 
and  inconsistent with it.
For $n_\gamma=3$, the reconstruction of at least one $\pi^0$ is required
and, for $n_\gamma=4$, the reconstruction of two $\pi^0$'s is required.   
The choice of the best association of the photons and  the $\pi^0$ reconstruction
are performed by a kinematic fit~\cite{perfo} which also improves the energy determination.

\begin{figure}[htp]
    \centerline{\epsfig{file=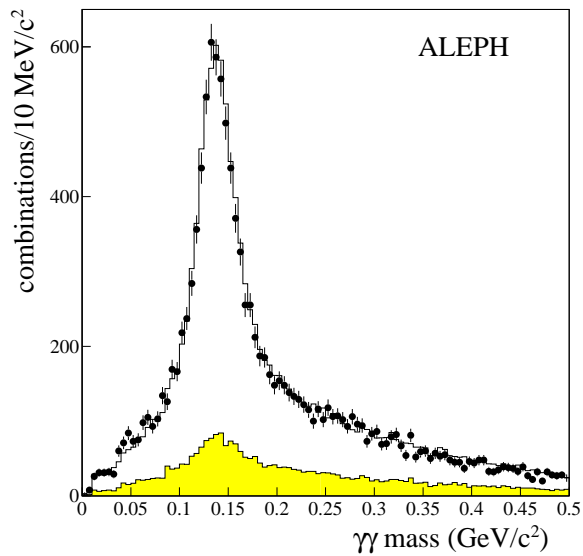,width=8cm}}
    \caption{\protect\small Distribution of the $\gamma \gamma$ mass
      in four photon hemispheres
 when the  effective mass of the two other photons
 is in the $\pi^0$ region. Data are represented by
 black dots and  Monte Carlo by histograms.
The shaded histogram shows the contribution of $\tau$ background.}
\label{fig:ep_2g2pi0}
\end{figure}

The purity of the selection and the quality of the reconstruction for the four photon sample
are shown   in Fig.~\ref{fig:ep_2g2pi0}
where the mass of  a $\gamma \gamma$ system is displayed  when the
mass of the two remaining photons is
consistent with the  $\pi^0$ hypothesis.

 \subsubsection{\boldmath   $\tau \to a_{1}\nu$,
$a_{1} \to 3h$}
  The presence of three charged tracks coming from the 
 interaction region is required. At least two
 of the three tracks 
 must be identified as  charged hadrons by the PID.
 The following cuts against the presence of $\pi^0$'s in the hemisphere 
are applied: no genuine photon 
 with energy larger than 1 GeV,   no  genuine $\gamma$ pair
 with $E_{\gamma\gamma}>\,$3\,GeV, and $m_{\gamma\gamma}$ compatible with  $m_{\pi^0}$.
 Hemispheres containing  a V$^0$, as defined by the 
  ALEPH V$^0$ package \cite{perfo},
 whose mass is compatible with a $K^0_{S}$  are rejected. 
 In addition, the effective mass of at least one of the $\pi^{-}
 \pi^{+}$ combinations is required to be larger than 0.55\,GeV/$c^2$.
 The non-$\tau$ background rejection  is the same as  in the
 other hadronic decay channels.
 \subsubsection{\boldmath Inclusive channel $\tau \to  h
\geq 0 \gamma\nu$}
 The charged track selection and the non-$\tau$ background rejection
 are those used in the selection of the $h\nu$ channel.
Only the cuts against the two-photon event background are tighter, 
because the pion momentum distribution used for the polarisation measurement
is very sensitive to this background.
Since the purpose of the study of the inclusive channel is  to cross-check
 the photon and $\pi^0$  treatment, no cut using  information from the
electromagnetic calorimeter in the selected hemisphere is applied.
\subsection{Systematic uncertainties}
\label{section:single_syst}
 The  sources of systematic errors  affecting the $\tau$ polarisation
 measurement are:
\begin{itemize}
\item 
 $\tau$ selection efficiency,
\item 
 non-$\tau$ background contamination,
\item 
$\gamma/\pi^0$ reconstruction,
\item
  energy/momentum calibration,
\item 
   modelling of a$_1$ decay and  radiative processes,
\item 
 $\tau$ decay channel cross-talk.
\end{itemize}
A summary of the errors in the single-$\tau$ method is given in 
Table~\ref{table:systot} and their estimation is discussed in the following. 

\begin{table}[htb]
\caption[systot]{\label{table:systot}\protect\small Summary of the systematic uncertainties  
(\%) on $A_\tau$  and $A_{e}$   in the single-$\tau$ analysis.
\label{systEP}}
\centerline{
\begin{tabular}{lccccccc} \hline
\multicolumn{8}{c}{\rule[0cm]{0cm}{0.4cm}${ A}_\tau$}\\
\rule[0cm]{0cm}{0.4cm}\rule[-.25cm]{0cm}{0.65cm}Source    & $h$ & $\rho$ &  3\,$h$&$h$\,2$\pi^0$&  $ e$   & $\mu$& Incl. $h$ \\ \hline 
selection & - &  0.01&    -&- & 0.14& 0.02& 0.08\\
tracking   & 0.06  & - &0.22&  -    & - &0.10 &  - \\ 
ECAL scale   & 0.15      &0.11 & 0.21 &1.10&0.47&  - & - \\ 
PID  & 0.15    &  0.06&0.04 &0.01  &0.07 &0.07&0.18 \\ 
misid.  & 0.05     &  -    & -     &  -   &0.08 &0.03 &0.05\\
photon    & 0.22&  0.24 &0.37 &0.22  &- & - & - \\ 
non-$\tau$ back.    & 0.19  & 0.08   &0.05 & 0.18  &0.54  &0.67 &0.15  \\
$\tau$  BR    & 0.09  & 0.04 & 0.10 & 0.26&0.03   & 0.03 &0.78\\ 
modelling  & - &-  &  0.70 &  0.70 &-& - &0.09\\ 
MC stat        & 0.30 & 0.26 &0.49& 0.63&0.61&0.63& 0.26\\
\hline
TOTAL                   &0.49     &0.38 &1.00 &1.52   & 0.96  &0.93 &0.87  \\ \hline
\hline 
\multicolumn{8}{c}{
\rule[0cm]{0cm}{0.4cm}${ A}_{e}$}\\ 
\rule[0cm]{0cm}{0.4cm}\rule[-.25cm]{0cm}{0.65cm}  Source    & $h$ & $\rho$ &  3\,$h$&$h$\,2$\pi^0$&  $ e$   & $\mu$& Incl. $h$ \\\hline
tracking &0.04&-&-&-&-&0.05&-\\
non-$\tau$ back.&0.11&0.09&0.04&0.22&0.91&0.24&0.17\\
modelling&-&-&0.40&0.40&-&-&-\\\hline
TOTAL&0.12&0.09&0.40&0.47&0.91&0.25&0.17\\\hline
\end{tabular}
}
\end{table}

 The dominant contributions are photon/$\pi^0$
reconstruction, ECAL calibration, non-$\tau$ background, and  modelling of the decay for
the $a_1$. This last effect is evaluated
\cite{polarb} by changing the model and its parameters within the limits
allowed by the data.

  The systematic effects are evaluated
 for each of the previously described tools using the difference between the data and Monte Carlo
plus one standard deviation as input.
The errors are propagated 
through all the steps of the analysis, taking into
account the fact that a given tool can be used in several places.
For that,  $A_\tau$ and $A_e$ are measured  by means of the
standard  $\hat W^{\pm}(\omega)$ functions on  Monte Carlo
events weighted for the change of the tool response. 
The shifts of the values of   $A_\tau$ and $A_e$ give  the
associated errors.
Cross-checks of the estimations are presented for some of the
important contributions.
\subsubsection{Charged particle identification }\label{sec:EP_sys_pid}
 The  comparison of
 the PID performances on  data and  Monte Carlo
 is  reported in Ref.~\cite{bib:aleph_ew}.
The uncertainty on the data/Monte Carlo efficiency-ratio computed for each year
of data taking is used to estimate systematic error. 
 
\subsubsection{Photon reconstruction}
\label{sec:EP_sys_gam}

\noindent {\it Clustering algorithm}

 The first step of photon reconstruction is
 the ``standard '' ALEPH clustering algorithm \cite{perfo}.
Its parameters
 are the distance to the closest charged track
 and the threshold energy. 
 It has been shown in Ref.~\cite{hbr} that the distance parameter 
can be changed   by 1~mm, while preserving  within one standard deviation
the agreement between data and Monte Carlo on the distance distribution.
This parameter shift introduced  in the simulation induces negligible effects
 on the polarisation, even in channels containing
 photon(s). 
 The possible effect of the threshold value
 is investigated by modifying
 the reconstructed  shower energy
  in the Monte Carlo 
 by the  clustering correction \cite{perfo}.
 The resulting variation  of the polarisation measurement
 is small but not completely negligible.
It  is part of the value given in the line ``photon'' of 
Table~\ref{table:systot}.

\noindent {\it Fake photon estimator}

\begin{figure}[ht]
\mbox{\epsfig{file=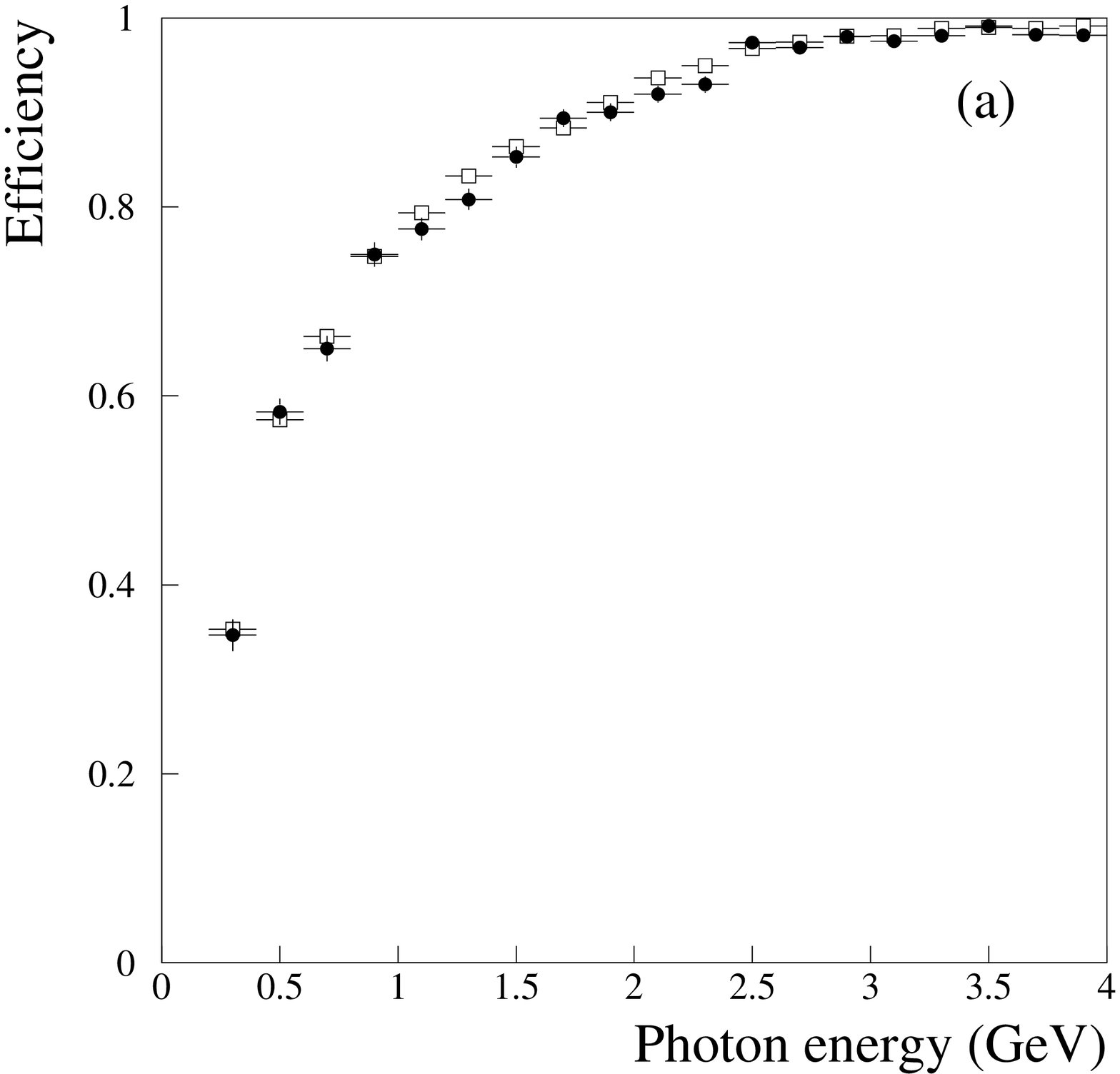,width=7.75cm}}\mbox{
\epsfig{file=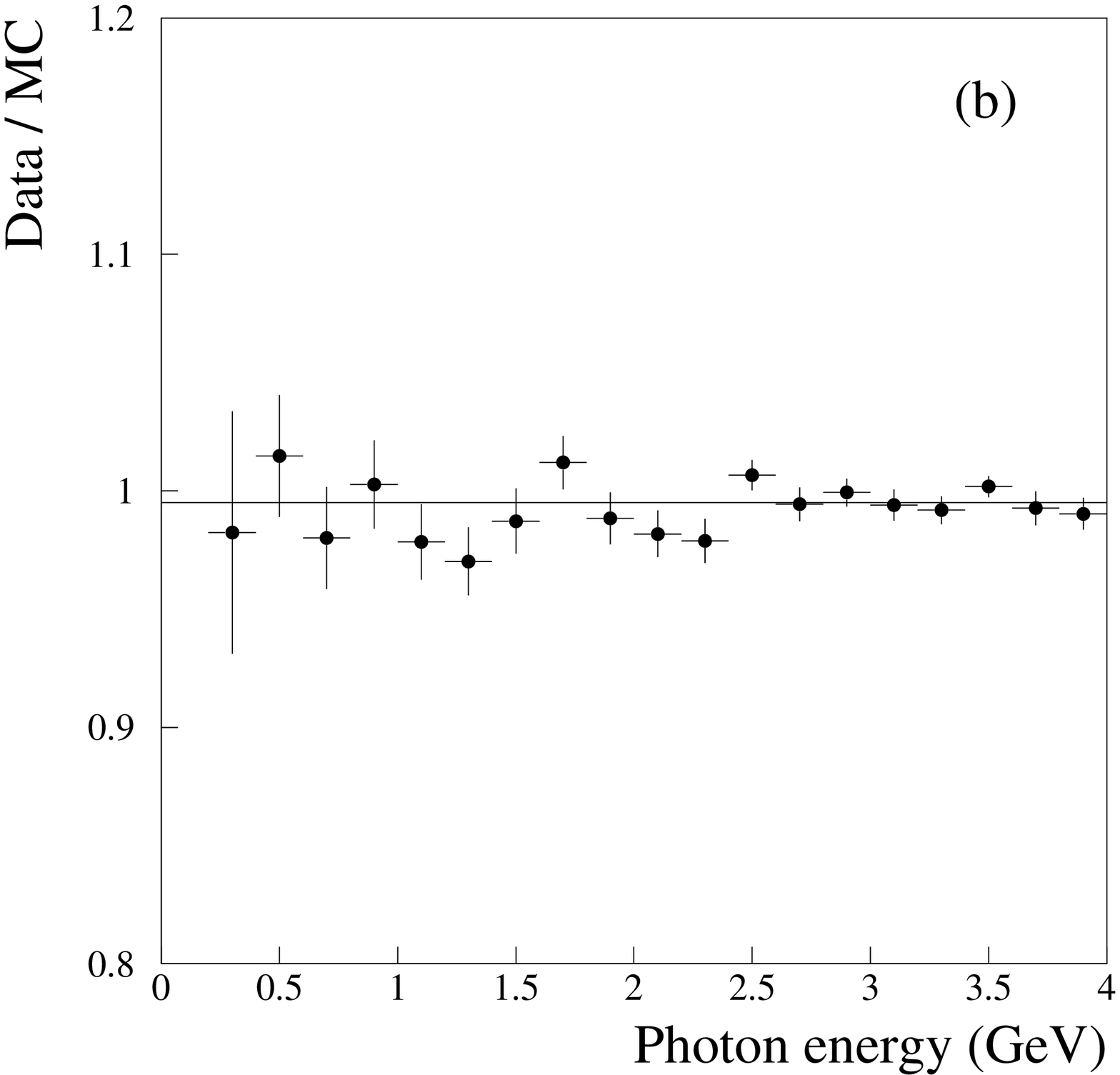,width=7.75cm}}
 \caption{\label{fake_eff}\protect\small (a) Efficiency of the fake photon 
estimator for genuine photons. The black dots
 are the data, the empty squares the $\tau^{+}\tau^{-}$ Monte Carlo.
(b) Data/Monte Carlo ratio
 for this efficiency.}
\end{figure}

The second step is the genuine/fake classification for which the
{\it P$_{fak}$} estimator is used.
In the first place, the efficiency of the identification of genuine
photons is measured on  a selected sample made of photons forming a
$\pi^0$ with one ``high quality''photon  (converted $\gamma$  or
isolated shower with 4\,GeV$<E<20$\,GeV).
For energies greater than 4\,GeV, the efficiency is not significantly
different from one. The efficiencies for $E<$\,4\,GeV are shown in 
Fig.~\ref{fake_eff}. The ratio between data and Monte Carlo is  
(99.50$\pm$0.19)$\times$10$^{-2}$ and an efficiency
variation  of 0.69$\times$10$^{-2}$  is used as input in the error computation.

\begin{figure}[hbt]
\centerline{
\epsfig{file=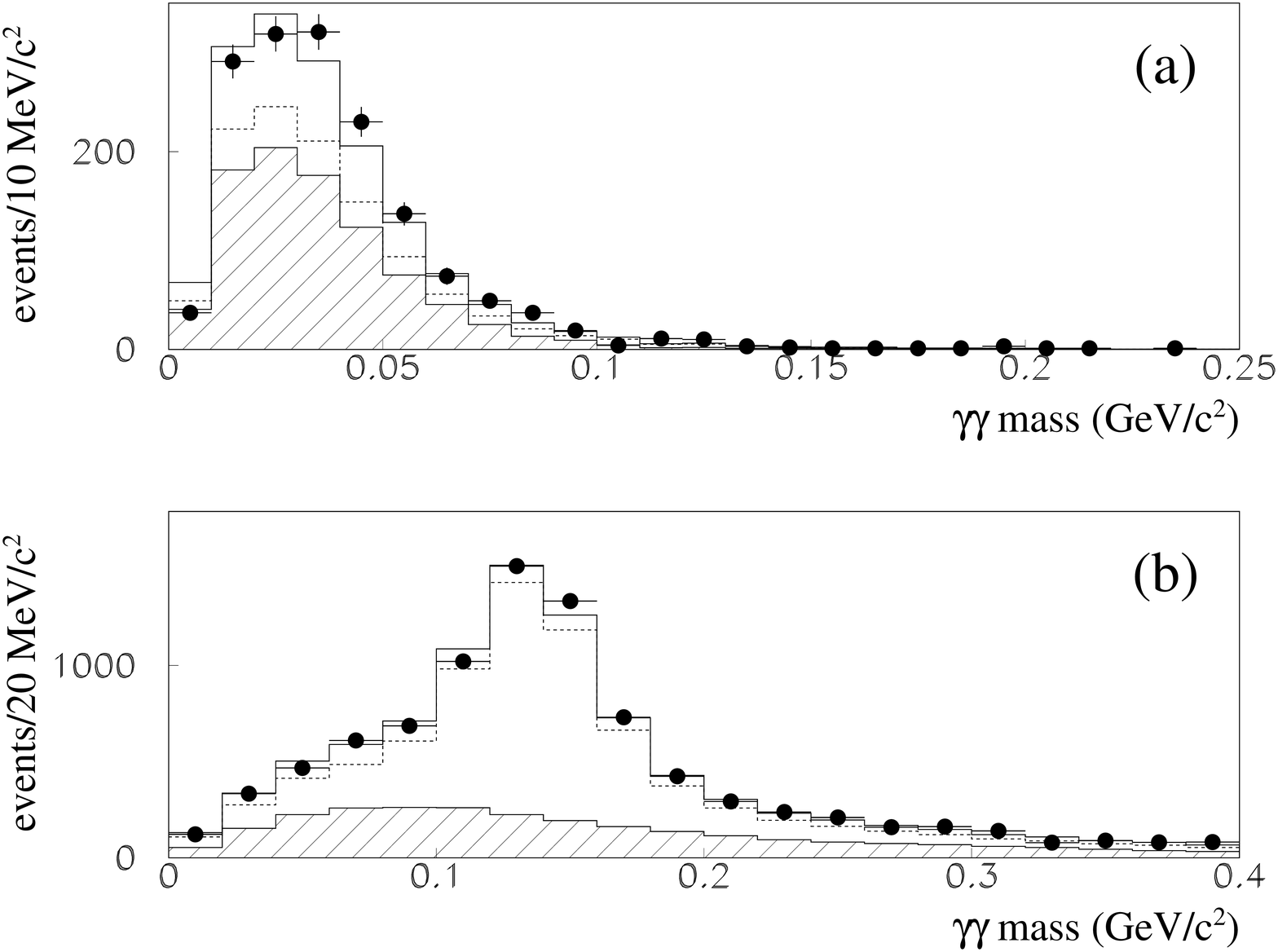,height=9cm}
} \caption{ \label {fake_multi}\protect\small  The $\gamma \gamma$ mass spectrum,
for a sample of $\pi\gamma\gamma$ hemispheres: (a)
 when the two photons are classified as fake, (b) when 1 photon
 is classified as fake and the other as genuine. 
 The black dots are the data,
 the dashed line histogram  the $\tau^{+}\tau^{-}$ Monte Carlo and the full line histogram
 the $\tau^{+}\tau^{-}$ Monte Carlo, with a weighted   fake photon
 contribution. The hatched histogram is the fake photon
 contribution in the $\tau^{+}\tau^{-}$ Monte Carlo, without weighting (see text). } 
\end{figure}

\indent
Since the Monte Carlo is able to reproduce the characteristics of the 
 fake photon production in a hadronic shower but not its precise normalization~\cite{hbr}, 
a weighting 
procedure is used to correct it. The weight for an
event is written $w(p)^n$, where $n$ is the fake photon multiplicity and
$p$ the hadron momentum. The function $w(p)$ is obtained by a fit of
the likelihood estimator distributions. 
A check of the procedure  is given by the comparison of  data and Monte Carlo
for the $\gamma\gamma$ mass spectra shown in Fig.~\ref{fake_multi}.
The related systematic error is
estimated in the standard way described at the beginning of Section~\ref{section:single_syst}, 
comparing  the measured weight to one. 

\subsubsection{\boldmath Bhabha, $\mu^+\mu^-$, and $q\bar q$ event backgrounds}
\label{sec:EP_sys_estim}
The ${\cal E}^{{e}^+{e}^-}$,  ${\cal E}^{\mu^+\mu^-}$, and 
${\cal E}^{{q}\bar{q}}$ estimators use  information from only one of the hemispheres of
an event, so correlations with the other hemisphere can only be 
 due to geometrical correlations in the detector construction (ECAL
cracks, overlap, etc.). The measurement of the estimator efficiencies in
the non-correlated regions is described first, and then an evaluation
of the residual background, which is not affected by the possible
correlations,  is presented. This last point  concerns
only the hadronic channels   because the selection
criteria for leptonic $\tau$ decays eliminate the correlated,
insensitive parts of the detector.

\noindent{\it Efficiency of the estimators}

\begin{figure}[p]
\centerline{
\epsfig{file=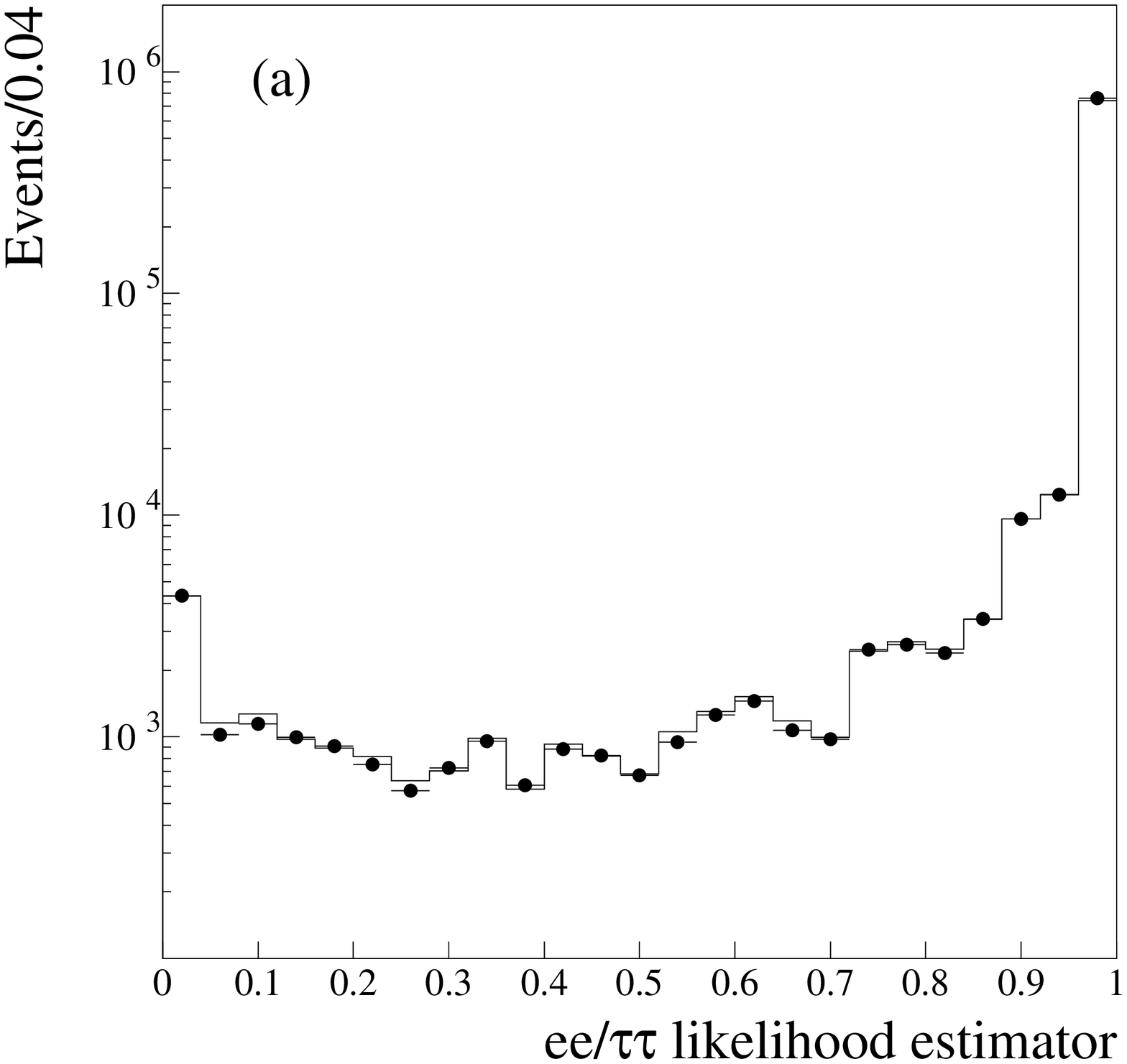,width=7.75cm}
\epsfig{file=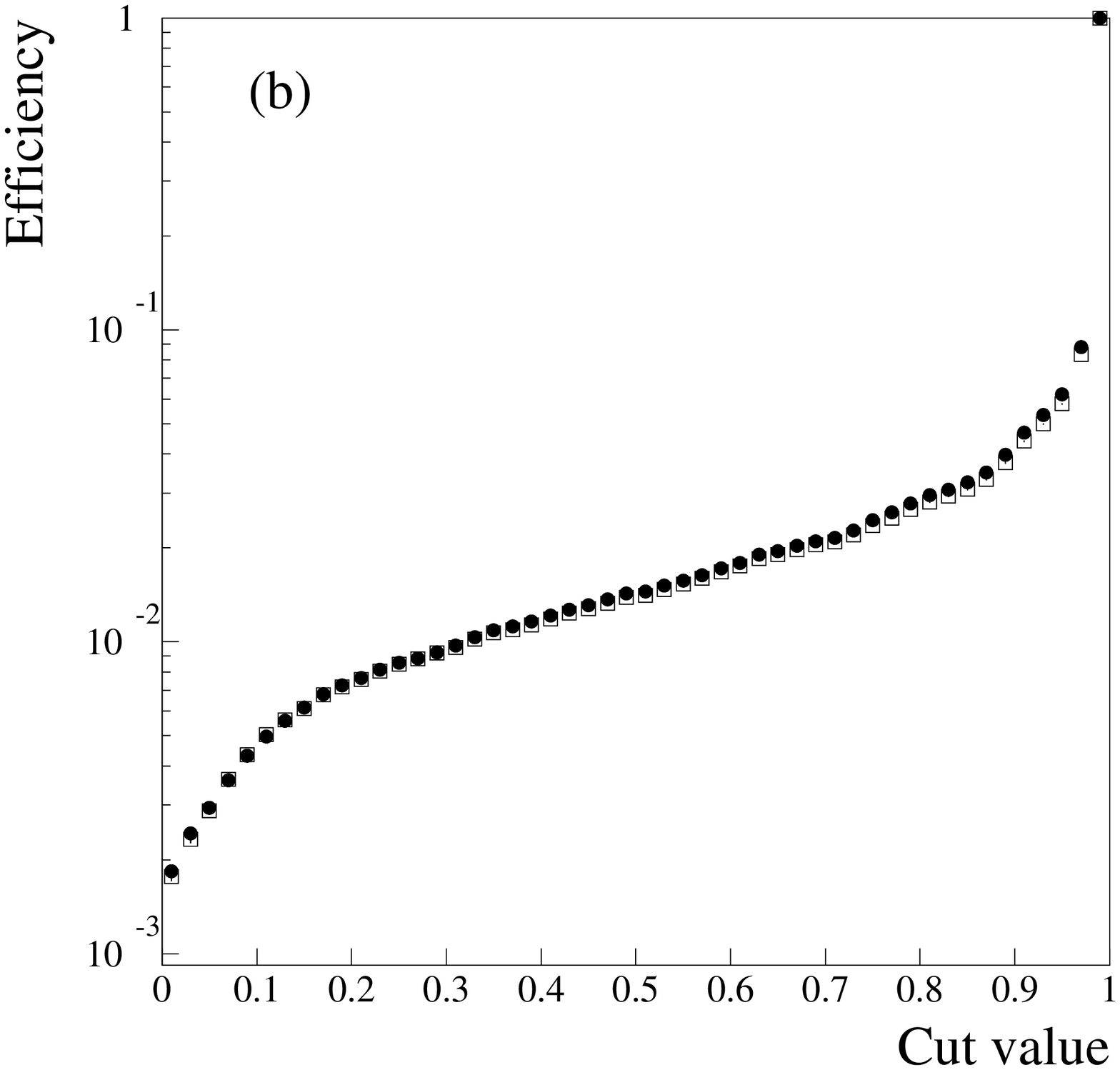,width=7.75cm}}
\caption{\protect\small (a) The   ${\cal{E}}^{e^{+}e^{-}}$  distribution
 after tagging an electron  in the opposite hemisphere. 
(b)~Efficiency on Bhabha events as function of the
 cut value. Black dots are the data,
 empty boxes and histogram, the Bhabha Monte Carlo.}
\label{unib-data}
\end{figure}

\begin{figure}[p]
\centerline{
\epsfig{file=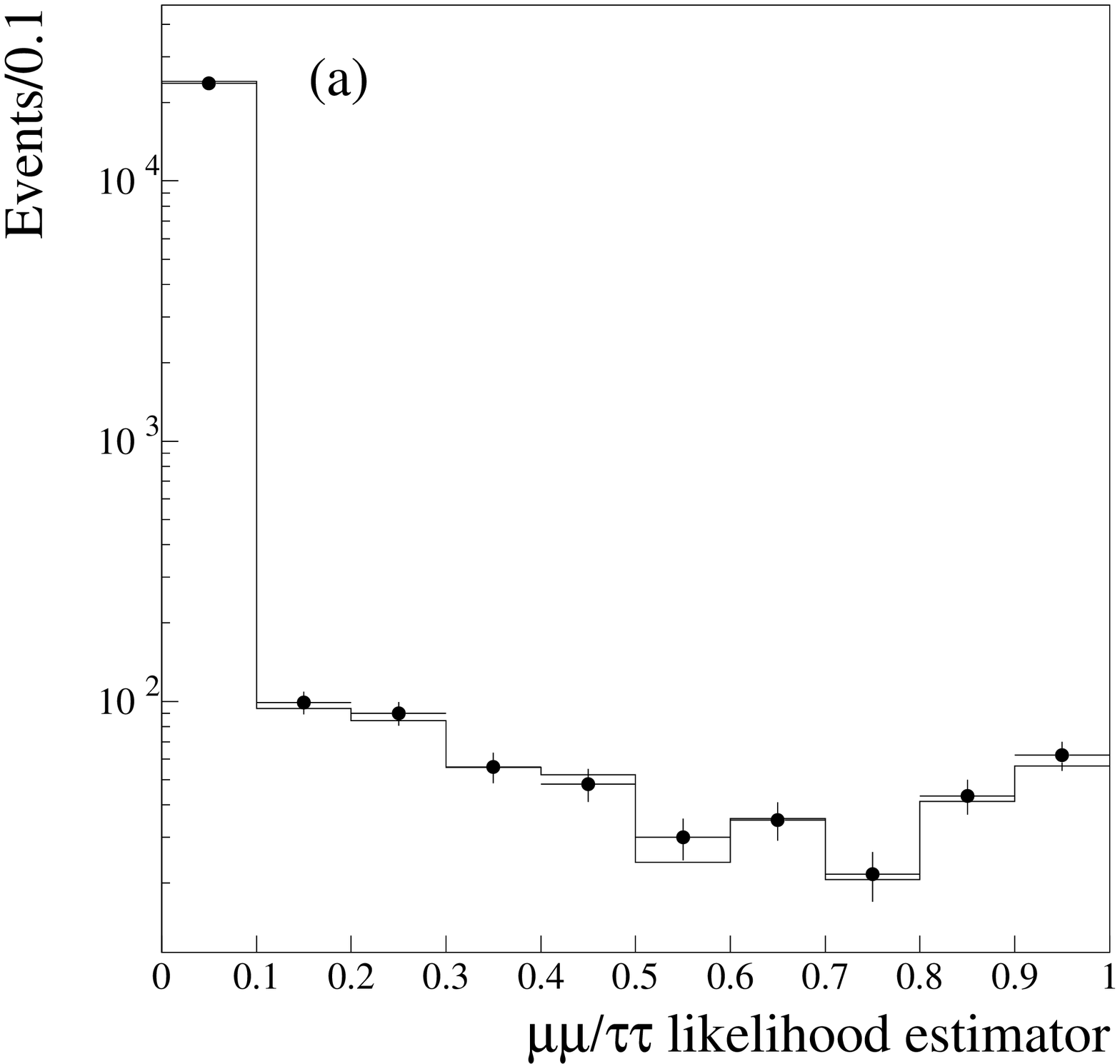,width=7.75cm}
\epsfig{file=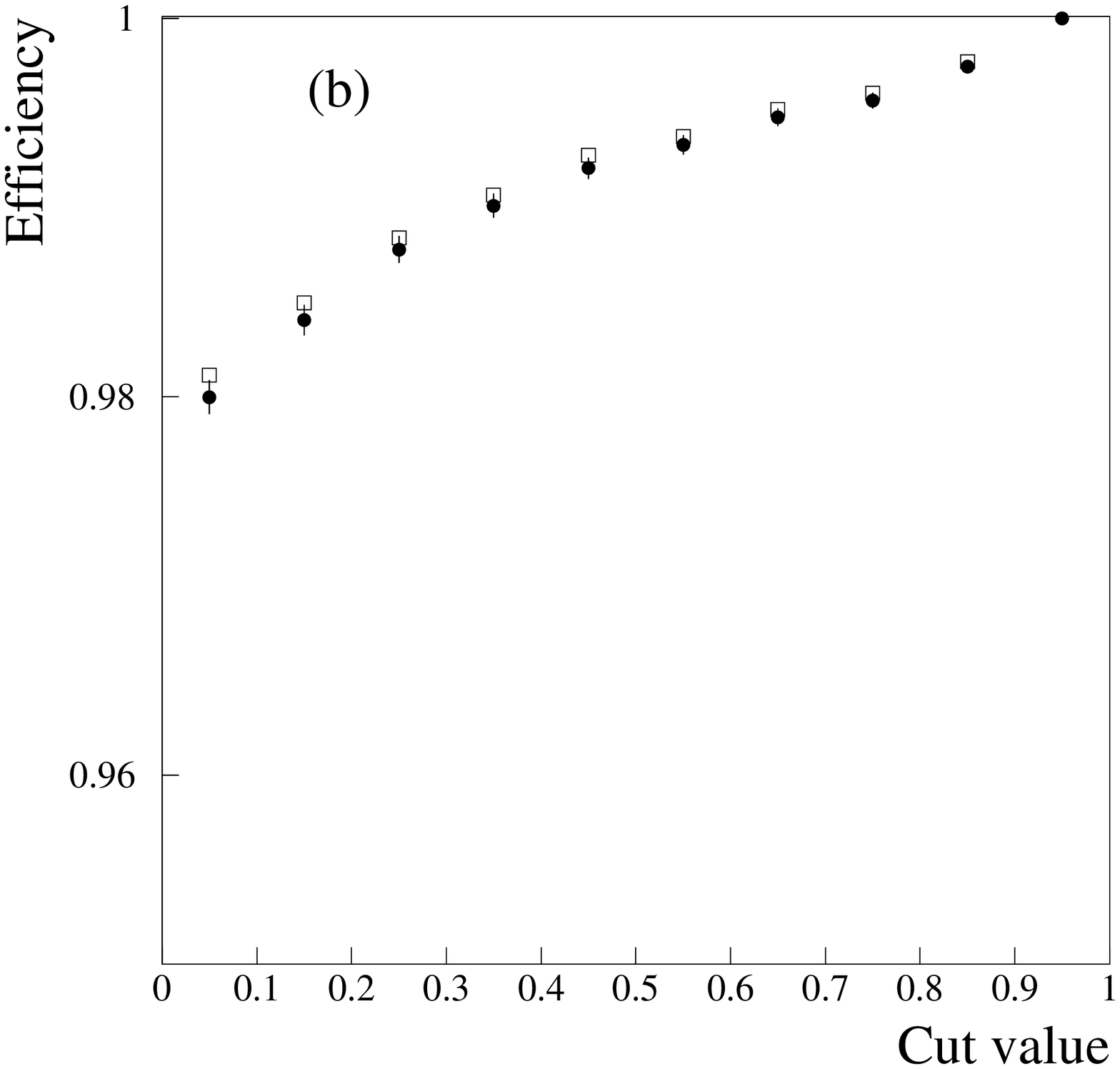,width=7.75cm}}
\caption{\protect\small (a) The
  ${\cal{E}}^{\mu^{+}\mu^{-}}$   distribution
  after tagging a  $\tau$ in the opposite hemisphere.
(b)~Efficiency on $\tau^+\tau^-$ events as function of the
 cut value. Black dots are the data,
 empty boxes and histogram, the $\tau^+\tau^-$ Monte Carlo.} 
\label{mumu1}
\end{figure}

In order to measure the efficiencies, almost pure samples of data and
Monte Carlo events are selected for the $\tau^+\tau^-$, ${e}^+{e}^-$,
 $\mu^+\mu^-$, and $q\bar q$ final states. A clean $\tau$ decay (three
charged $\pi$'s or  a $\rho$), a high energy electron, a high energy muon
or a hadronic jet is required in one hemisphere and the opposite one
is used to study the estimators. Examples of their responses in the
case of ${\cal E}^{{e}^+{e}^-}$ and  ${\cal E}^{\mu^+\mu^-}$ are shown in 
Fig.~\ref{unib-data} and~\ref{mumu1}. 
\begin{table}[b]
\caption{\label{table:estimeff}\protect\small Efficiencies of the 
  ${\cal E}^{{e}^+{e}^-}$ and  ${\cal E}^{\mu^+\mu^-}$ estimators (in \%) measured 
on data and Monte Carlo samples}
\centerline{
\begin{tabular}{lcc}\hline
&\multicolumn{2}{c}{${\cal E}^{{e}^+{e}^-}$}\\
&DATA&M.C.\\\hline
$\tau^+\tau^-$&$98.92\pm 0.03 $&$98.95\pm 0.01$\\
$e^+e^-$&$2.77\pm 0.02$&$2.66\pm 0.02$\\\hline\hline
&\multicolumn{2}{c}{${\cal E}^{\mu^+\mu^-}$}\\
&DATA&M.C.\\\hline
$\tau^+\tau^-$&$99.45\pm 0.03$&$99.48\pm0.01$\\
$\mu^+\mu^-$&$0.17\pm 0.04$&$0.16\pm 0.05$\\\hline
\end{tabular}
}
\end{table}

A cut at a value of 0.8 on these two estimators allows a good
separation between $\tau$'s and leptonic background. The measured
efficiencies, taking into account the small residual contamination of
the samples, are given in Table~\ref{table:estimeff}.
These numbers are used to determine the  systematic uncertainties which contribute
to the line ``selection'' of Table~\ref{systEP} and to the line
``non-$\tau$ background'' for the two leptonic channels.

A similar analysis, already described in
Ref.~\cite{bib:aleph_ew}, is performed for the ${\cal E}^{{q}\bar{q}}$ estimator.

\noindent{\it Residual backgrounds}
 
The estimation of the residual background, chiefly Bhabha events, in
the hadronic channels is an essential task on account of its impact on
the ${\cal A}_{e}$ measurement. For that reason, procedures were
developed to  measure directly the background contributions from the data themselves.

\begin{figure}[htp]
\centerline{
\epsfig{file=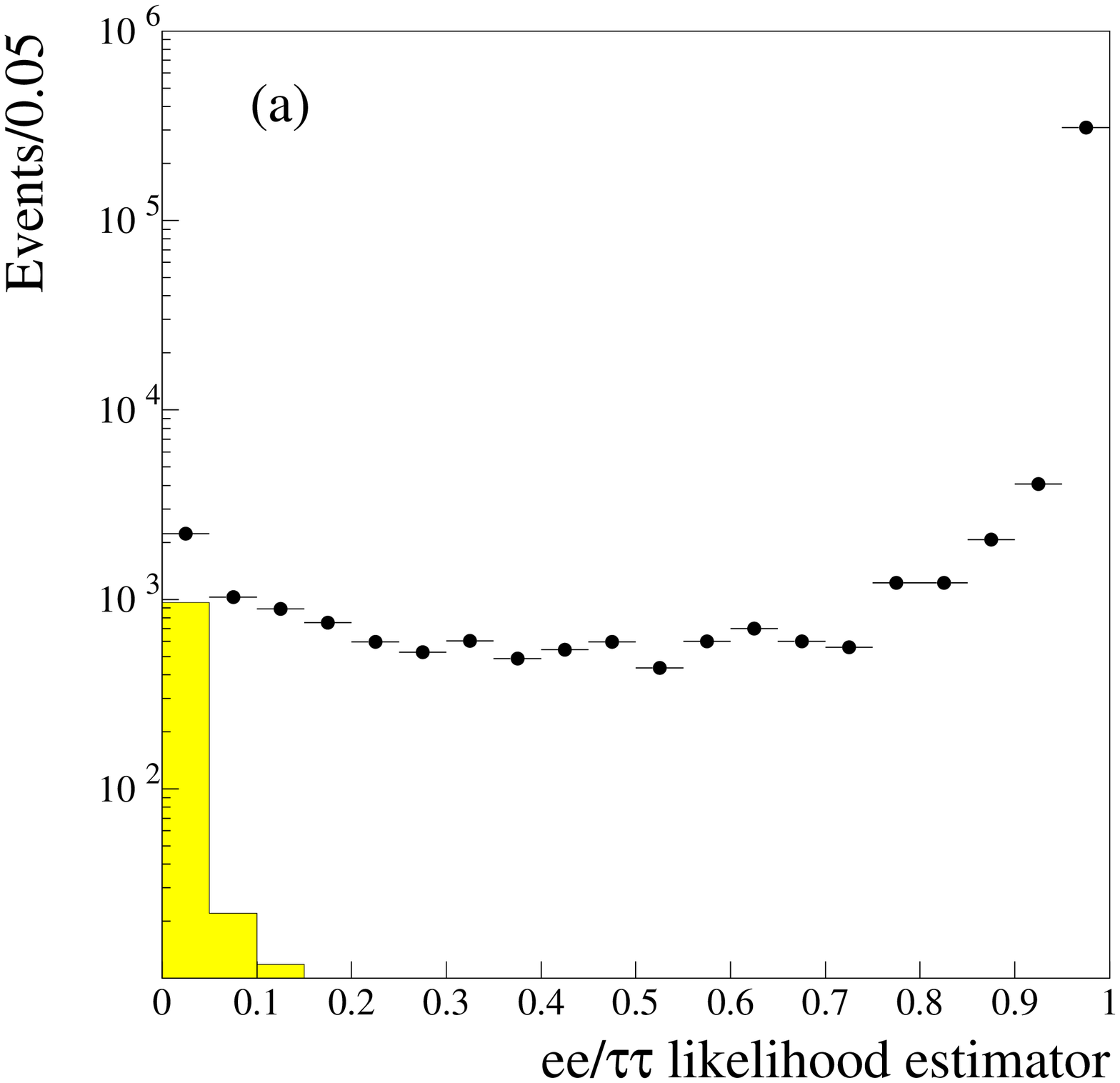,width=7.75cm}
\epsfig{file=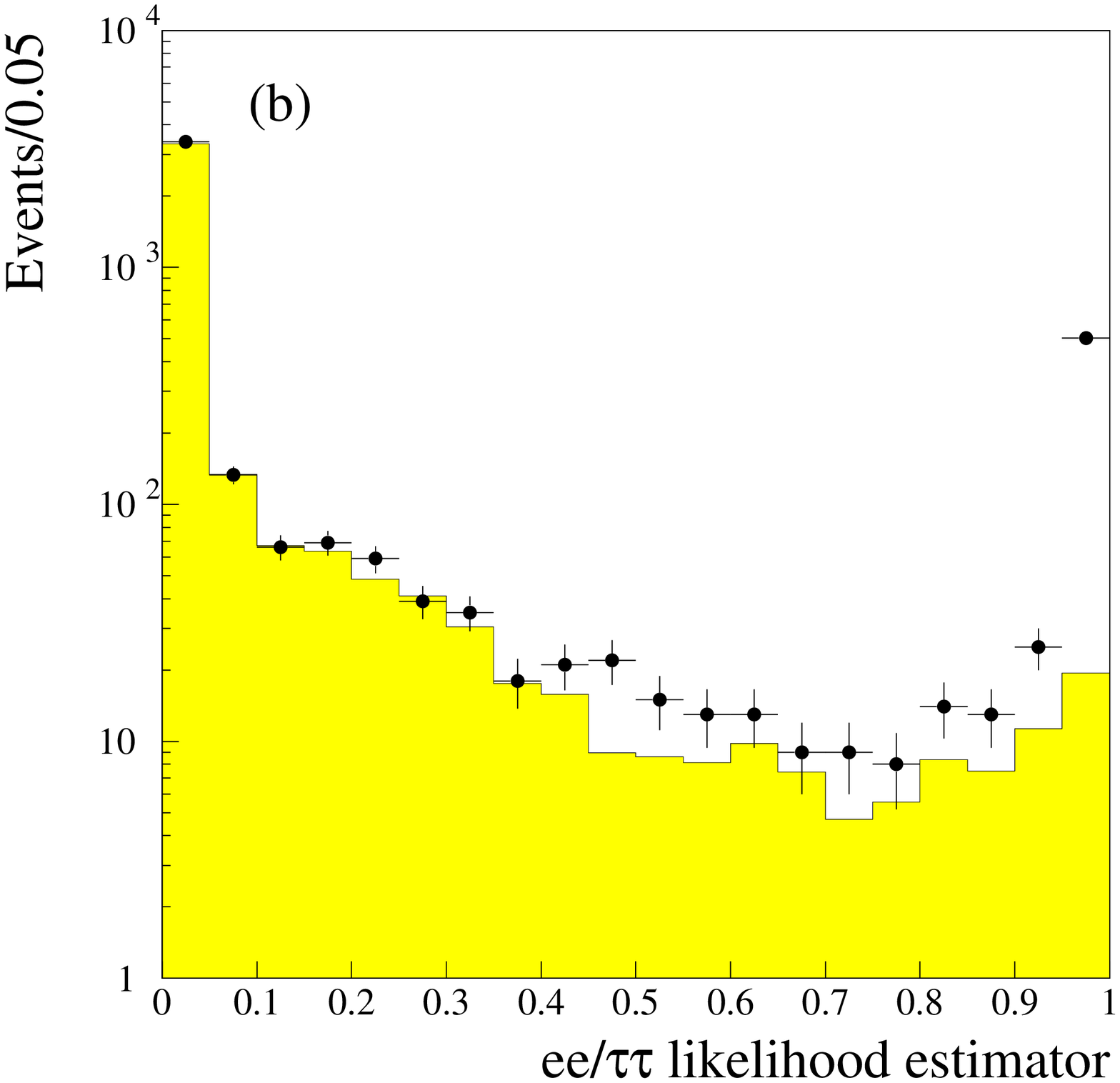,width=7.75cm}}
\caption{\protect\small The distribution of the  ${\cal{E}}^{e^{+}e^{-}}$ estimator
   in the $ h       \nu       $ channel, 
  (a) for events with an electron identification, (b)
for a pion identification. 
 The shaded histograms are the $\tau^{+}\tau^{-}$ Monte Carlo predictions.}
\label{fig:ep_hchan}
\end{figure}

The method is exemplified  by the case of
the $\tau\to h\nu$ channel. Applying all the cuts used for the
channel selection but the $e/\pi$ identification cut on the $\tau$ side, and the
${\cal E}^{{e}^+{e}^-}$ cut on the opposite one, a set of data is
constituted which is divided in two subsets by the response of the
$e/\pi$ PID: (a) for $e$ PID and (b) for $\pi$ PID.  
The distributions of ${\cal E}^{{e}^+{e}^-}$ in
the two subsets and the $\tau^+\tau^-$ Monte Carlo expectations shown
 in Fig.~\ref{fig:ep_hchan} clearly display the Bhabha event background.            

The events with  ${\cal E}^{{e}^+{e}^-}\!\!>0.8$ in sample (b) are used to
determine the kinematic characteristics of this
background. The normalization of the residual background after the 
${\cal E}^{{e}^+{e}^-}$ cut is given by
the number of events with  ${\cal E}^{{e}^+{e}^-}\!\!>0.8$ in sample (b)
multiplied by the ratio of the numbers of events with   ${\cal
  E}^{{e}^+{e}^-}\!\!<0.8$ and ${\cal E}^{{e}^+{e}^-}\!\!>0.8$ in sample (a).
This background is subtracted for the polarisation measurement and the
statistical error on its estimation enters the systematic
error. The small difference of the efficiencies of the cut on  ${\cal
  E}^{{e}^+{e}^-}$ in samples (a) and (b), mainly due to
geometrical effects and initial state radiation,  can be computed by
Monte Carlo and  is added to the systematic uncertainty.

As a cross-check, an independent study of the Bhabha event background has been
performed, using the tools developed for the measurement of the
$\tau$ leptonic branching ratios
 \cite{lbr}. Its conclusions are 
consistent with the above presented results.

Similar procedures are used to study the $\mu^+\mu^-$
contamination. They show that such a background is reduced to a
negligible level by the selection process.

\subsubsection{Other backgrounds}
\longpage

\noindent{\it Two-photon background } 

In a similar way, the residual $\gamma\gamma$ background can be
studied in the data from the excesses of events in the distributions of the
variables (acollinearity, visible energy, etc.) used for its rejection. 

\noindent{\it  $\ell^{+}\ell^{-}{\rm V}$ background}

Such a background  appears mostly in the leptonic channels. 
The contaminations evaluated by a study of the opposite hemispheres
are $(0.2\pm 3.2)\times10^{-4}$ for the electron channel and    
$(5.5\pm 3.6)\times10^{-4}$ for the muon. To compute the tiny associated uncertainties 
the  FERMISV Monte Carlo \cite{bib:4ferm} is used.

\noindent{\it  Cosmic ray background}

The contaminations of the two relevant channels, $\tau\to h\nu$ and
$\tau\to\mu\nu\bar\nu$ are estimated  using the independent
information from the track parameters and the number of ITC hits.
Their values  are $(6.9\pm 1.9)\times10^{-4}$ and $(1.5\pm 0.5)\times10^{-4}$
respectively, leading to very small contributions to the
uncertainties. Out-of-time cosmic rays are normally identified as hadrons, owing to
the  HCAL and muon chamber inefficiency in such events~\cite{lbr}.
 
\noindent{\it $\tau$ background}\label{sec:EP_sys_taubk}

 Uncertainties due to  the crosstalk between different decay channels
 are evaluated by varying within their errors the different branching ratios
 used in the Monte Carlo.

\subsubsection{Detector calibration}
\label{sec:EP_sys_cali}

\noindent {\it Momentum calibration}

 The uncertainty on the momentum 
 calibration described in Section~\ref{sec:calib} is used, for each
 $\cos\theta$ bin,
to estimate the  contribution  to the polarisation systematic errors. 

\noindent{\it  ECAL  calibration }

 The ECAL energy calibration  (Section~\ref{sec:calib})
 is obtained from  electrons taken in Bhabha events, $\gamma \gamma$
 processes, or $\tau$ decays to electrons.
 Uncertainties coming from the statistics of these samples and the
 handling of the ECAL saturation are 
 introduced as systematic errors, leading to an uncertainty of 2.5$\times$ 10$^{-3}$ on the 
 calibration, which is used as input for the computation of
 polarisation errors.

\begin{figure}[ht]
\centerline{
\mbox{\epsfig{file=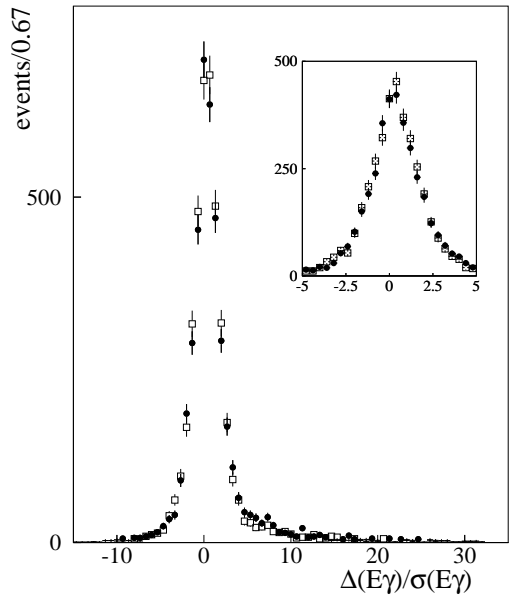,height=8.5cm}}}
\caption{\protect\small  Difference between 
 the reconstructed and calculated photon energy, 
 $\Delta E\gamma / \sigma_{E\gamma}$ for $\mu^{+}\mu^{-}\gamma$ events.
 The insert is a zoom of the region [-5,5]. The black dots represent the data and the empty squares  the 
 $\mu^{+}\mu^{-}\gamma$ Monte Carlo.} 
\label{fig:eg_mumu}
\end{figure}

 An independent check of the ECAL calibration has been performed using the  $\mu^+\mu^-\gamma$ events
 for which the photon energy is deduced, with a small error, from the angles,  by the formula
 $$
 {E}_\gamma ^{\mathrm{calc.}} = \frac {\displaystyle \sin\theta_{\mu^{+} \mu^{-}} \,\sqrt{s} }
 {\displaystyle \sin\theta_{\mu^{+} \mu^{-}}+ \sin\theta_{\mu^{+} \gamma }    
 +  \sin\theta_{\gamma \mu^{-}} }~.
 $$
 The distributions of $\Delta {E}_\gamma/\sigma_{E_\gamma}$,
 where  $\Delta {E}_\gamma$ is the difference $ {E}_\gamma
 ^{\mathrm{calc.}} -{E}_\gamma^{\mathrm{rec.}}$ between the kinematically calculated 
and ECAL energies, and $
 \sigma_{{E}_\gamma}$
 the expected resolution, 
 are shown in Fig.~\ref{fig:eg_mumu} for data and Monte Carlo and
 a very good agreement  can be observed.

Since some calorimetric cuts are used in the selection of the
two hadronic final states without photons, a tuning of the ECAL
response to charged hadrons is also performed. The related
uncertainties are independently evaluated.

 \subsection{Results}

  Table~\ref{EP_stat} gives the numbers of selected data events, 
 the selection efficiencies in the angular acceptance, and the
 contaminations  from $\tau$  and  non-$\tau$ backgrounds.
\begin{figure}[p]
\begin{center}
\mbox{~}\\[-1.5cm]
\mbox{
\epsfig{file=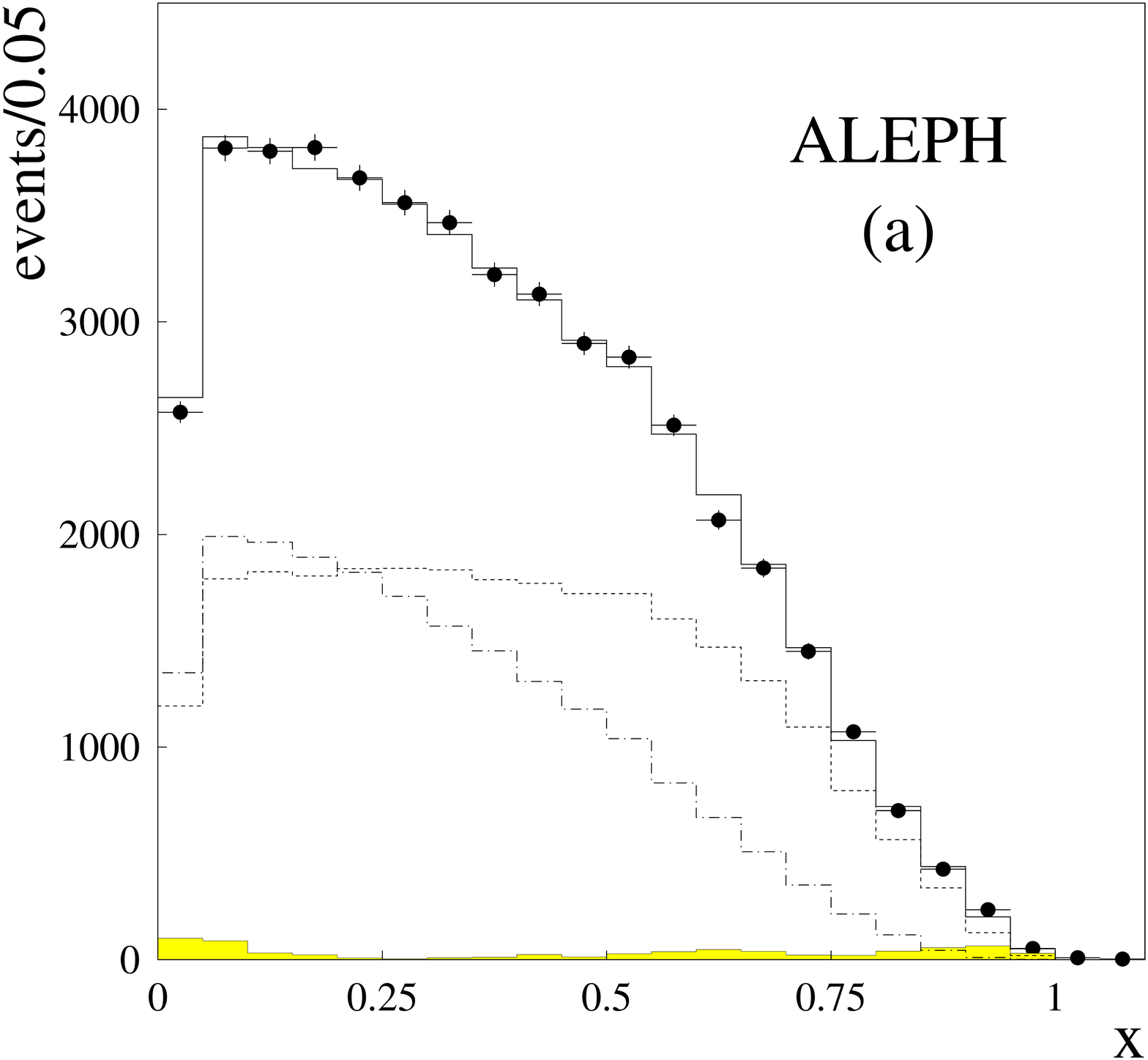,width=7.5cm} 
\epsfig{file=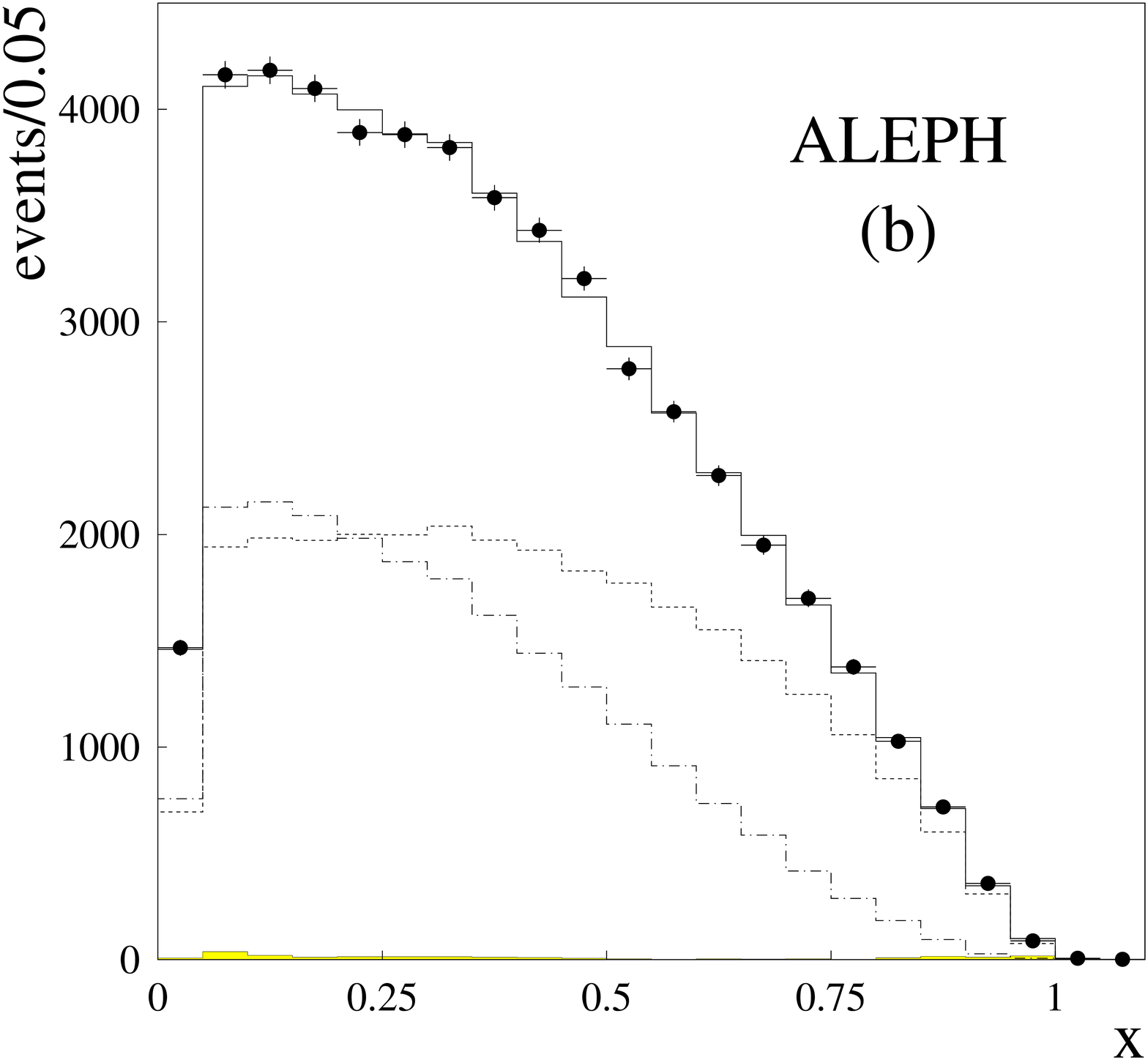,width=7.5cm}}
\mbox{
\epsfig{file=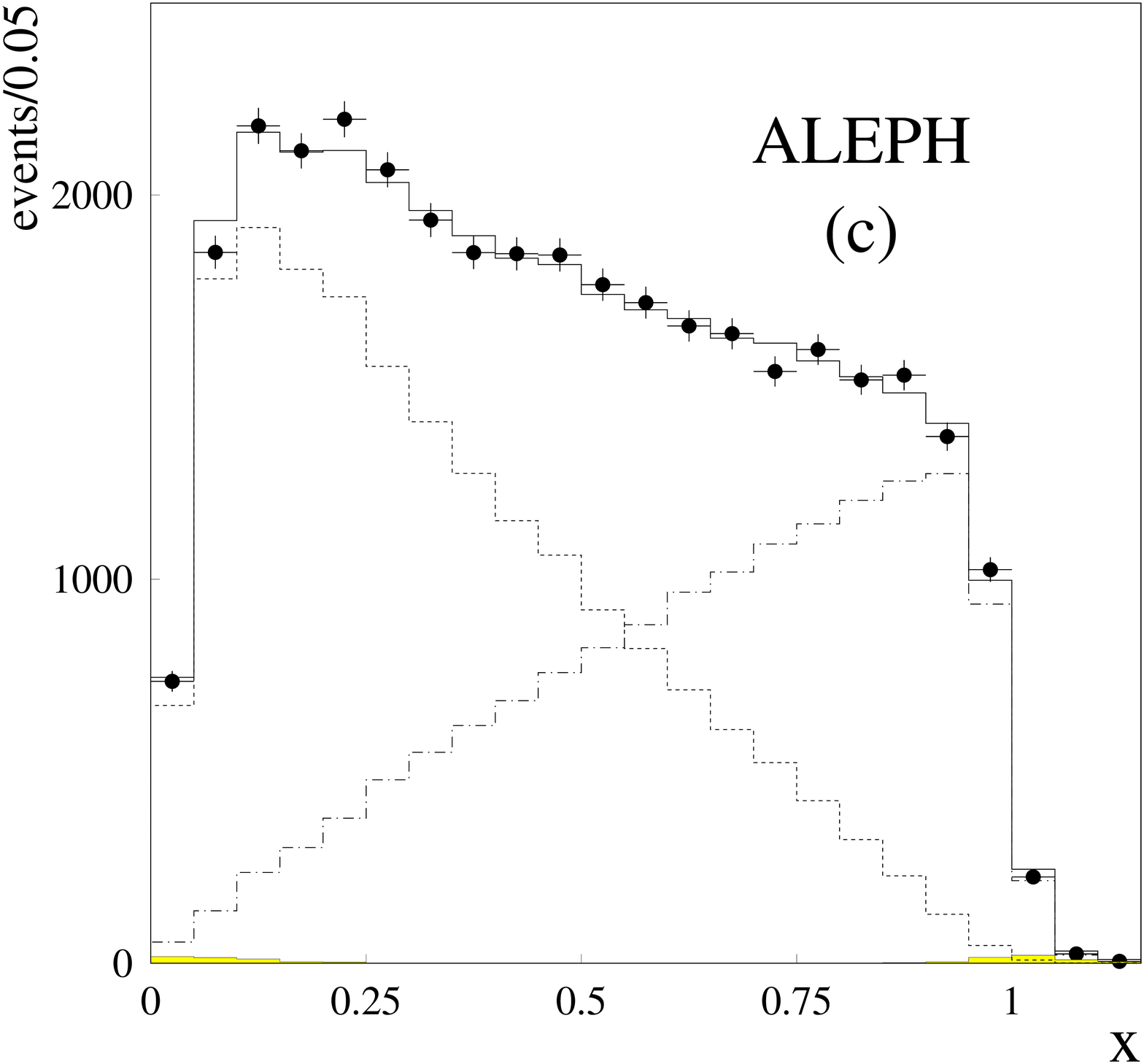,width=7.5cm}
\epsfig{file=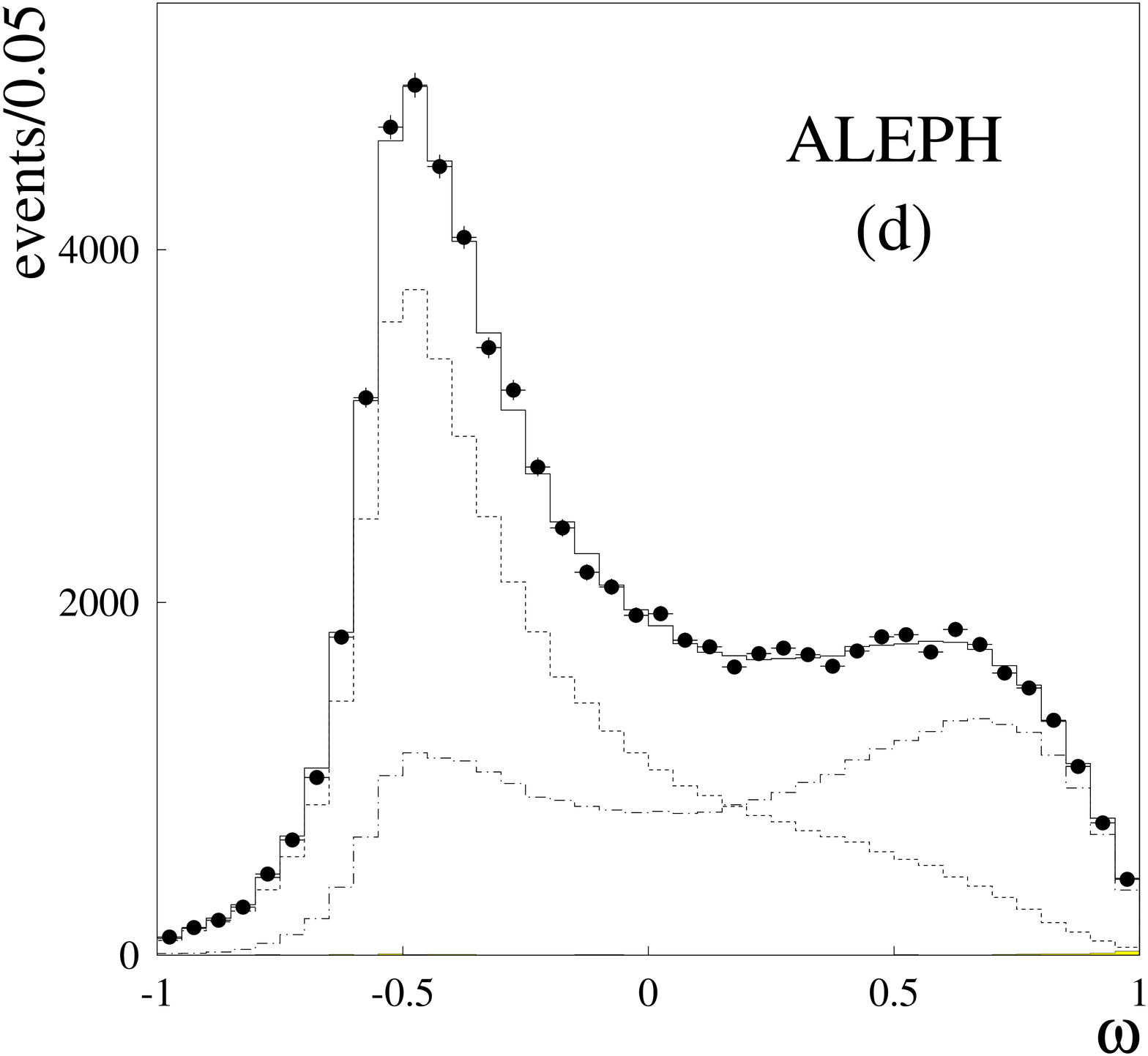,width=7.5cm} }
\mbox{
\epsfig{file=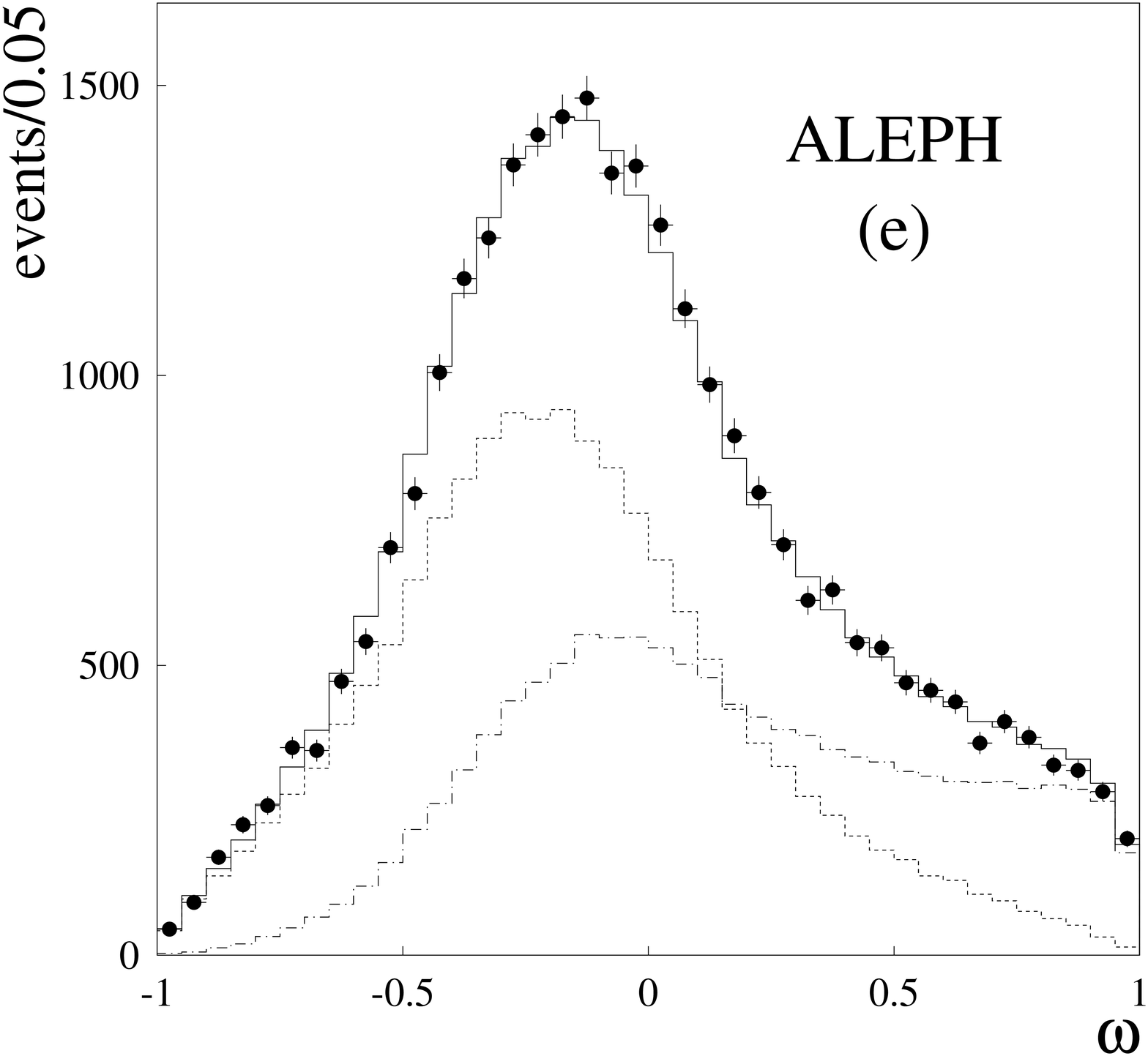,width=7.5cm} 
\epsfig{file=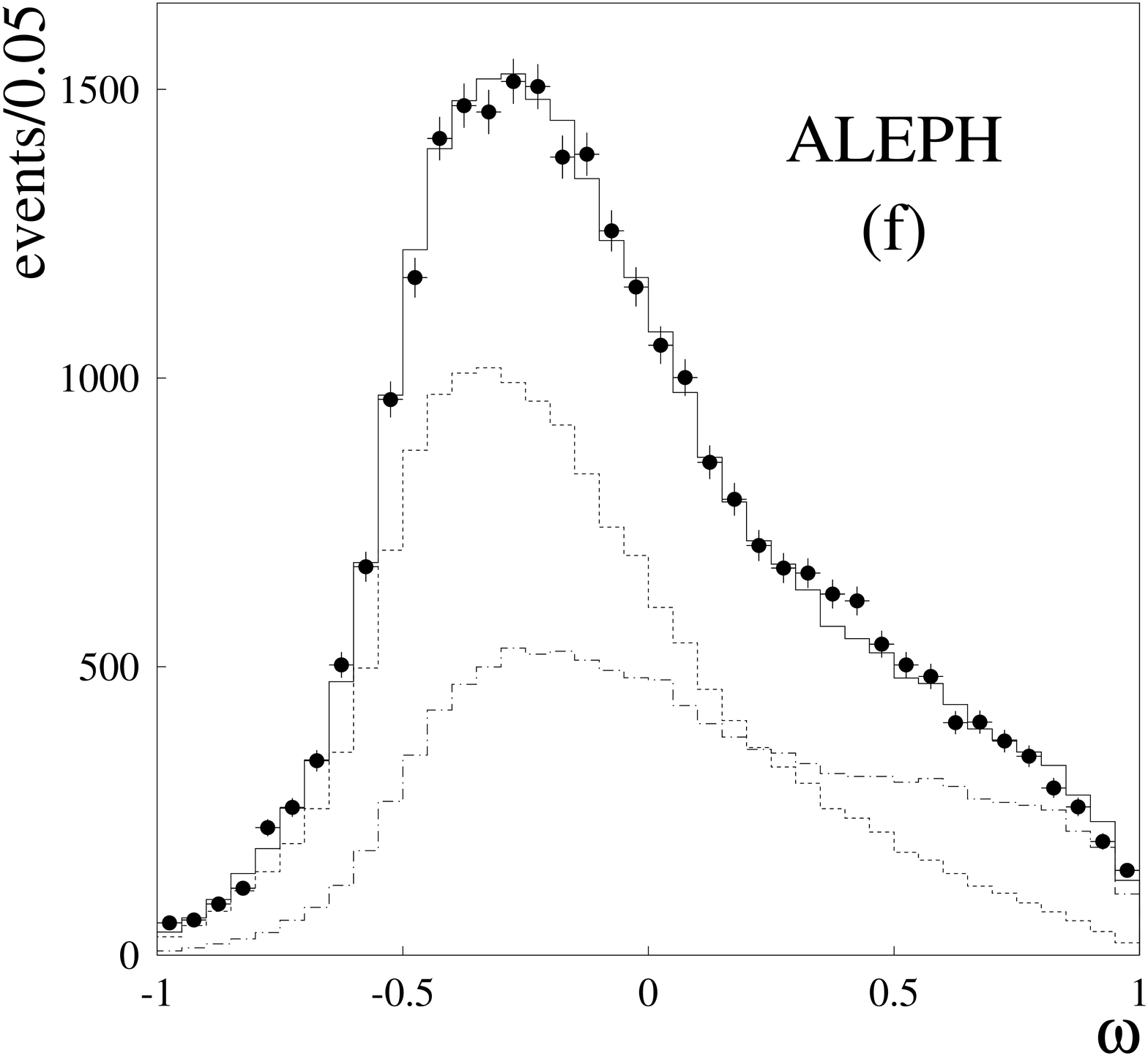,width=7.5cm} }
\caption{\protect\small Distributions of the  $x$ variable
 for $ \tau \to e  \nu   \bar\nu$ (a),
 $ \tau \to \mu \nu   \bar\nu$  (b),
 and $ \tau \to h  \nu$
 (c).   Distributions of the $\omega$ variable for
 $\tau \to \rho  \nu$ (d) and $\tau\to
 a_1\nu$: $a_1\to 3h$ (e)  and
 $a_1\to h2\pi^0$ (f). The dotted and dash-dotted lines correspond to the contributions
of left- and right-handed $\tau$'s respectively. The shaded area is the non-$\tau$ background contribution.}
\label{fig:X_all}
\end{center}
\end{figure}
\begin{table}[p]
\caption{\protect\small  Number of selected events in the
 data, selection efficiency within the angular acceptance, and 
 contamination from the $\tau$ background and the non-$\tau$ background.  
In the $3h$ mode, the $KK\pi$ and $K\pi\pi$ modes are 
treated as $\tau$ background.
\label{EP_stat}}\vspace{.5cm}

\begin{center}
\begin{tabular}{lcccc} \hline
   Channel  & $h$ & $\rho$ & $3h$ & $h 2\pi^0$   \\ \hline
 candidates & 34249  & 75296 & 27854  & 27924    \\
 efficiency (\%) & 76.6 & 78.1 & 77.2 & 66.9   \\
 $\tau$ back. (\%) & 10.6 & 10.9 & 10.9 & 24.3   \\
 Bhabha back. (\%) & $0.16\pm 0.03$  & $0.07\pm 0.02$  & $0.06\pm 0.05$   
& $0.10\pm 0.08$ \\
 other non-$\tau$ back. (\%) & 0.24  & 0.11  & 0.11  & 0.09  \\ \hline
\end{tabular}\\[.3cm]
\begin{tabular}{lccc} \hline
   Channel  &  $e$ & $\mu$& Incl. $h$ \\ \hline
 candidates &  47106 & 50585  & 134108\\
 efficiency (\%) &  75.9 & 86.4 & 63.3 \\
 $\tau$ back. (\%)  & 0.8 & 0.7 & 0.5 \\
 Bhabha back. (\%) &  $1.00\pm 0.03$ & -&$0.07\pm 0.01$ \\ 
 other non-$\tau$ back. (\%) &  0.50 & 0.30 & 0.40 \\ \hline
\end{tabular}\\[.3cm]
\end{center}
\end{table}

The polarisation is  measured independently for each channel and
for each year of data taking. Some of the systematic errors are common
to all the years and some are common to different channels.

In a first step, the measurements for each channel in the different LEP operation periods
 are  combined, taking care  of the common
systematic errors and adding the small corrections dependent on the
centre-of-mass energy described in  Section~\ref{sec:zfitt}. 
The values obtained are given in the corresponding lines of Table~\ref{EP_results}.
\begin{table}[p]
\caption{\protect\small Results for ${\cal A}_{\tau}$ and ${\cal A}_{ e}$ obtained with the single-$\tau$ analysis.
The
first error is statistical, the second systematic.
\label{EP_results}}\vspace{.5cm}
\begin{center}
\begin{tabular}{ccc} \hline
\rule[0cm]{0cm}{0.5cm}\rule[-.25cm]{0cm}{0.5cm}   Channel  & ${\cal A}_{\tau}$ (\%) & ${\cal A}_{ e}$ (\%) \\ \hline 

\rule[0cm]{0cm}{0.5cm} hadron  & 15.21 $\pm$ 0.98 $\pm$ 0.49 & 
                                 15.28 $\pm$ 1.30 $\pm$ 0.12 \\ 
\rule[0cm]{0cm}{0.5cm} rho  & 13.79 $\pm$ 0.84 $\pm$ 0.38 &
                                 14.66 $\pm$ 1.12 $\pm$ 0.09 \\ 
\rule[0cm]{0cm}{0.5cm} a1(3h)    
                               & 14.77 $\pm$ 1.60 $\pm$ 1.00 & 
                                 13.58 $\pm$ 2.11 $\pm$ 0.40 \\ 

\rule[0cm]{0cm}{0.5cm} a1(h2$\pi^0$)
                               & 16.34 $\pm$ 2.06 $\pm$ 1.52 & 
                                 15.62 $\pm$ 2.72 $\pm$ 0.47 \\ 

\rule[0cm]{0cm}{0.5cm} electron  & 13.64 $\pm$ 2.33 $\pm$ 0.96 & 
                                 14.09 $\pm$ 3.17 $\pm$ 0.91 \\ 
\rule[0cm]{0cm}{0.5cm} muon                      & 13.64 $\pm$ 2.09 $\pm$ 0.93 & 
                                 11.77 $\pm$ 2.77 $\pm$ 0.25 \\ 

\rule[0cm]{0cm}{0.5cm}\rule[-.25cm]{0cm}{0.5cm}  pion inclusive                                     
                               & 14.93 $\pm$ 0.83 $\pm$ 0.87 & 
                                 14.91 $\pm$ 1.11 $\pm$ 0.17 \\  
 \hline
\rule[0cm]{0cm}{0.5cm}\rule[-.25cm]{0cm}{0.5cm}  Combined   & 14.44 $\pm$ 0.55 $\pm$ 0.27 &  14.58 $\pm$ 0.73 $\pm$ 0.10 \\ \hline
\end{tabular}
\end{center}
\end{table}

Figure~\ref{fig:X_all} 
shows the $x$ distributions for the hadron, muon, and electron
channels; and the $\omega$ distributions  for the $\rho \nu$
and $a_1\nu$ channels ($3 h $ and  $ h 2\pi^0$).
The distributions for data and  the contributions (Monte Carlo)
from the two $\tau$ helicities and  non-$\tau$ background are superimposed.

The measurements from all the channels are then combined taking into account
their correlations evaluated  by Monte Carlo. 
The inclusive hadronic analysis has been devised as a cross-check of the
exclusive ones since their systematic uncertainties are essentially uncorrelated.
The values given in Table~\ref{EP_results}  show that this test is satisfactory.
Statistically, on the contrary, the exclusive and inclusive analyses are strongly
 correlated so that the inclusive measurement brings only little new information
 in the final result.

The final values of ${\cal A}_e$ and ${\cal A}_\tau$  are given
in  the last line of Table~\ref{EP_results}.
 Their errors  are corrected for the correlation between opposite hemispheres.

\begin{figure}[p]
\begin{center}
\mbox{
\epsfig{file=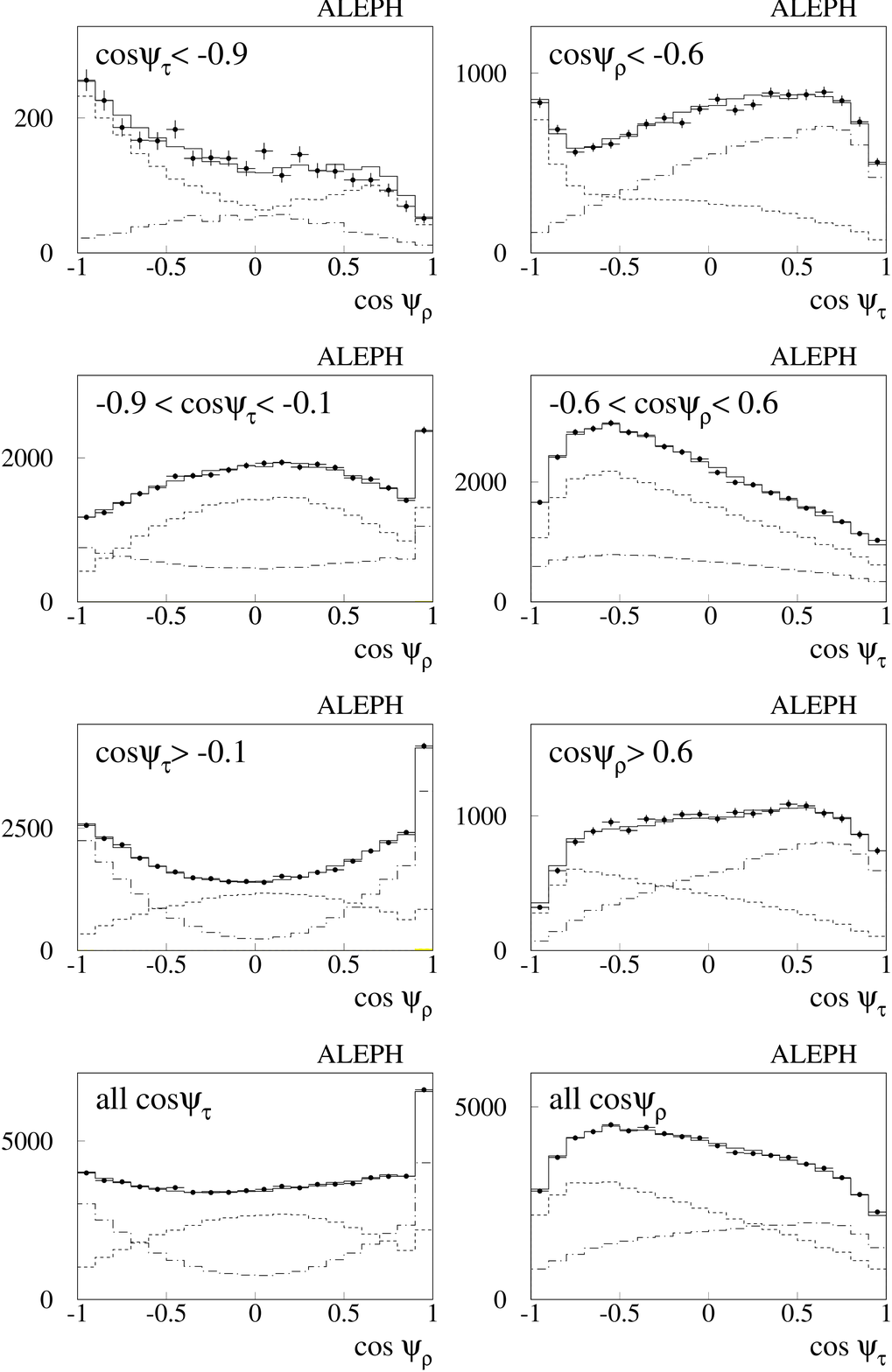,width=15cm} }
\caption{\protect\small Distributions of the two decay angles in the $\tau\to\rho\nu$
 channel. The vertical axes are in unit of events/0.1.  The dotted and dash-dotted lines correspond to contributions
of left- and right-handed $\tau$'s respectively. The 
non-$\tau$ background contribution is negligible. }
\label{fig:rhoang_all}
\end{center}
\end{figure}

As an illustration of the understanding of the $\rho$ channel,
 Fig.~\ref{fig:rhoang_all}
shows the $\tau$ and $\rho$ decay angle distributions
with good agreement
  between data and Monte Carlo, even for low energy $\rho$'s ($\cos\psi_\tau<-0.9$).

\section{Analysis with the \mbox{\boldmath $\tau$} direction
method}
\label{sec:taudir}
\subsection{Philosophy of the method}

The $\tau$ polarisation analysis described in this section is based on
two ideas: use of proven analysis tools and enhancement of the 
measurement sensitivity. On one hand, event selection, particle 
identification, photon and $\pi^0$ reconstruction, and 
decay classification are taken from previous ALEPH 
analyses of $\tau$ leptonic and hadronic branching 
ratios~\cite{hbr,lbr}, and of the hadronic spectral 
functions~\cite{spectrala,spectralb}. On the other hand, the maximum
sensitivity to the $\tau$ polarisation can be achieved in the hadronic
decay channels through a complete set of observables including the 
determination of the initial $\tau$ direction. This new information,
as compared to standard analyses, is obtained through a kinematic
reconstruction complemented by the precision measurement of the
charged particles using the microvertex detector in order to remove part
of the intrinsic ambiguity of the method. 

\subsection{Event selection and decay classification}

\subsubsection{\boldmath$\tau$-pair selection}

This analysis is based on a global event selection that retains 
$\tau$ pair candidates from Z decays. Each event is divided into two
hemispheres along the thrust axis. The $\tau$ event selection is described
in~\cite{hbr} and references therein. The method is based on the
removal of non-$\tau$ background identified as Bhabha, mu-pair, hadronic,
photon-photon induced, and cosmic ray events. A well-tested procedure of
reconstructing the spatial energy flow~\cite{perfo} is used in order 
to include all the energy measurements and their topology in the detector.

However, to enhance the selection efficiency and to reduce the biases 
in the polarisation measurement, some minor modifications of the usual 
cuts~\cite{hbr} are introduced:
\begin{itemize}
\item cuts against $\gamma \gamma$ events and the cut on the energy 
of the leading tracks in both hemispheres against Bhabha and di-muon events 
are not applied when both hemispheres are identified to be hadronic.  
\item in order to reduce the contamination from $\gamma \gamma
\rightarrow \tau \tau $ and $\gamma \gamma \rightarrow q \bar{q} $
background, events with one hemisphere with a single charged hadron and no $\pi^0$ are
removed if $|\cos\theta| > 0.792 $ and $x < 0.088$.
\item hadronic cuts are  loosened taking advantage of the
more precise hadronic mass determination after $\pi^0$ reconstruction,
rather than relying on the energy-flow variables used in the selection.
\end{itemize}

\subsubsection{Particle identification}
\label{pidsect}

A likelihood method for charged particle identification
incorporates the relevant information from the detector.
In this way, each charged particle is assigned a set of probabilities
from which a particle type is chosen. A detailed description of the
method can be found in Refs.~\cite{lbr} and~\cite{brold}.
\begin{figure}[p]
\begin{center}
\mbox{
\epsfig{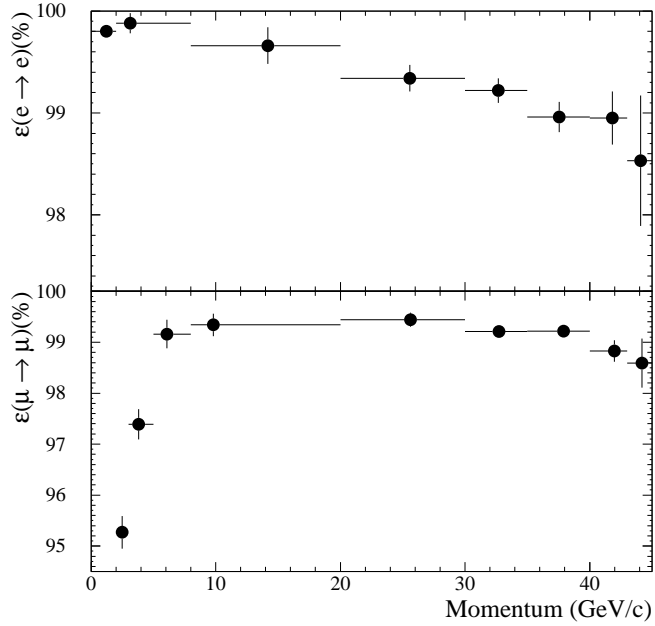}}
\end{center}\vskip-.5cm
\caption[.]{\protect\small The identification efficiencies for electrons and muons as 
a function of momentum for 1994-1995 data (in percent). 
The efficiencies simulated by the Monte Carlo for
the $\tau$ sample are corrected by measurements performed on control
samples as discussed in the text.}
\label{pid_l_9495}
\end{figure}
\begin{figure}[p]
\begin{center}
\mbox{
\epsfig{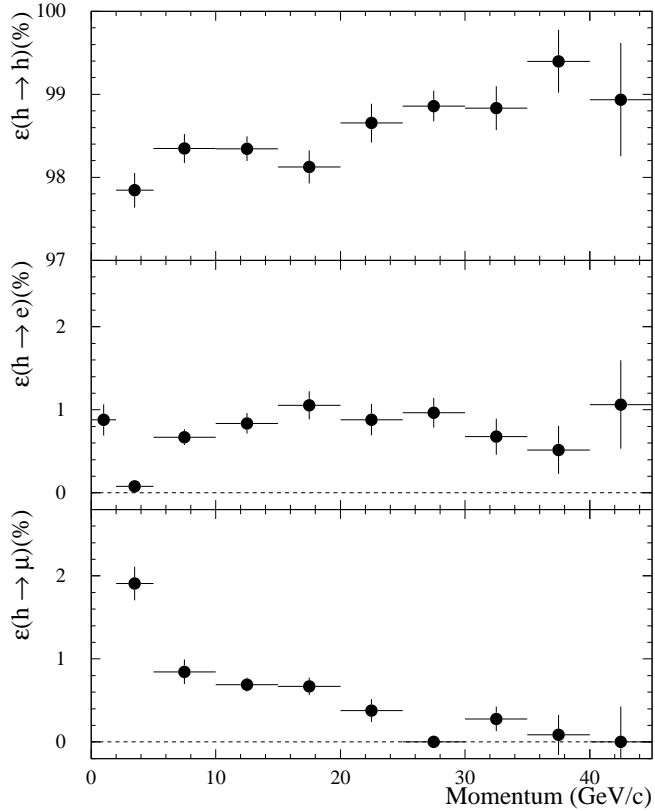}}
\end{center}\vskip-.5cm
\caption[.]{\protect\small The identification efficiency and misidentification
probabilities for hadrons as a function of momentum for 1994-1995 data 
(in percent). The values are obtained directly from data using 
a $\pi^0$-tagged hadron sample.}
\label{pid_h_9495}
\end{figure}

Briefly, the particle identification is based on the following information:
{\it (i)} the dE/dx in the TPC,
{\it (ii)} the longitudinal and transverse shower profile in  the ECAL
near the extrapolated track, and
{\it (iii)} the energy and the average shower width in the HCAL, 
together with the number of  planes, out of the last ten of the
HCAL, which had fired and the number of hits in the muon chambers. Because of the
calorimetric requirements, the complete identification method is only 
applied to particles with momenta larger than 2 GeV/$c$.

The performance of the identification method has been studied in detail using
samples of leptons from Bhabha, $\mu$-pair, and two-photon  events, 
and hadrons from $\pi^0$-tagged hadronic $\tau$ decays.
All data sets cover the full angular and momentum range~\cite{hbr,
lbr}. Ratios of particle identification efficiencies in the data
and the corresponding Monte Carlo samples are obtained as a function of
momentum. They are then used to correct the Monte Carlo efficiencies in the
$\tau^+ \tau^-$ sample, thus taking into account the different momentum 
distributions and particle environment. Misidentification probabilities
are obtained in the same way. The results of these analyses were already given
for the 1990-1993 data~\cite{lbr}. 
Figures~\ref{pid_l_9495} and~\ref{pid_h_9495} show
the corresponding efficiencies and misidentification probabilities measured
in the 1994-1995 data.
\begin{table}[b]
\caption{\protect\small   Particle identification efficiencies and
misidentification probabilities (in percent) in one-prong $\tau$ decays
as corrected from control data samples. The values are integrated on the
corresponding momentum spectra ($h$ refers to the single hadron channel)
above 2~GeV/$c$, excluding the cracks between ECAL modules.
\label{pid}}
\begin{center}
\begin{tabular}{lrrr} \hline
\multicolumn{1}{c}{Id.$\downarrow$ \hspace{1cm} True $\rightarrow$} &
\multicolumn{1}{c}{\rule[0cm]{0cm}{0.5cm} $e$} &
\multicolumn{1}{c}{$\mu$} &
\multicolumn{1}{c}{$h$}  \\ \hline 
\ $e$   & 99.57 $\pm$ 0.07 & $\le$0.01        & 0.71 $\pm$ 0.04  \\ 
\ $\mu$ & $\le$0.01        & 99.11 $\pm$ 0.08 & 0.72 $\pm$ 0.04  \\ 
\ $h$   & 0.32 $\pm$ 0.06  & 0.88 $\pm$ 0.08  & 98.45 $\pm$ 0.06 \\ 
\ no id.& 0.09 $\pm$ 0.04  & 0.01 $\pm$ 0.01  & 0.12 $\pm$ 0.03 \\ \hline
\end{tabular}
\end{center}
\end{table}
\longpage

Table~\ref{pid} gives the efficiency matrix for one-prong $\tau$ decays
used for this analysis, i.e., for particles with a momentum larger
than 2~GeV/$c$ and not in a crack region between ECAL modules.
The 2~GeV/$c$ cut is not applied to electron candidates
because of the good dE/dx separation at low momentum, while the
crack veto is not required for muons which are essentially identified
in the HCAL. The numbers refer to the  1991-1995 data sample.

\subsubsection{\boldmath Non-$\tau$ background}
\label{nontausect}

A new method is developed to directly measure in the final data samples the
contributions of the major non-$\tau$ backgrounds: Bhabhas, $\mu^+ \mu^-$ 
pairs, and $\gamma \gamma \rightarrow e^+ e^-, \mu^+ \mu^-$ 
events. The procedure does not rely on an absolute normalization 
for the simulated channels, which is difficult to obtain reliably after 
the large necessary rejection is applied. It only makes use of
a qualitative description of the distribution of the discriminating
variables. The basic idea is to apply cuts on the data in order to reduce
as much as possible the $\tau \tau$ population while keeping a high
efficiency for the background source under study. The background fraction
is directly fitted in the cut data sample and then extrapolated to the
standard sample.

For Bhabha background, different event topologies are considered in turn. For
$e\mbox{-}e$ and $e\mbox{-}crack$ topologies, the acoplanarity angle 
(defined by the two tracks azimuthal difference) is required to be 
larger than $179^\circ$ or the acollinearity angle (i. e. 180$^\circ$ 
minus the angle between the directions of the two tracks)  should be 
less than $175^\circ$ 
with the difference of transverse energies being less than 3 GeV. The
corresponding Bhabha efficiency (compared to the normal sample) is $88\%$ as
determined from the simulation. The angular distribution of the
restricted sample is then fitted to $\tau \tau$ and Bhabha components 
(also including a small contribution from the other non-$\tau$ backgrounds)
from the simulation. Therefore, the Bhabha Monte Carlo input is only used to determine 
the (large) selection efficiency and the $\cos\theta$ shape inside the final
sample, not relying on any determination of the absolute Monte Carlo normalization. 
The derived Bhabha contribution has a statistical uncertainty which is
assigned as a systematic error. Several combinations of variables have been 
tried, showing a good stability of the result within its error.
For the more numerous $e\mbox{-}h$ and $h\mbox{-}h$ samples, a total
energy cut of 55 GeV is applied and the main reduction of $\tau$ events
is achieved by suppressing true hadrons as compared to electrons 
misidentified as hadrons: this is done by restricting the opposite 
hadron in an $e\mbox{-}h$ event to have an electron identification probability larger 
than 0.01 (most of the true hadrons are below this value). A similar
approach is used for the $h\mbox{-}h$ samples with even tighter
identification probability cuts for hemispheres with only one charged
hadron (in this case $86\%$ of the hadrons have an electron probability 
smaller than 0.0002). The Bhabha signal then appears on a distribution
of the acollinearity angle between the two hemispheres ($\delta \theta$)
at values close to $180^\circ$.

A similar technique is used to estimate the $\mu$-pair background.  
An efficiency of $90\%$ is obtained by requiring an acoplanarity 
angle larger than $178^\circ$ for $\mu \mu$ events in the final sample. 
Here, the fitted distribution is
that of the calculated photon energy along the beam for a postulated 
$\mu \mu \gamma \gamma_{\mathrm{beam}}$ kinematics to take the most general case
compatible with only the information on the two muons. For $\mu\mbox{-}h$ and
$h\mbox{-}h$ final states, the procedure follows the one used for Bhabha background.

The measurement of the  remaining background induced by $\gamma \gamma$ interactions
is based on
cuts on the total event energy (typically $E<35$~GeV, 
depending on the channels)
and the acollinearity angle between the two hemispheres 
($\delta \theta < 175^\circ$). The $\gamma \gamma$ background shows up clearly
at small overall transverse momentum well above the small $\tau$ surviving 
contribution.

\begin{figure}[htb]
\begin{center}
\mbox{
\epsfig{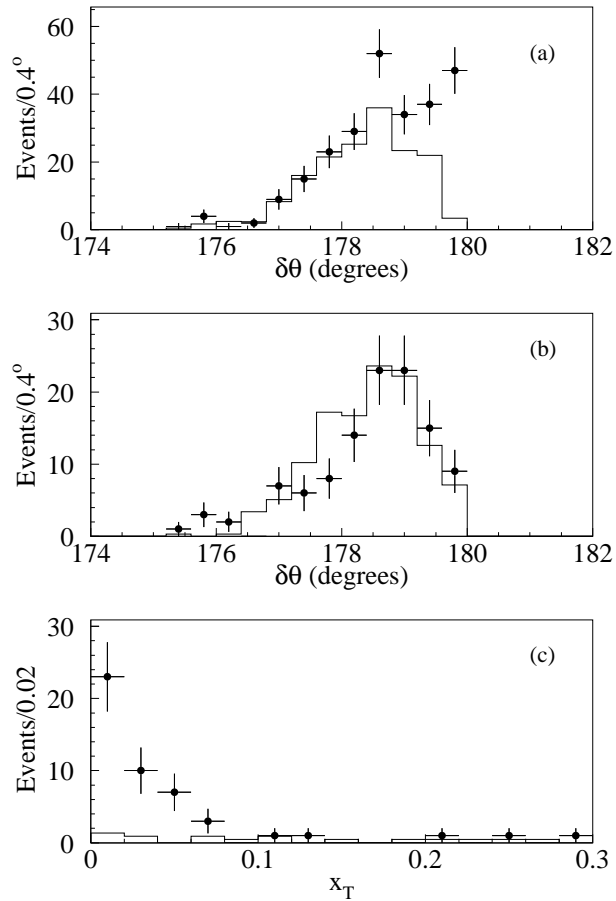}}
\end{center}\vskip-.5cm
\caption[.]
{\protect\small Distributions in the final $\rho$ sample after applying cuts to reduce the
$\tau$ contribution and enhance non-$\tau$ background (see text for details):
(a) acollinearity distribution to enhance Bhabha background;
(b) acollinearity distribution to enhance $\mu \mu$ background;
(c) total transverse momentum normalized to the centre-of-mass energy to
enhance $\gamma \gamma \rightarrow hadrons$ background. The histograms
give the respective $\tau$ contributions as expected from the simulation.}
\label{nontau_meth}
\end{figure}

All the above non-$\tau$ background contributions are determined on the
polarisation data sample with the help of special cuts. The corresponding
Monte Carlo generators are only used to compute the efficiency of the cuts.
This method is illustrated in Fig.~\ref{nontau_meth} showing the
evidence for Bhabha and 
$\gamma \gamma \rightarrow \tau \tau ~\mbox{or}~ q \bar{q}$ backgrounds
in the $\rho \nu$ channel, the contribution from $\mu \mu$ being 
negligible in this case.

\begin{figure}[htb]
\begin{center}
\mbox{
\epsfig{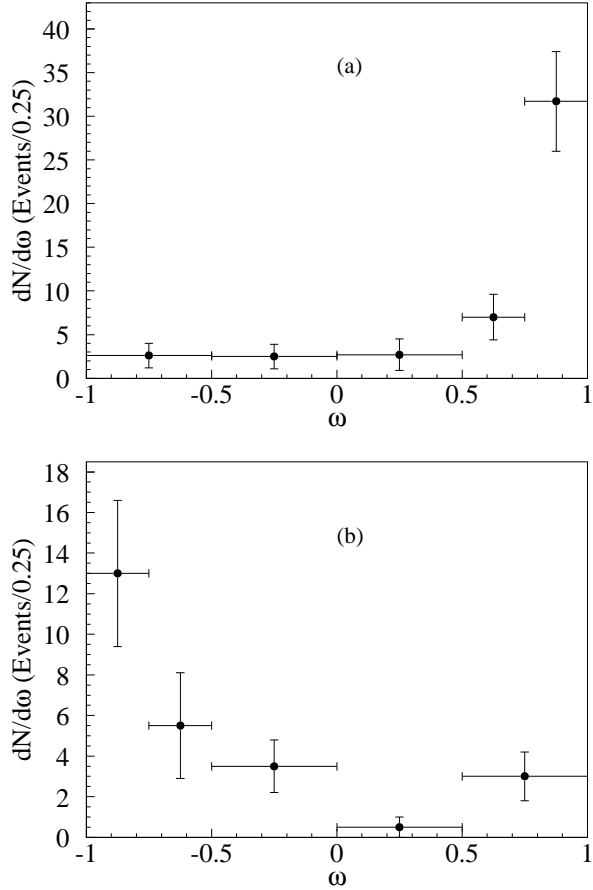}}
\end{center}\vskip-.5cm
\caption[.]
{\protect\small Distributions of the $\omega$ variable in the 
final $\rho$ sample after applying cuts to reduce the
$\tau$ contribution and enhance non-$\tau$ background (see text for details):
(a) Bhabha background;
(b) background from $\gamma \gamma \rightarrow hadrons$.
In both cases the small remaining $\tau$ contribution estimated from the 
simulation has been subtracted out.}
\label{nontau_om}
\end{figure}

As an important bonus, this procedure leads to a determination from data
alone of the background distributions which are relevant for the polarisation
measurement. This method is preferable to using the Monte Carlo predictions
which are questionable in view of the large rejection factors achieved.
Distributions of the $\omega$ variable obtained 
in the $\rho \nu$ channel are given in Fig.~\ref{nontau_om}.

As in the analysis of Section~\ref{section:singletau}, cosmic ray background is determined to be very small after the additional 
cuts~\cite{lbr}. The contamination from hadronic Z decays affects 
only the hadronic channels. Its small contribution is estimated with the JETSET~7.4
generator~\cite{lund}. Tests were previously made to ascertain the reliability
of the prediction for the rate of surviving low multiplicity events by
comparing to data~\cite{hbr}. A $30 \%$ uncertainty was derived from 
these tests.

The non-$\tau$ background contributions are given in Table~\ref{channel_stat},
separately for the considered channels.
\begin{table}[ht]
\caption{\protect\small Channel identification efficiencies and contaminations as measured
on the data for the $\tau$ direction method.
The efficiency includes a geometrical acceptance factor of~0.86.
``acol'' refers to the sample of events with a leptonic $\tau$ decay, where  
information from the acollinearity angle \cite{polarb,acol} is used.
\label{channel_stat}}
\begin{center}
\begin{tabular} {lcccc}          \hline
Channel &
 $h$  & $\rho$ & $3h$ & $h2 \pi^0$ \\
\hline 
 candidates     & 33350 & 78553 & 25287 & 28757\\
 efficiency (\%)    &  63.1 &  69.1 & 62.9  & 57.5\\
 $\tau$  back. (\%) &  
         7.3    &  9.1   &   3.9  &   20.7 \\
 Bhabha         &  57 $\pm$ 10 & 118 $\pm$ 40 & -     &  9 $\pm$ 5 \\
 $\mu\mu$       &  20 $\pm$ 5 & 16 $\pm$ 14  & -    &  2 $\pm$ 2  \\
 $q\overline{q}$  &  61 $\pm$ 30  & 205 $\pm$ 69 & 72 $\pm$ 35  &  37 $\pm$ 15\\
 $\gamma\gamma ee$    
                &  -       &   -         & -     & 2 $\pm$ 3 \\
 $\gamma\gamma\mu\mu$
                & 7 $\pm$ 7  &   -         & -     &  -      \\
 $\gamma\gamma\tau\tau  +  \gamma\gamma q\overline{q}$
                &15 $\pm$ 10 &  31 $\pm$ 11  &0 $\pm$ 4& 3 $\pm$ 5
\\\hline
\end{tabular}
\\[.3cm]
\begin{tabular} {lccc}          \hline
Channel &
  $e$ & $\mu$ & $acol$\\ 
\hline 
 candidates     &  52952 & 50249 & 85035 \\
 efficiency (\%)    &   68.7 & 74.8 & 74.4 \\
 $\tau$  back. (\%) &    0.7    &  1.0     & 0.7\\
 Bhabha         &   383 $\pm$ 66 & -    & 187 $\pm$ 37 \\
 $\mu\mu$       &  10 $\pm$ 3 & 212 $\pm$ 55 & 94 $\pm$ 24 \\
 $q\overline{q}$  &  8 $\pm$ 4  &  -    &  8 $\pm$ 4  \\
 $\gamma\gamma ee$    & 187 $\pm$ 45 &  -        & 91 $\pm$ 20 \\  
 $\gamma\gamma\mu\mu$ &  7  $\pm$ 5 &  241 $\pm$ 57 &106 $\pm$ 24 \\
 $\gamma\gamma\tau\tau  +  \gamma\gamma q\overline{q}$&  22 $\pm$
6 &   22 $\pm$ 6  & 28 $\pm$ 7  \\ \hline 
\end{tabular}
\end{center}\vskip-.5cm
\end{table}
\subsubsection{Photon and \mbox{\boldmath $\pi^0$} reconstruction}
\label{photsect}\longpage
As in the previous analysis,
the photon and $\pi^0$ reconstruction is performed according to 
a likelihood method in which several estimators and probabilities are 
computed in order to distinguish between genuine and fake photons
produced by hadronic interactions in ECAL or by electromagnetic
shower fluctuations~\cite{hbr}.

\begin{figure}[p]
\begin{center}
\mbox{
\epsfig{file=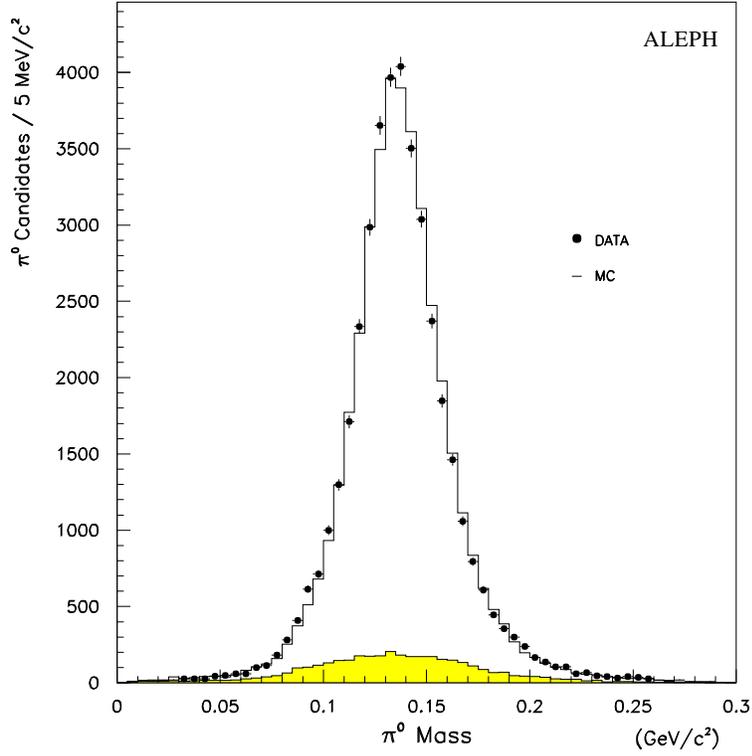,height=10.0cm
}}
\vskip -.5cm
\end{center}
\caption[.]
{\protect\small Photon-pair mass distribution for reconstructed $\pi^0$ candidates.
Data (points) are compared to the simulation (histogram). The shaded
histogram gives the simulated contribution where at least one of the
photons is fake.}
\label{pi0_mass}
\end{figure}

An algorithm is used to form photon pairs in each 
hemisphere~\cite{hbr} using a $\pi^0$ estimator 
$D^{\pi^0}_{i,j} = P_{\gamma_i} P_{\gamma_j} P_{\pi^0_{i,j}}$ for two
photons $i$ and $j$, where $P_{\pi^0_{i,j}}$ is the probability
from a kinematic $\pi^0$-mass constrained fit.  
A good agreement between data and Monte Carlo is observed for 
the distributions of $P_{\pi^0_{i,j}}$ and $D^{\pi^0}_{i,j}$. 
Figure~\ref{pi0_mass} shows the corresponding 
$\pi^0$ mass distributions with the expected contribution from
photon pairs including at least one fake photon.

\begin{figure}[p]
 \begin{center}
\mbox{
\epsfig{file=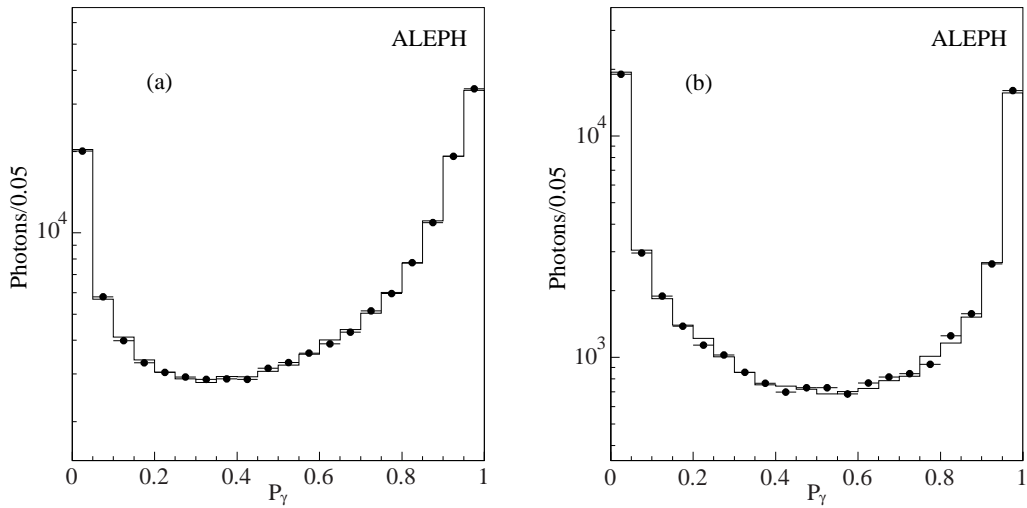,width=15.0cm
}}
\end{center}
\vskip-5mm
\caption{\label{photon_estim}\protect\small Photon estimators used to distinguish between genuine and fake 
photons ($P_\gamma = 1$ corresponds to the case of genuine photons):
(a) estimators for photons used in resolved $\pi^0$ reconstruction (the photon energy
is not used),  (b) estimators for residual photons (using the photon energy).
Data (points) are compared to the simulation (histogram).}
\end{figure}

High-energy $\pi^0$'s with overlapping photon showers are reconstructed
through an analysis of the spatial energy deposition in the ECAL
towers (``unresolved $\pi^0$'s''). All the remaining single 
(``residual'') photons are then considered and those with $P_\gamma > 0.5$ 
are selected as $\pi^0$ candidates. The $P_\gamma$ distributions are given
in Fig.~\ref{photon_estim} for all photons and for the residual ones after
renormalizing the amount of fake photons in the simulation by a factor
depending on the $\tau$ decay hadronic final state and ranging from
1.17 to 1.28. The shapes in data and Monte Carlo are in good agreement.
The energy dependence of the data distributions is also well reproduced
by the simulation: hence the photon efficiencies resulting from the
$D_{ij}^{\pi^0}$ and $P_\gamma$ cuts are adequately described by the
Monte Carlo, as already observed in Section~\ref{sec:EP_sys_gam} and Fig.~\ref{fake_eff}.

\begin{figure}[ht]
\begin{center}
\mbox{
\epsfig{file=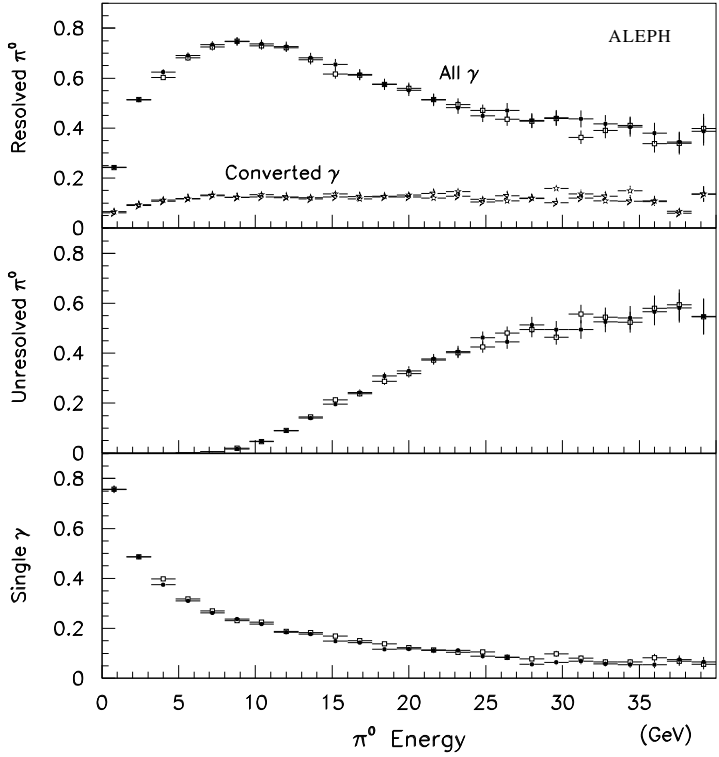,height=10.0cm
}}
\end{center}
\caption[.]
{\protect\small Fractions of different $\pi^0$ types as a function of the $\pi^0$ energy.
The points represent the data and the open squares the simulation. Also
indicated with stars are the fractions for resolved $\pi^0$ containing at 
least one converted photon.}
\label{pi0_frac}
\end{figure}

Finally, photon conversions are identified following the procedure
described in Ref.~\cite{hbr}. They are added to the list of 
good photons and included in the $\pi^0$ reconstruction.
The fractions of resolved, unresolved and residual $\pi^0$'s
are presented in Fig.~\ref{pi0_frac} as a function of the $\pi^0$
energy. It is seen that the data distributions are well reproduced
by the simulation. The fraction of resolved ``calorimetric'' $\pi^0$'s remains significant
out to large energies in this analysis as compared to the previous one
(see Section~\ref{sec:EP_sel_rho} and Fig.~\ref{fig:epi0_rho}); this effect is the result of a looser
$\pi^0$ definition here, where highly asymmetric $\gamma \gamma$ 
pairs are more likely to be counted as resolved $\pi^0$'s,
whereas they would be identified as unresolved in the other method
because of the tight invariant mass cut.

\subsubsection{Decay classification}
\label{addcutsect}

$\tau$ decays are classified according to the number of
charged tracks and their identification, and the number of reconstructed
$\pi^0$'s. 
Beyond the classification achieved in the previous 
analyses~[17,19-21],
some additional cuts have been developed 
in order to reduce the $\rho$ and $K^*$ background level in the $h$ channel,
as explained below. 

In order to reject $\rho$ decays  where the $\pi^0$ is close to the
charged pion and undetected as such, 
the sum of the energies in the first two stacks of  
ECAL  is required to be less  than $75\%$  of the charged pion momentum.
The energy is measured  including all the pads  
in a cone of $30^\circ$ around the track. 
To further reduce the $\rho$ contamination, the hemispheres with one 
charged hadron and a single residual photon with a probability
$P_\gamma$ greater than 0.2 are rejected. 

\begin{figure}[htb]
\begin{center}
\mbox{
\epsfig{file=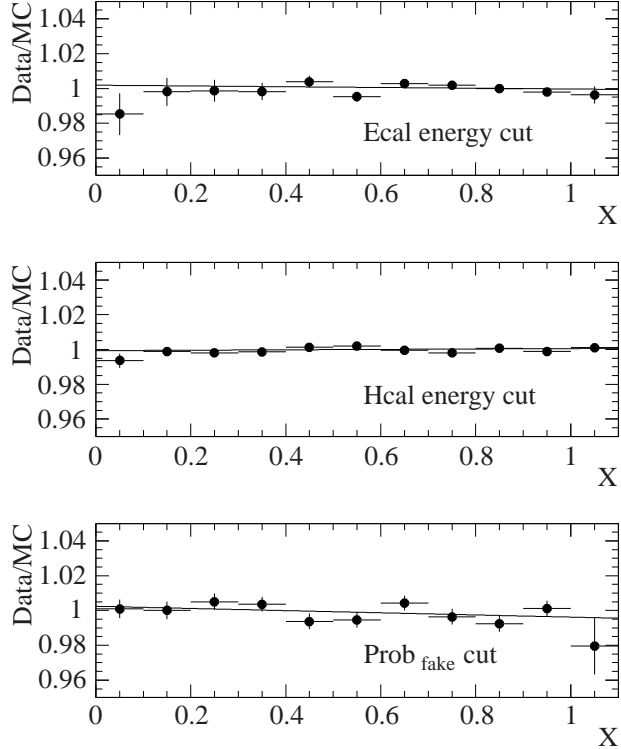,height=10.0cm}}
\end{center}
\caption{\protect\small The energy dependence of the ratio of data to Monte Carlo 
efficiencies for the additional  cuts in  the pion channel in the
1991-1995 data.}
\label{f_pi_cuts}
\end{figure}

The $K^*$ (decaying into $K^0_L\, \pi$) component  
is reduced by a cut on the HCAL  
energy deposit ($E_{\mathrm{hcal}}$) and the  azimuthal offset  ($\delta\phi$)
between the track impact and the energy weighted barycentre of the energy deposit 
in HCAL~\cite{kzero}. The cut is applied on  the  following variables: 
\begin{eqnarray}
   \delta_E &=& \frac{E_{\mathrm{hcal}} - P_h}{\sigma_h}\\
   \delta \phi &=& \xi~| \phi_{\mathrm{barycentre}} -\phi_{\mathrm{track~impact}} |   
\end{eqnarray}
where $P_h$ is the hadron momentum and $\sigma_h$  the corresponding HCAL energy resolution.
The value of the parameter $\xi$ is $ +1 $ if  $\phi_{\mathrm{barycentre}}$ 
is in the direction of the charged hadron  bending,  and 
 $ -1 $  otherwise. 
The following region in the $( \delta_E \ ,\delta \phi)$  plane is cut : 
$$ \left\{ \begin{array}{ll}
           \delta_E >    &  0 \\
           \delta \phi < &  -1^\circ \\
           \delta \phi < & (\frac{2}{3}\delta_E - 3)^\circ 
        \end{array}
        \right. $$
In order to verify that those supplementary cuts do not introduce 
biases into the measurement of the $\tau$ 
polarisation, the ratio of the efficiencies for each 
of the three cuts 
on data and on  Monte Carlo, $\epsilon_{\mathrm{data}} / \epsilon_{\mathrm{MC}}$,  is 
studied as a function of the pion momentum  (Fig.~\ref{f_pi_cuts}).   
The computation of these efficiencies and of their statistical errors 
takes into account the non-pion background estimated by Monte Carlo. 
The small bias introduced by these cuts is evaluated by a linear fit and corrected in
the final polarisation measurement. The corresponding statistical
error is taken as a systematic uncertainty.

For the $\rho$ channel, a cut on the invariant mass is 
required, $m_{\rho} > 0.4~\mathrm{GeV}/c^2$, to reduce the feedthrough from 
the $h$ mode. For the  $a_1$ channel, the invariant mass is required to be 
less than the $\tau$ mass. In the case of a decay into three charged
hadrons ($3h$), three good tracks are required in the hemisphere.

The channel identification efficiencies, the
$\tau$ background, and the non-$\tau$ background contributions as measured
in the data (except for $q\bar{q}$) are shown in Table~\ref{channel_stat}.

\subsection {The \mbox{\boldmath $\tau$} direction}

\subsubsection{\boldmath
Kinematic reconstruction of the $\tau$ direction}

For events where both $\tau$'s decay into hadrons (called hadronic events),
it is possible to reconstruct the $\tau^+ \tau^-$ direction, 
thereby enhancing the sensitivity of the polarisation measurement 
in the $\rho$ and $a_1$ channels as seen in Section~\ref{sec:method}. 
This method is used here for the first time.

In hadronic events, although the momentum of each neutrino is
unknown, the $\tau$ direction can be determined up to a twofold
ambiguity. The two solutions lie at the intersection of the cones
having for axis the hadronic direction in each hemisphere and
an opening angle $\phi_0$ computed from the measured hadron momentum $P_h$
and  energy $E_h$, assuming the pion mass for each individual hadron,
\begin{equation}
 \cos \phi_0 = \frac{ 2 E_\tau E_h - {m_\tau} ^2 - {m_h} ^2 }
             {2 P_\tau P_h }~, 
\label{eq_phi}
\end{equation}
where $P_\tau$ and $E_\tau$ are the energy and the momentum of the
parent $\tau$ assumed to be produced at the beam energy, and $m_h$ the hadronic mass.

A procedure is used for events where the cones do not intersect, 
mainly because of detector resolution effects.
The hadron momenta are allowed to fluctuate within the expected
resolutions of charged and neutral pions. The $\tau$ direction is taken
as the average of the found solutions among the random trials, provided
at least 10 solutions are found for 500 trials. This procedure attributes 
a $\tau$ direction to half of the events where the two cones are not 
initially intersecting.

As a result, $80 \%$ of the hadronic events are available for the 
polarisation analysis using the $\tau$ direction. The remaining 
$20 \%$ without a reconstructed direction are analysed in the standard way.
On the whole $52\%$ of the $\tau \rightarrow \rho \nu$ decays benefit 
from this improvement.

\subsubsection{\boldmath
Handling the two $\tau$ directions}

Since the typical $\tau$ decay length is 2~mm at LEP~I, the precise
determination of the secondary charged tracks brings some 
information allowing in principle to lift the twofold ambiguity
and to choose the actual $\tau$ direction. This procedure stems from
the 3-dimensional method developed for the measurement of the $\tau$
lifetime~\cite{3dip}. In practice one can only separate the two solutions
on a statistical basis with the help of an estimator, $D_h$, and assign to
each of the two directions a probability $P_{1,2}$ to be the true one.
This procedure is explained in Appendix A, while more details can be found in Ref.~\cite{irena}.

Reference distributions for the estimator are set up with Monte Carlo
for various event configurations ($\pi\mbox{-}\pi$, $\rho\mbox{-}\rho$, $\rho\mbox{-}\pi$, etc.) 
and probabilities  $P_{1,2}$ for each of the two found $\tau$ directions to
be the true one are obtained.

If the highest probability is retained, it is found by simulation that
the closest of the two directions to the true one is chosen in 
$65 \%$ of the cases for all channels combined. A more relevant way to
characterize the procedure is to consider the precision achieved on the
$\tau$ direction. While the mean value of the angle between the two
directions is 24 mrad, the angle between the direction with the higher
probability and the true one is 12 mrad on average. The corresponding
value for the direction closest to the true one is 7 mrad (dominated by
resolution effects) and a value of 15 mrad is obtained if a random
choice between the two solutions is made. In practice, the two solutions
are retained, with their respective probabilities. 

\subsubsection{\boldmath
Analysing the polarisation with the $\tau$ direction}

The polarisation is analysed in each hemisphere using
the proper optimal variables, $\omega_{1,2}$, calculated for the
observed decay for each choice of direction using
the description of the hadronic system given in Ref.~\cite{kuhn}.
Both directions are entered into the expected decay distribution 
depending on $P_{\tau}$:
\begin{eqnarray}
 W &=& P_1 F(\omega_1) (1 + P_{\tau} \omega_1)\nonumber
        + P_2 F(\omega_2) (1 + P_{\tau} \omega_2)\\
   &=& \tilde{F} (1 + P_{\tau} \omega)
\end{eqnarray}
with the new optimal observable
\begin{equation}
 \omega = \frac {P_1 F(\omega_1) \omega_1 + P_2 F(\omega_2) \omega_2}
                {P_1 F(\omega_1) + P_2 F(\omega_2)}~.
\end{equation}

The ideal sensitivity given in Table~\ref{table:sens} is naturally 
degraded by detector resolution effects and mostly 
by the imperfect determination of the $\tau$
direction. The expected sensitivities are given in Table~\ref{sensi}
from a Monte Carlo study. The gain in sensitivity using the calculated
probabilities instead of affecting each direction with an {\it a priori}
$50 \%$ weight is quite small, but it provides the ultimate gain in precision
achievable within the ALEPH detector capabilities.

For events where one of the $\tau$'s decays leptonically and for which the
information on the $\tau$ direction cannot be retrieved, the acollinearity
angle between the decay products in the two hemispheres contains additional
information as compared to the separate analysis of single 
decays. This method~\cite{polarb,acol} is also used in this analysis.

\begin{table} [htp]
\caption{\protect\small Sensitivities obtained in the $\rho$ and the $a_1$ channels, for various
methods using the true or the reconstructed direction. Only 
events with two reconstructed directions are considered here. 
\label{sensi}}
\begin{center}
\begin{tabular} {cccc}          \hline
     method     &  $\rho $-$ had $    &
  $ a_1(3\pi) $-$ had $  & $ a_1(\pi2\pi^0) $-$ had $  \\   
  \hline
 $\omega_{\mathrm{true}}$, without ${\vec{\tau}}$    & 0.48 $\pm$0.01 & 0.41 $\pm$0.01
& 0.41 $\pm$0.01 \\ 
 $\omega_{\mathrm{true}}$, with ${\vec{\tau}}$       & 0.57 $\pm$0.01
& 0.57 $\pm$0.01 & 0.56 $\pm$0.01 \\ \hline 
 $\omega_{\mathrm{rec}}$, without ${\vec{\tau}}$     & 0.46 $\pm$0.01 &
 0.40 $\pm$0.01  & 0.37 $\pm$0.01 \\ 
 $\omega_{\mathrm{{rec}}}$, with $P_1=P_2=0.5$
      & 0.51 $\pm$0.01 & 0.47 $\pm$0.01   & 0.42 $\pm$0.01 \\ 
 $\omega_{\mathrm{rec}}$, with $P_{1,2}$
      & 0.51 $\pm$0.01 & 0.48 $\pm$0.01  & 0.43 $\pm$0.01 \\ \hline
\end{tabular}
\end{center}

\end{table}

\subsection {Summary of systematic effects} 
\begin{table} [ht] 
\caption{\protect\small Summary of  systematic uncertainties (\%) for $A_\tau$ and $A_e$
with the $\tau$ direction method.
\label{t_syst1}}
\begin{center}
\begin{tabular} {lccccccc}          \hline 
\multicolumn{8}{c}{
\rule[0cm]{0cm}{0.4cm} 
$A_\tau$} \\ 
\rule[0cm]{0cm}{0.4cm}\rule[-.25cm]{0cm}{0.65cm}Systematic effect    & $h$ & $\rho$ 
 & $3h$ & $h2\pi^0$ &$e$ & $\mu$ & $acol$    \\   \hline 
eff.  $h \rightarrow  h$ id.    &  0.17  &  0.06 & -    & 0.06 & 0.20 & 0.35
  & 0.01   \\ 
misid. $(e,\mu) \rightarrow h $  &  0.24  &  0.05 & -    & 0.09 & 0.13 & 0.25
  & 0.57   \\ 
$\tau\tau$ selection            &  0.13  &  0.03 & 0.01 & 0.01 & 0.03 & 0.04
  & -      \\ 
$\tau$ BR and background       &  0.04  &  0.05 & 0.03 & 0.09 & 0.01 & 0.02
  & 0.02   \\ 
tracking                       &  0.08  & 0.07  & 0.22  & - &  -   & 0.21
  & 0.30   \\ 
$\gamma$-reconstruction & -    & 0.22 & 0.29 & 0.66 & -  & -
  & - \\ 
$\pi^0$-reconstruction  & 0.11 & 0.29 & 0.68 & 0.62 & -  &  - 
  & - \\ 
fake photons                  & 0.31 & 0.17 & 0.28 & 0.75 & - & -
  & - \\ 
ECAL scale                    & -    & 0.20 & 0.33 & 0.63 & 0.15 & -
  & - \\ 
ECAL + HCAL cut                &  0.22  & -     &  -    & -    & -    & -
  & -      \\ 
modelling             & -  & -  &  0.68 & 0.68 & - & -
  & -  \\ 
non-$\tau$ background           &  0.24  & 0.16  & 0.07  & 0.05 & 0.73 & 0.50
  & 0.60    \\ 
$\tau$  MC statistics           &  0.34  & 0.30   & 0.61   & 0.77  & 0.73
 & 0.80 & 1.44  \\ \hline 
TOTAL                           & 0.66   & 0.57   & 1.30   & 1.70  & 1.07
 & 1.06 & 1.69  \\ \hline \hline
  \multicolumn{8}{c}{
\rule[0cm]{0cm}{0.4cm} $A_e$} \\ 
\rule[0cm]{0cm}{0.4cm}\rule[-.25cm]{0cm}{0.65cm}Systematic effect    & $h$ & $\rho$ 
 & $3h$ & $h2\pi^0$ &$e$ & $\mu$ & acol    \\   \hline
tracking & 0.04 & - & - & - & - & 0.05 & - \\
non-$\tau$ background       &  0.13   & 0.08  & 0.02  & 0.07 & 1.23 & 0.24
  & 0.24    \\ 
modelling                     & -      & -     &  0.40  & 0.40  & - & -
  & -  \\ \hline
TOTAL                           & 0.13   & 0.08   & 0.40   & 0.41  & 1.23
 & 0.24 & 0.24  \\ \hline 
\end{tabular}
\end{center}
\end{table}

Possible sources of systematic effects come from the $\tau$ selection, particle
identification, and $\pi^0$ reconstruction. The dominant systematic uncertainties
are due to  photon identification and $\pi^0$ reconstruction, the non-$\tau$
background, and, in the case of the $a_1$ modes, the dynamics of the decay.
Table~\ref{t_syst1} shows the summary of the main components of the systematic
uncertainties for $A_\tau$ and $A_e$. 

The particle identification efficiency matrix is measured as a function
of momentum using data samples of kinematically identified particles,
as described in Section~\ref{pidsect}. The procedure yields both the
relevant efficiencies and their associated uncertainties which are
dominated by the statistics of the control samples. The effect of this
systematic uncertainty on the polarisation measurement is significant
only for the $e$ or $\mu$ channels.

To study the effect of the $\tau$ pair event selection, the difference
in the selection efficiency between data and the Monte Carlo
is analysed for each cut, as discussed in detail in Refs.~\cite{hbr,
lbr}. The systematic error  on the selection has a significant effect 
only on the $h$ channel where the nearly flat momentum distribution is
distorted at low values by the $\gamma \gamma$ cuts and for large
momenta by the Bhabha and $\mu$-pair cuts.

\longpage
The branching ratios used in the Monte Carlo are corrected using the 
measured values of ALEPH~\cite{hbr,lbr} and the 
uncertainty in the amount of feedthrough in a given $\tau$ decay channel
from the other modes is obtained by varying the branching ratios within 
the errors given in Refs.~\cite{hbr,lbr}, taking into 
account the correlations between them. The impact on the $\tau$ 
polarisation measurement is negligible.

The tracking resolution affects only the $a_1$ decay since three reconstructed
charged hadrons are required. In this decay the tracks 
are generally close to each other and therefore
partially overlap. A reconstruction problem could occur if the two
same-sign tracks have nearly the same transverse momentum. Data and Monte 
Carlo distributions of the angle between the two tracks at the vertex are
in fact in agreement within statistics and a systematic uncertainty for the
polarisation is derived. The systematic error on the momentum calibration
is also estimated with the largest effect observed on the $\mu$ channel.

\longpage
A complete investigation of the observables entering photon and $\pi^0$
reconstruction is undertaken. For each variable 
a comparison between data and Monte Carlo is performed in order to 
search for possible systematic deviations. These differences are
then injected in the complete reconstruction procedure, affecting both the
number of events in a given channel and the corresponding $\omega$
distributions. Each data-Monte Carlo comparison yields a systematic 
uncertainty limited by the statistics of the data. As discussed in full 
detail in Ref.~\cite{hbr}, these studies include the effect of the
clusterization algorithm, the photon probabilities,
the threshold energy cut for low energy photons or the minimal distance
between a charged track and a photon in the electromagnetic calorimeter.
Similarly, the effect of the cuts on the $\pi^0$ estimators is also
determined. All the contributions to the total systematic uncertainties
for this source are added in quadrature. 

Another source of uncertainty in the photon and $\pi^0$ treatment
is due to the problem of fake photons generated by hadron interactions
in ECAL or electromagnetic shower fluctuations. This source of photon
candidates is underestimated in the Monte Carlo simulation with respect
to the data. This deficit produces a bias on the polarisation measurement
and needs to be corrected for with the systematic uncertainty carefully
estimated. In general, the effect has been studied by removing fake
photons in the simulation and repeating the analysis: decay
classification and polarisation determination in all channels. This
method establishes the sensitivity of the measurement to the problem.
Then the fake photon deficit in the simulation is measured in all the
needed hadronic channels by fitting probability distributions in data
and Monte Carlo as discussed in Section~\ref{photsect} and the corresponding
systematic uncertainties are deduced. 

A more direct method is used in
the $h$ channel for the events with a residual photon where a tighter
probability cut was applied to reject more $\rho$ background 
(see Section~\ref{addcutsect}). Since the $\rho$ contribution can be
well estimated by Monte Carlo, it is possible to use the results of this
cut both on data and Monte Carlo to estimate the relative effect of fake 
photons as a function of momentum. Any momentum dependence of the Monte Carlo
deficit would produce a bias for the polarisation. Such a bias is measured
(Fig.~\ref{f_pi_cuts}) and the statistical uncertainty in the linear
momentum fit is taken as systematic error. This procedure in the $h$ channel
does not take into account the loss of events with multiple fake photons
entering the definition of $\pi^0$ candidates. A special study was made to
estimate this effect as the simulation tends to underestimate the fake
photon multiplicity, with the result that the effect on the polarisation
is small compared to that of residual photons.
 
Since some additional calorimetric cuts are introduced in order to further
reduce the level of the $\rho$ and $K^*$ background in the hadron channel,
it is important to study their effect on the hadron momentum distribution. 
The results have been presented in Section~\ref{addcutsect} and the 
corresponding systematic uncertainties derived.

Finally, the effect on the ${ A}_\tau$ and ${ A}_e$ measurements from the
different sources of non-$\tau$ background has been studied in detail
directly on the data with the method presented in Section~\ref{nontausect}.
This procedure allows  the energy distribution
of the remaining Bhabha events in the selected sample to be investigated and 
a direct measurement of their $\cos\theta$ distribution to be made. 
The latter information is particularly relevant for the $A_e$ determination
because of the asymmetric Bhabha contamination. Using as input the 
distributions derived from data with their statistical and systematic
uncertainties, their impact on $A_e$ and $A_\tau$ is assessed. It is
important to note that the dominant statistical uncertainty affects both
the normalization of the background and the shape of the $\omega$ and
$\cos\theta$ distributions. The most seriously affected channel is 
clearly that of the electron, both for $A_\tau$ and $A_e$. 
\subsection{Results}

Table~\ref{t3}  gives the results for  ${\cal A}_e$ and ${\cal A}_\tau$ in the seven
channels considered in this analysis. The statistical and systematic
uncertainties are given separately. The last line presents the combination of these results 
taking into account their correlations, including the polarisation correlation between the
 two hemispheres in an event.
 
\begin{figure}[p]
 \begin{center}
  \begin{tabular}{cc}
   \subfigure[{\protect\small Events without $\tau$ direction.}]{\epsfig{          %
      file=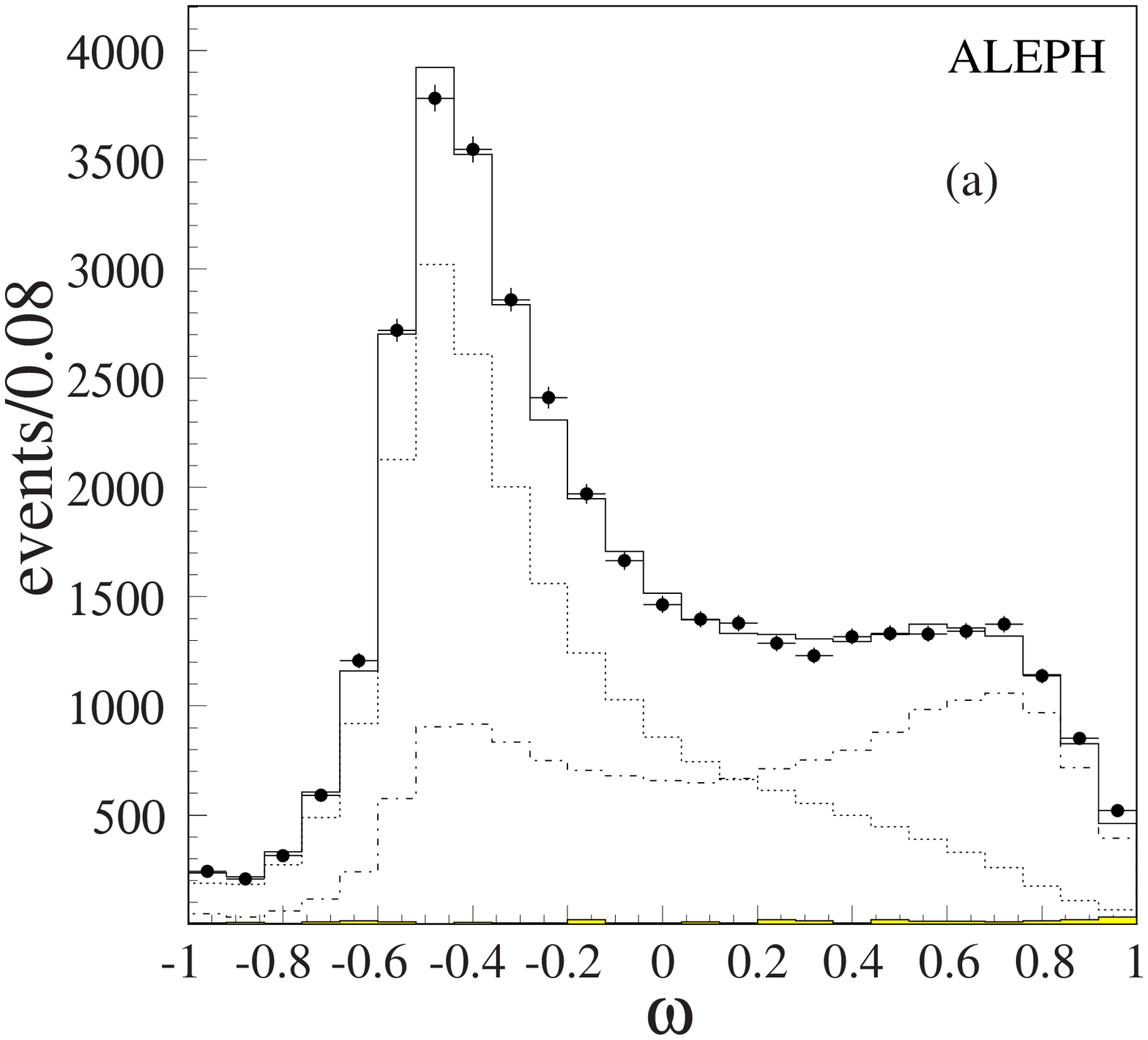,width=75mm}} & 
   \subfigure[{\protect\small Events with $\tau$ direction.} ]{\epsfig{          %
      file=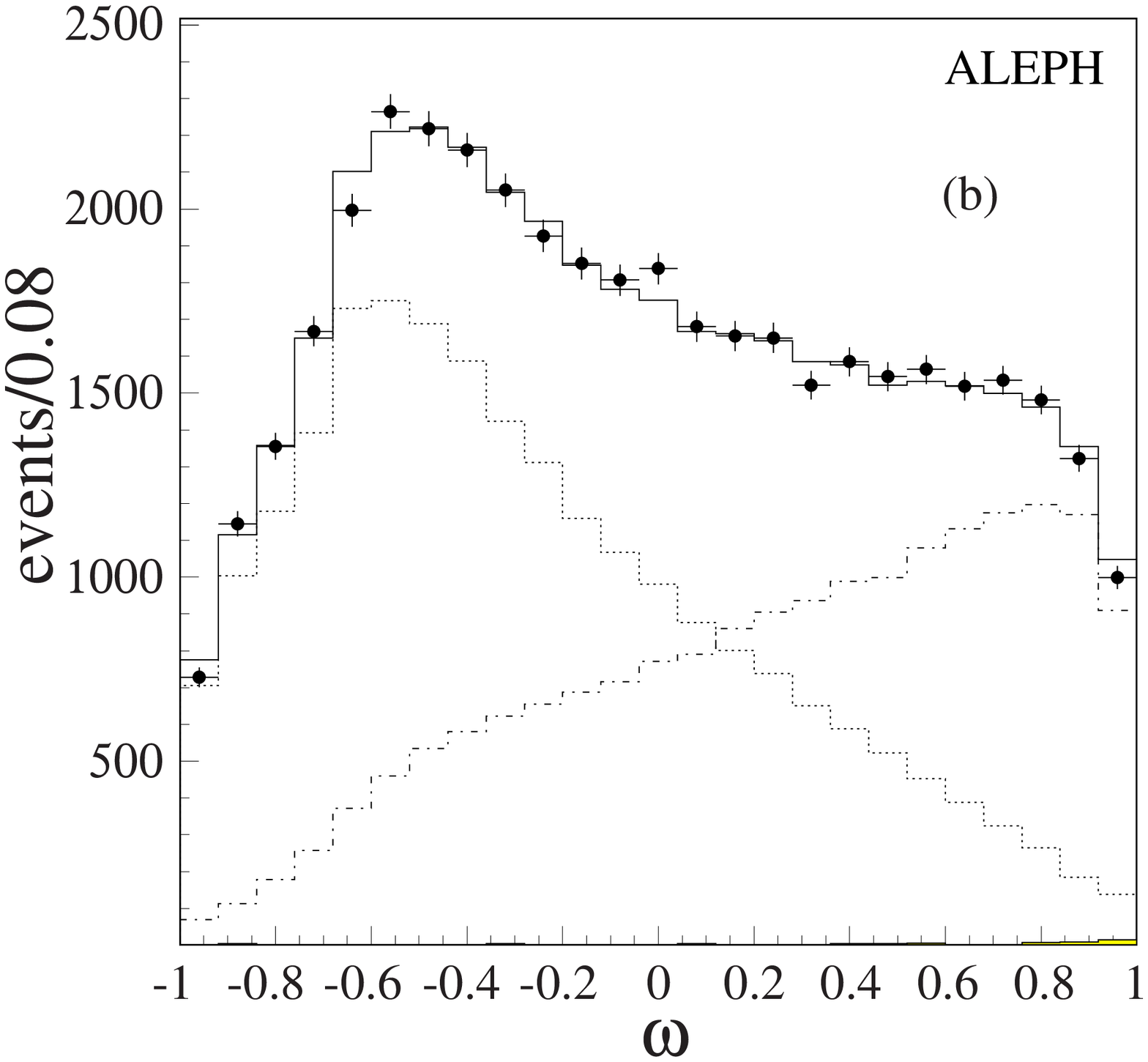,width=75mm}}  \\
\multicolumn{2}{c}{
   \subfigure[{\protect\small Standard $\omega$ for all events.} ]{\epsfig{file=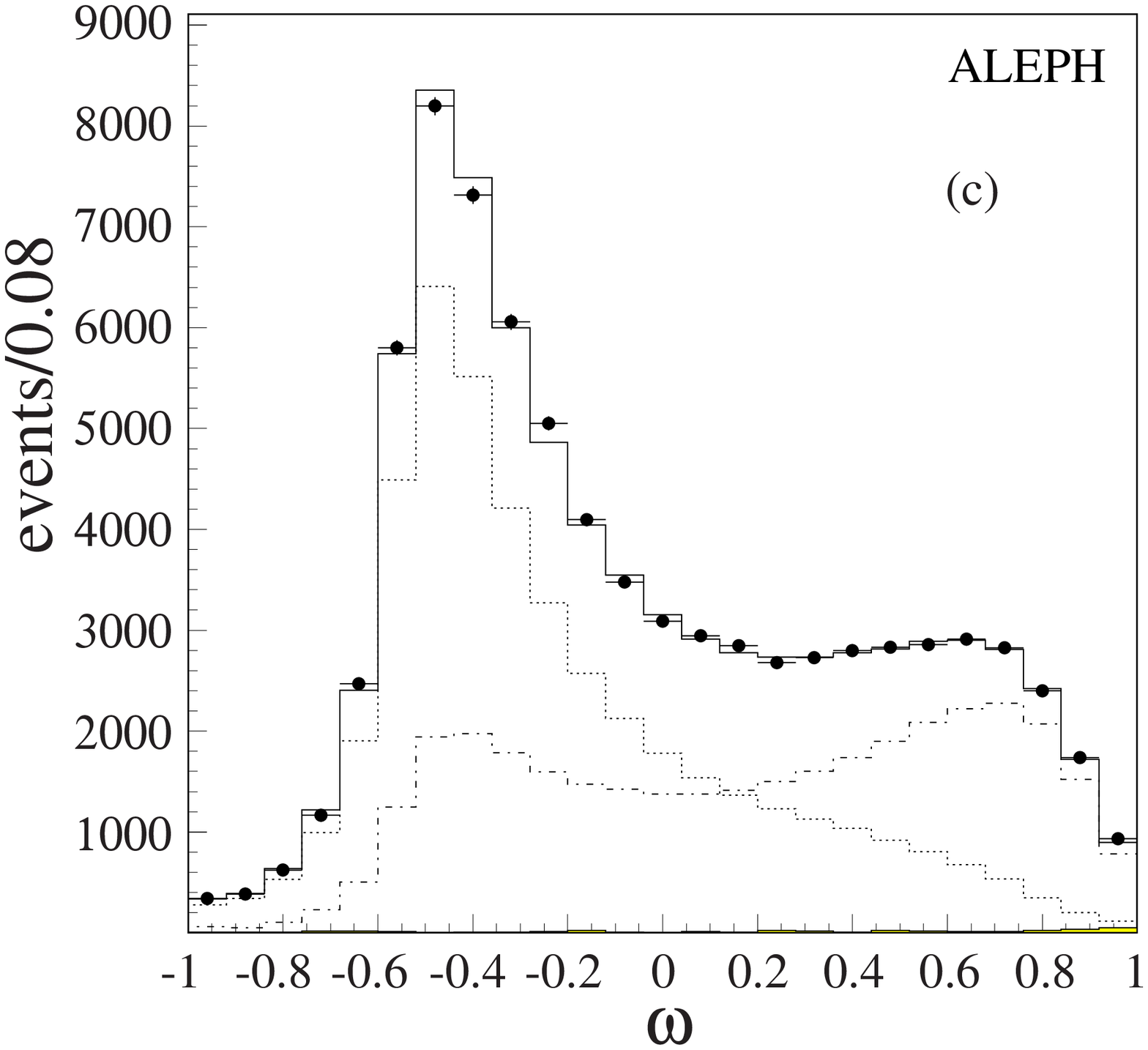,width=75mm}}}
 \end{tabular}
\end{center}
\vskip-5mm
\caption[b3]{\label{fig:diromega3}\protect\small Distributions of the $\omega$ observable used in the
polarisation fit of the $\rho$ channel.
The data are shown by points with statistical errors bars.
The dotted and dash-dotted lines corresponds to the contributions 
of left- and right-handed $\tau$'s respectively, as fitted in the data. 
The shaded area shows the non-$\tau$ background contribution.
The solid line corresponds to the sum of all simulated
contributions. }
\end{figure}

\begin{figure}[p]
 \begin{center}
  \begin{tabular}{ccc}
   \subfigure[{\protect\small Events without $\tau$ direction.}]{\epsfig{          %
      file=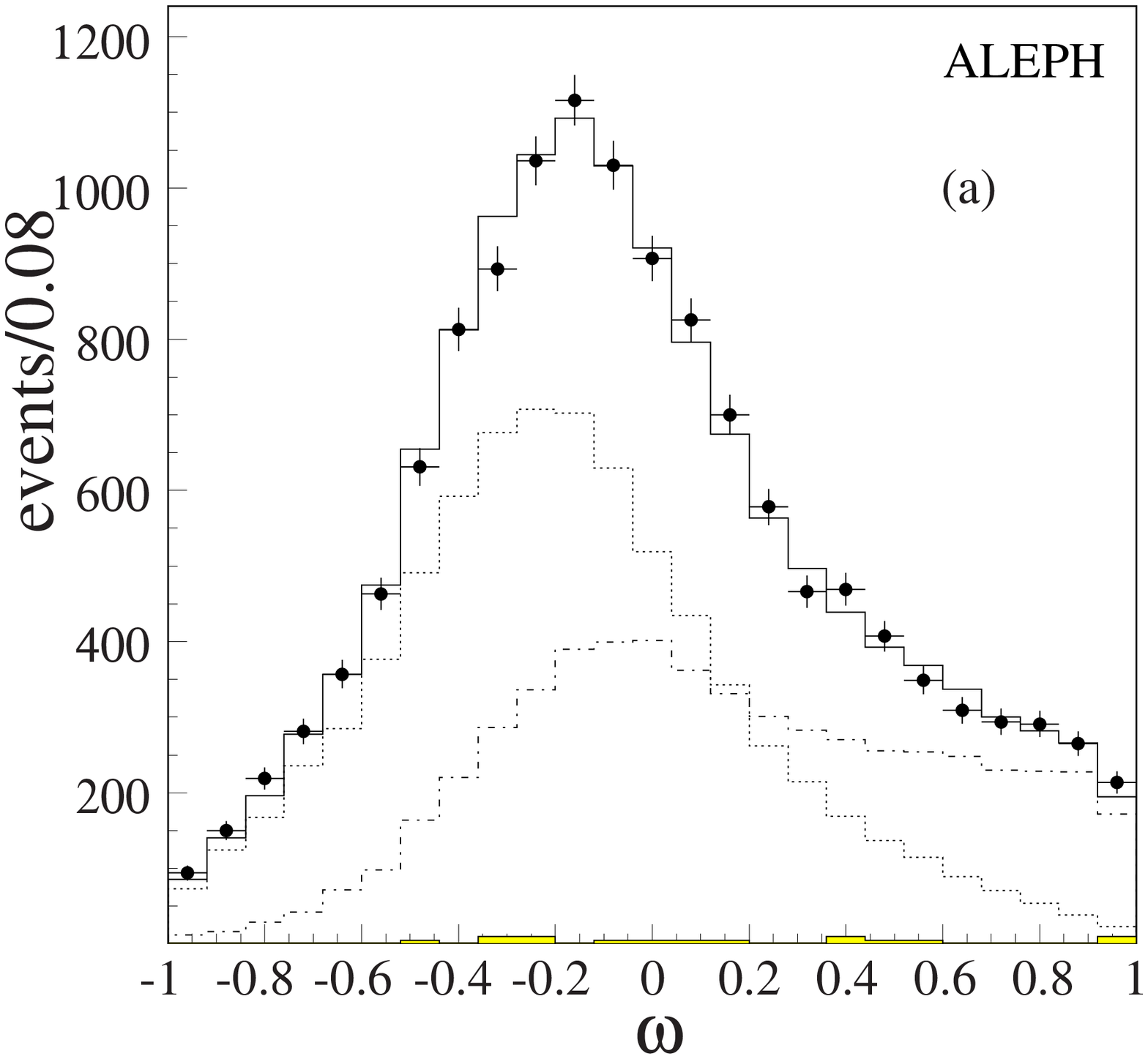,width=75mm}} &
   \subfigure[{\protect\small Events with $\tau$ direction.}]{\epsfig{          %
      file=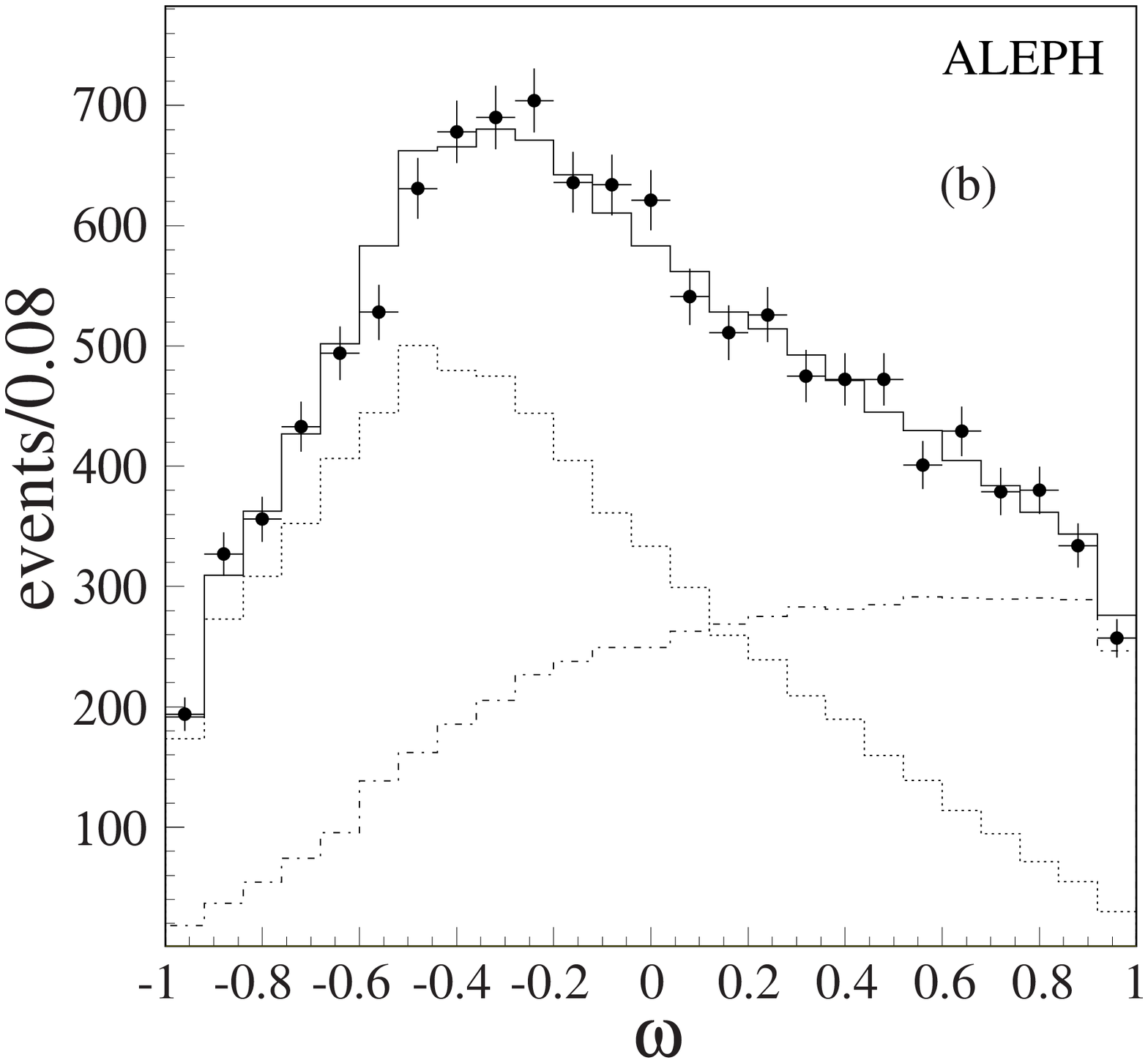,width=75mm}} \\
\multicolumn{2}{c}{
   \subfigure[{\protect\small Standard $\omega$ for all events.}]{\epsfig{file=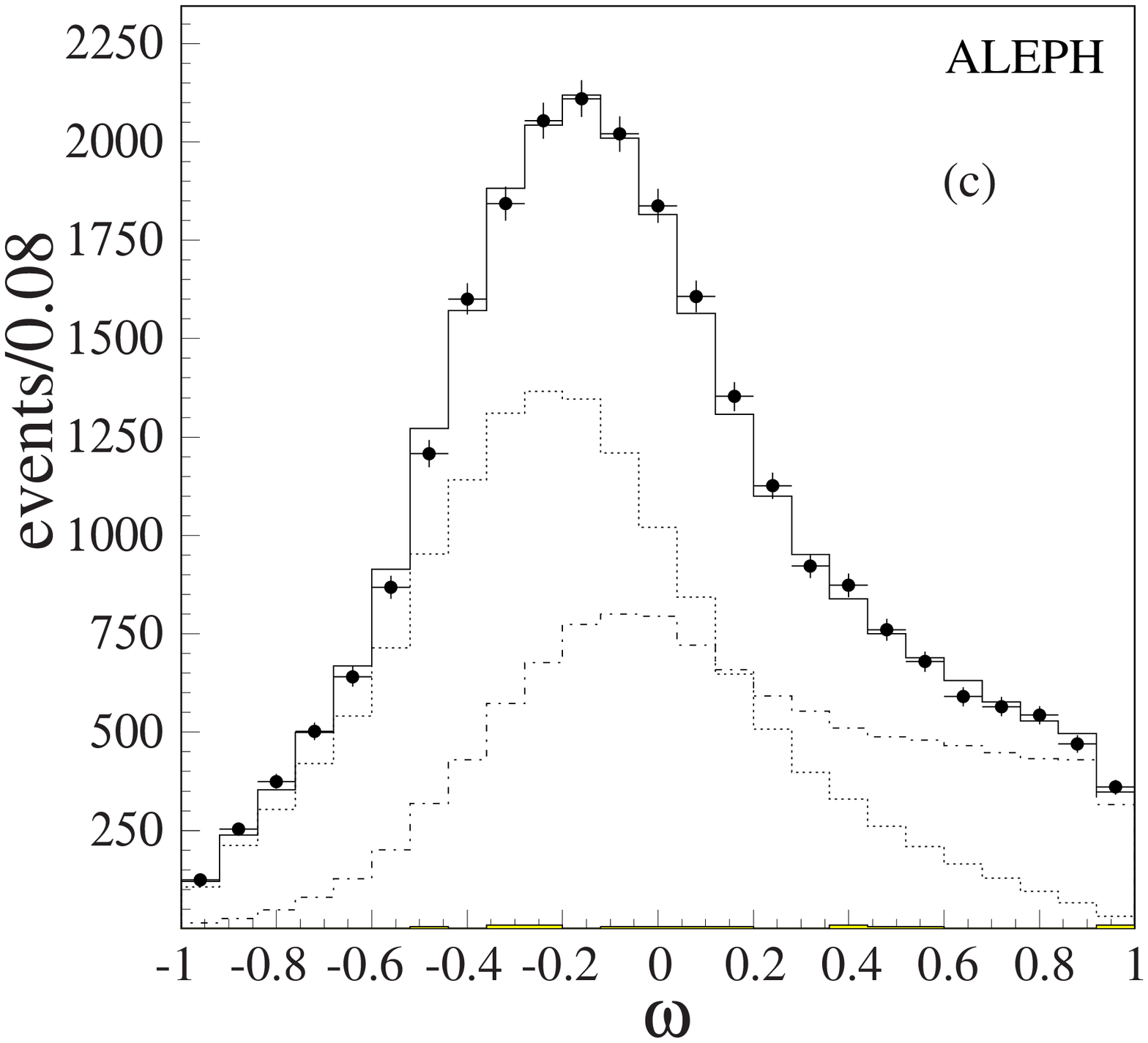,width=75mm}}}
 \end{tabular}
\end{center}
\vskip-5mm
\caption[b4]{\label{fig:diromega4}\protect\small Distributions of the $\omega$ observable used in the
polarisation fit of the $a_1 \rightarrow 3 \pi^{\pm}$ channel. 
The data are shown by points with statistical errors bars.
The dotted and dash-dotted lines corresponds to the contributions 
of left- and right-handed $\tau$'s respectively, as fitted in the data. 
The shaded area shows the non-$\tau$ background contribution.
The solid line corresponds to the sum of all simulated
contributions. }
\end{figure}

\begin{figure}[p]
 \begin{center}
  \begin{tabular}{ccc}
   \subfigure[{\protect\small Events without $\tau$ direction.} ]{\epsfig{          %
      file=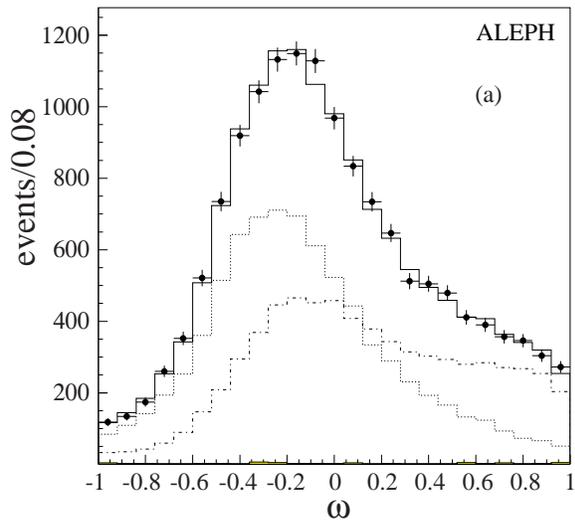,width=75mm}} &
   \subfigure[{\protect\small Events with $\tau$ direction.} ]{\epsfig{          %
      file=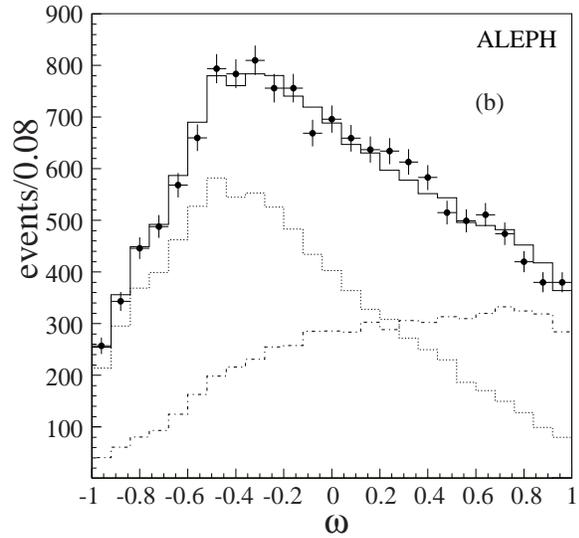,width=75mm}} \\
\multicolumn{2}{c}{
   \subfigure[{\protect\small Standard $\omega$ for all events.} ]{\epsfig{file=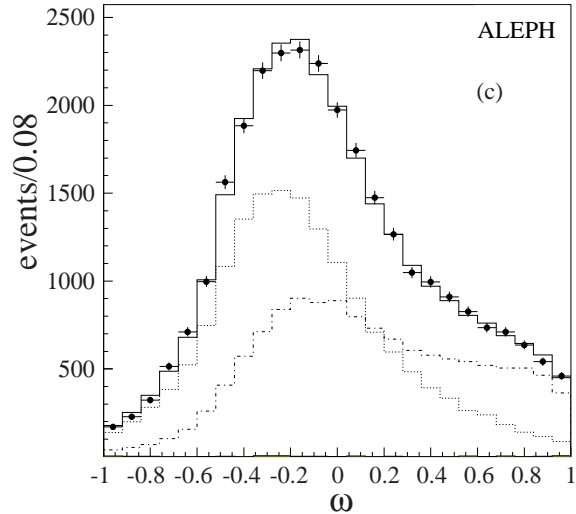,width=75mm}}}
 \end{tabular}
\end{center}
\vskip-5mm
\caption[b5]{\label{fig:diromega5}\protect\small Distributions of the $\omega$ observable used in the
polarisation fit of the $a_1 \rightarrow \pi^{\pm} 2 \pi ^0 $ channel. 
The data are shown by points with statistical errors bars.
The dotted and dash-dotted lines corresponds to the contributions 
of left- and right-handed $\tau$'s respectively, as fitted in the data. 
The shaded area shows the non-$\tau$ background contribution.
The solid line corresponds to the sum of all simulated
contributions. }
\end{figure}

Figures~\ref{fig:diromega3}, \ref{fig:diromega4}, and~\ref{fig:diromega5} show the $\omega$
distribution for the $\rho$ and $a_1$ decay modes. The $\omega$ distributions
for the decays
where it was possible to reconstruct the $\tau$ direction are also shown 
separately.

\begin{table} [h] 
\caption{\protect\small  ${\cal A}_e$ and ${\cal A}_\tau$  results with statistical  and
systematic uncertainties for the 1990-1995 data
with the $\tau$ direction method.
\label{t3}}\vspace{.5cm}
\begin{center}
\begin{tabular} {ccc}          \hline
\rule[0cm]{0cm}{0.5cm}\rule[-.25cm]{0cm}{0.5cm} Channel & ${\cal A}_\tau$ $(\%)$  & ${\cal A}_e$ $(\%)$ \\ 
\hline 
\rule[0cm]{0cm}{0.5cm} hadron          
        &  $15.49 \pm 1.01  \pm 0.66 $ 
        &  $17.36 \pm 1.35  \pm 0.13 $ \\ 

\rule[0cm]{0cm}{0.5cm} rho
        &  $13.71 \pm 0.79  \pm 0.57  $ 
        &  $15.04 \pm 1.06  \pm 0.08 $ \\ 

\rule[0cm]{0cm}{0.5cm} a1(3h) 
        &  $15.01 \pm 1.55  \pm 1.30  $ 
        &  $15.78 \pm 2.07  \pm 0.40  $ \\ 

\rule[0cm]{0cm}{0.5cm} a1(h2$\pi^0$) 
        &  $15.94 \pm 1.73  \pm 1.70   $ 
        &  $12.65 \pm 2.31  \pm 0.41  $ \\ 

\rule[0cm]{0cm}{0.5cm} electron 
        &  $14.58 \pm 2.18  \pm 1.07 $ 
        &  $16.67 \pm 2.92  \pm 1.23 $ \\ 
\rule[0cm]{0cm}{0.5cm} muon
        &  $14.45 \pm 2.13  \pm 1.06 $ 
        &  $12.05 \pm 2.78  \pm 0.24 $ \\ 

\rule[0cm]{0cm}{0.5cm}\rule[-.25cm]{0cm}{0.5cm} acollinearity 
        &  $13.34 \pm 3.83  \pm 1.80  $ 
        &  $19.41 \pm 5.02  \pm 0.24 $ \\ \hline 

\rule[0cm]{0cm}{0.5cm}\rule[-.25cm]{0cm}{0.5cm} Combined
        &  $14.58 \pm 0.53  \pm 0.39 $ 
        &  $15.50 \pm 0.71  \pm 0.11 $ \\ \hline

\end{tabular}
\end{center}
\end{table}

\section{Combined results}
\label{section:final}
\subsection{Standard model corrections}
\label{sec:zfitt}
The $A_\tau$ and $A_e$ parameters are obtained by fitting the integral, 
in each $\cos\theta$ bin, of the
function (\ref{eq:p-theta}) to the measured polarisation. To obtain the combinations of
effective couplings (${\cal A}_l$) related to $\sin^2\theta_{\mathrm{W}}^{\mathrm{eff}}$,
 corrections have to be applied
which take care of the fact that the polarisation was not measured at
the Z pole
and that there are radiative corrections and Z-$\gamma$ interference
effects.
These corrections are computed
by the  ZFITTER program \cite{zfit} in the following way. 

For a set of the standard model parameters
(Z mass, top mass,  Higgs mass, $\alpha_s$),
ZFITTER provides  the value of the effective couplings in the
framework of the standard model.
Cuts are defined on acollinearity and momentum which reproduce
the
ones used in the analysis. Then 
 the ${\cal A}_{fb}$ asymmetry,
 and the
polarisation
in each $\cos\theta$ bin  are calculated at the given energy. 
The same fitting procedure as for the data
is applied and
the difference between the effective couplings and the fitted parameters
is
taken as the correction.

The study of the stability of the corrections
against the variation of the input parameters within their errors
  shows that the related error
 is negligible compared to the systematic errors involved in the
analyses.
The same is true for the values chosen for the cuts.
With this procedure, the inaccuracy of (\ref{eq:p-theta})
 in describing the
polarisation does not introduce any bias but may simply reduce very slightly
the sensitivity.
The level of these corrections is 0.04\% with an uncertainty of 0.01\%.
It has been checked that using as input the measured values of ${\cal A}_{fb}$
gives the same results.

\subsection{Consistency and combination of the results}
To investigate the consistency of the two
analyses, both statistical and systematic effects have to be taken 
into account.

As a result of  the different selection procedures, there is a
significant number of events which belong to only one of the two samples. 
The fractions of unshared events in the
different channels, averaged over the full angular distribution,
are 24\%~($\pi$), 27\%~($\rho$),
24\%~($a_1 \to 3 \pi$), 47\%~($a_1 \to \pi 2 \pi^0$),
24\%~($e$), and 13\%~($\mu$). Typically, these fractions are by a factor
1.6 larger in the small-angle region ($|\cos \theta| > 0.7$) than in
the central part. The distribution of the   events has to be taken into account
in the evaluation of the consistency  and in the combination of the measurements
since this
small-angle region has a significant contribution to the ${\cal A}_e$ determination.
Another statistical effect comes from the channels
where the $\tau$ direction is taken into account
since different information is used in the two analyses.
Finally, the Monte Carlo samples
do not completely overlap.

The systematic effects of the two methods have in common the detector behaviour and the
basic event reconstruction; however the analysis tools used are
largely independent. Furthermore, the procedures developed to
measure efficiencies, and various corrections are different in the two analyses
and thus a  part
of the estimated systematic uncertainties is uncorrelated.

The  ${\cal A}_{\tau}$ values given by the two analyses - for each channel and
globally -  are in  agreement within the estimated uncertainties.
The combination of  all channels, gives
$\Delta {\cal A}_{\tau} = (0.14 \pm 0.52)\%$, where the statistical and
systematic part of the error are comparable. The
agreement is not as good in the case of the ${\cal A}_{e}$ measurement
where the deviation between the two analyses, mainly due to the hadron channel, is
$\Delta {\cal A}_{e} = (0.92 \pm 0.48)\%$.
However, 
the  systematic uncertainties on  ${\cal A}_{e}$, dominated by the
effect of the non-$\tau$ background (chiefly  Bhabha events), are
smaller by a factor of 5-10 than the statistical
errors, and the many cross-checks performed show that
the  background levels and
distributions are known within the quoted uncertainties.
 This lead to the conclusion  that
the observed difference is the result of a statistical
fluctuation.

\subsection{Final results}

The consistency of the two sets of results having been checked, they
are combined  to obtain the final results. Since the two
analyses have comparable statistical and systematic uncertainties,
the combination procedure is not a critical one. It was chosen
to average the results using equal weights. For the computation of the final 
errors  the statistical correlation
between the two samples is taken into account
 together with the systematic uncertainty correlations.
\begin{figure}[h]
\begin{center}
\mbox{
\epsfig{file=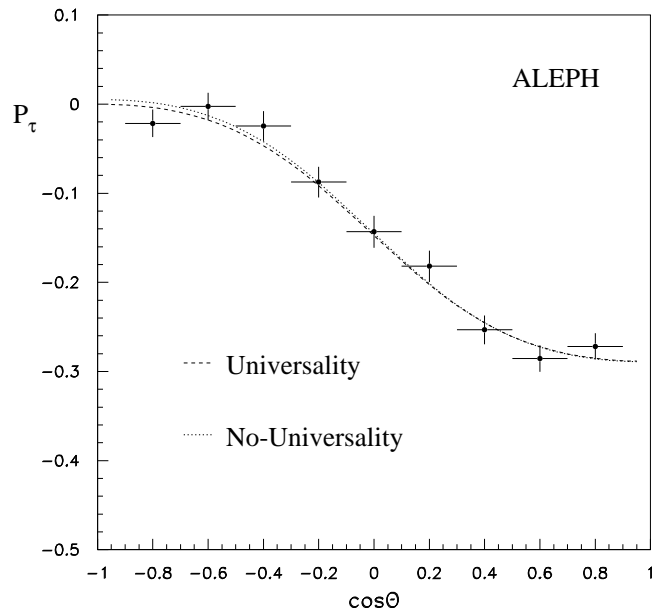,height=8.0cm}
}
\end{center}
\caption[.]
{\protect\small Polarisation dependence on $\cos\theta$ for the LEP~I data.
The curves corresponding to Eq.~\ref{eq:p-theta} for parameters given
 by fits with (dashed line) and without (dotted line)
the universality constraint are superimposed.}
\label{fig_20}
\end{figure}

The results  are given in Table~\ref{t_fin} for all the analysed modes and
in Table~\ref{t_fin_c2} for the global values.
Figure~\ref{fig_20} shows the average tau polarisation as a function
of $\cos\theta$.

\section{Discussion}

The ${\cal A}_\tau$ and ${\cal A}_e$ parameters are measured
by means of similar methods
by the three other LEP collaborations~[27-29].
 The  ${\cal A}_e$ asymmetry is also measured 
by the SLD collaboration \cite{sld} using the beam polarisation. 

\begin{figure}[htbp]
\begin{center}
\mbox{\epsfig{file=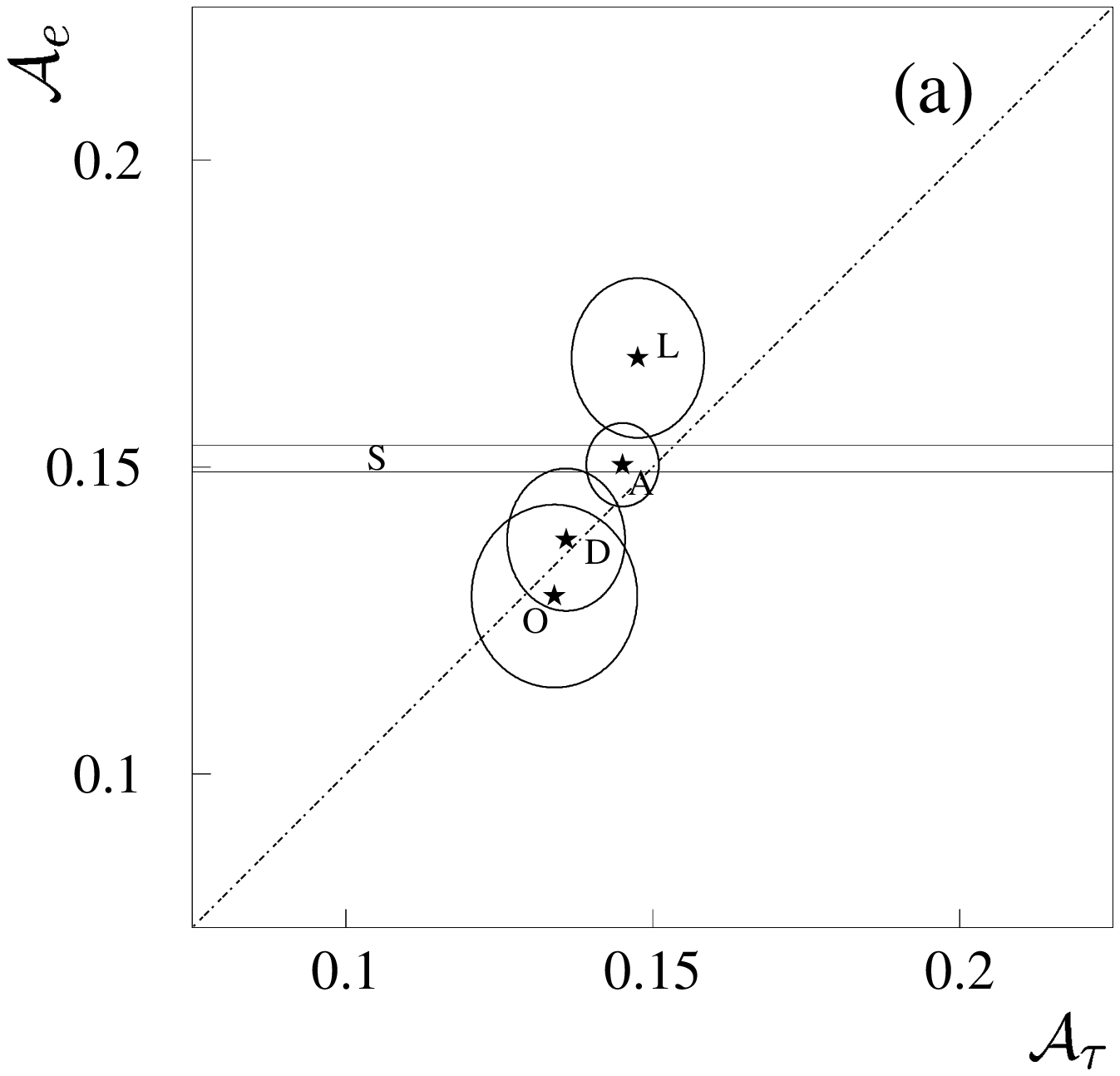,width=7.4cm}}\hspace{.2cm}\mbox{\epsfig{file=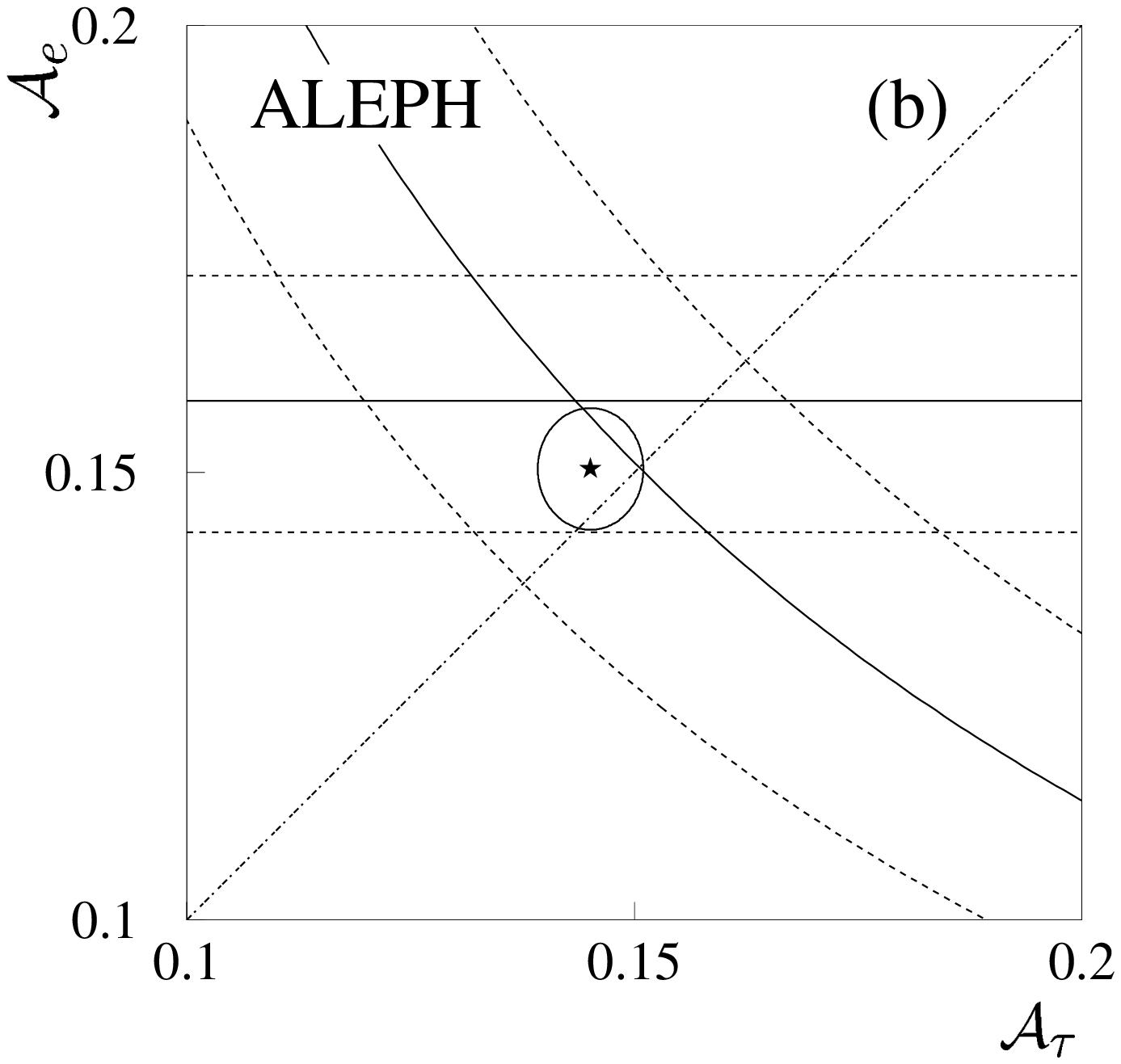,width=7.4cm}}
\caption{ \label{fig:adlos}\protect\small (a) Comparison  of  the  ${\cal A}_e$ and ${\cal A}_\tau$
measurements by the ALEPH (A), DELPHI (D), L3 (L), and OPAL (O)   collaborations.
The ellipses are one standard error contours (39\% CL).
The horizontal lines represent the SLD (S) ${\cal A}_e$ measurement through the left-right 
asymmetry of the cross-sections plus and minus one standard deviation. 
(b) Comparison  of ALEPH  ${\cal A}_\tau$, ${\cal A}_e$, and 
$A^0_{fb}$ measurements. The ellipse is the one standard error contour. The hyperbolas and horizontal lines represent the ${ A}_{fb}^{0,\tau}$ and ${ A}_{fb}^{0,e}$
measurements respectively (plus and minus one standard error).}
\end{center}
\end{figure}

All the results are shown in Fig.~\ref{fig:adlos}~(a) where a reasonable
general agreement can be observed.

The ${\cal A}_\tau$, ${\cal A}_e$, and 
${ A}_{fb}^{0}$ asymmetries at the Z pole are related by the equations:
\begin{equation}
{ A}_{fb}^{0,e}=\frac{3}{4}({\cal A}_e)^2~,\hspace{1.5cm}
{ A}_{fb}^{0,\tau}=\frac{3}{4}{\cal A}_e{\cal A}_\tau~.
\end{equation} 
The comparison of the present measurements of ${\cal A}_e$ and ${\cal A}_\tau$ with 
the forward-backward asymmetries measured by the ALEPH collaboration \cite{bib:aleph_ew} is 
presented in Fig.~\ref{fig:adlos}~(b).

Assuming universality, the leptonic forward-backward asymmetry
measured by \break
ALEPH, using all the leptonic Z decay channels, is 
${ A}_{fb}^{0,l}=0.0173\pm 0.0016$ \cite{bib:aleph_ew}, to be compared with 
$\frac{\textstyle 3}{\textstyle 4}({\cal A}_{e\mbox{-}\tau})^2=0.0163\pm 0.0010$.

\begin{figure}[htb]
\centerline{\epsfig{file=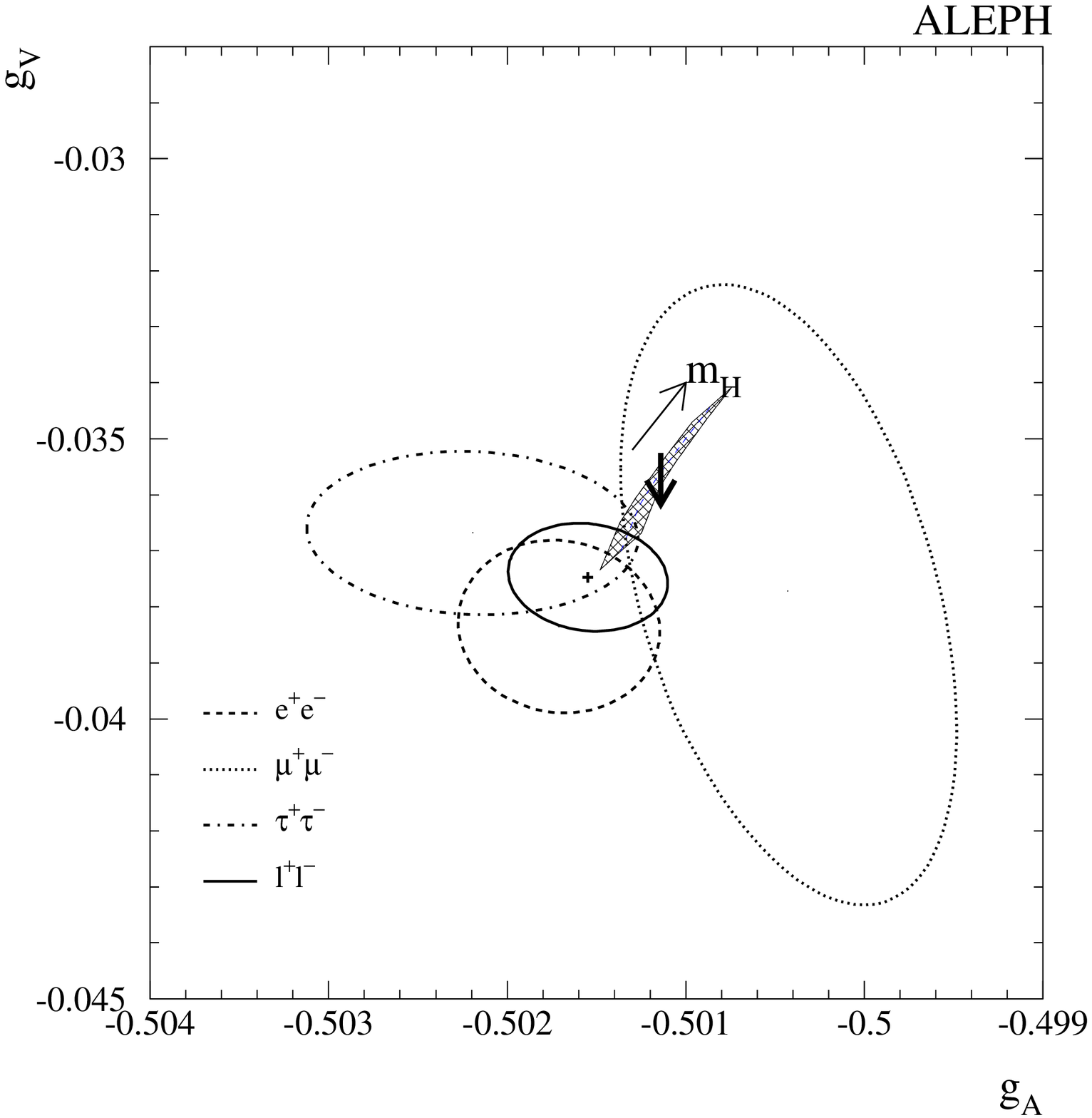,width=11cm}}
\caption{\label{fig:gv-ga}\protect\small Effective lepton couplings. The ellipses are one standard deviation contours
(39\% CL). The shaded area indicates the Standard Model expectation for
$M_{\mathrm{t}}= 174\pm 5~{\mathrm{Gev}}/c^2$ and $90 <M_{\mathrm{H}}\, ({\mathrm{GeV}}/c^2)<1000$;
the vertical arrow shows the change  if the electromagnetic coupling constant is 
varied within its error.}
\end{figure}

The information given by the $\tau$ polarisation measurement can be used, 
together with  the forward-backward asymmetries and the partial widths of
the Z decays into lepton pairs~\cite{bib:aleph_ew}, to determine the Z leptonic couplings.
The results obtained by the procedure of Ref.~\cite{bib:aleph_ew} are given in
Fig.~\ref{fig:gv-ga} and Table~\ref{tab:gv-ga}.   
\begin{table} [p] 
\caption{\protect\small Combined ${\cal A}_{\tau}$ and ${\cal A}_e$  results with statistical and
systematic uncertainties for the 1990-1995 data. The pion inclusive channel
is highly correlated with the hadronic channels.
\label{t_fin}}\vspace{.5cm}
\begin{center}
\begin{tabular} {ccc}          \hline
\rule[0cm]{0cm}{0.5cm}\rule[-0.25cm]{0cm}{0.5cm} Channel & ${\cal A}_{\tau}$ $(\%)$ & ${\cal A}_e$ $(\%)$ \\ 
\hline 
\rule[0cm]{0cm}{0.5cm} hadron         
               &  $15.35 \pm 0.96  \pm 0.50 $
               &  $16.32 \pm 1.27   \pm 0.09 $ 
 \\ 

\rule[0cm]{0cm}{0.5cm} rho          
               &  $13.75 \pm 0.78  \pm 0.42  $
               &  $14.85 \pm 1.03  \pm 0.06 $ 
 \\ 

\rule[0cm]{0cm}{0.5cm} a1(3h)
               &  $14.89 \pm 1.50   \pm 1.05   $ 
               &  $14.68 \pm 1.98  \pm 0.40 $ 
\\ 

\rule[0cm]{0cm}{0.5cm} a1(h2$\pi^0$)
               &  $16.14 \pm 1.74  \pm 1.44   $
               &  $14.13 \pm 2.30  \pm 0.42 $ 
\\  

\rule[0cm]{0cm}{0.5cm} electron
               &  $14.11 \pm 2.18  \pm 0.84 $
               &  $15.38 \pm 2.92  \pm 0.77 $ 
\\  

\rule[0cm]{0cm}{0.5cm} muon
               &  $14.05 \pm 2.07  \pm 0.82 $
               &  $11.91 \pm 2.71  \pm 0.17 $
\\  

\rule[0cm]{0cm}{0.5cm} acollinearity  
               &  $13.34 \pm 3.83  \pm 1.80  $ 
               &  $19.41 \pm 5.02  \pm 0.24 $ 
\\  

\rule[0cm]{0cm}{0.5cm}\rule[-.25cm]{0cm}{0.5cm} pion inclusive
               &  $14.93 \pm 0.83  \pm 0.87 $ 
               &  $14.91 \pm 1.11  \pm 0.17 $ 
\\  \hline

\end{tabular}
\end{center}
\end{table}

\begin{table} [p] 
\caption{\label{t_fin_c2}\protect\small Combined ${\cal A}_e$, ${\cal A}_\tau$ and ${\cal A}_{e-\tau}$ results with statistical 
 and systematic uncertainties for the 1990-1995 data.}\vspace{.5cm}
\begin{center}
\begin{tabular} {cc}          \hline
\rule[0cm]{0cm}{0.5cm}\rule[-0.25cm]{0cm}{0.5cm}
  ${\cal A}_\tau$ (\%)     &  $14.51 \pm 0.52  \pm 0.29  $ \\ 
\rule[0cm]{0cm}{0.5cm}
  ${\cal A}_{e}$  (\%)      &  $15.04 \pm 0.68  \pm 0.08  $ \\ 
\rule[0cm]{0cm}{0.5cm}\rule[-0.25cm]{0cm}{0.5cm}
  ${\cal A}_{e\mbox{-}\tau}$ (\%) &  $14.74 \pm 0.41  \pm 0.18  $ \\ \hline 
\end{tabular}
\end{center}
\end{table}

\begin{table}[p]
\caption{\label{tab:gv-ga}\protect\small 
Effective lepton vector and axial vector couplings.}\vspace{.5cm}

\begin{center}
\begin{tabular}{lrlr}
\hline
\rule[0cm]{0cm}{0.5cm}\rule[-0.25cm]{0cm}{0.5cm}
$g_{V}^{e}$&$-0.0384\pm 0.0015$&
$g_{A}^{e}$&$-0.50171\pm 0.00056$\\
\rule[0cm]{0cm}{0.5cm}
$g_{V}^{\mu}$&$-0.0377\pm 0.0055$&
$g_{A}^{\mu}$&$-0.50043\pm 0.00094$\\
\rule[0cm]{0cm}{0.5cm}\rule[-0.25cm]{0cm}{0.5cm}
$g_{V}^{\tau}$&$-0.0367\pm 0.0015$&
$g_{A}^{\tau}$&$-0.50219\pm 0.00093$\\\hline
\rule[0cm]{0cm}{0.5cm}\rule[-0.25cm]{0cm}{0.5cm}
$g_{V}^{l}$&$-0.03747\pm 0.00096$&
$g_{A}^{l}$&$-0.50155\pm 0.00045$\\\hline
\end{tabular}
\end{center}
\end{table}

Comparing these numbers with the values of the couplings presented in 
Ref.~\cite{bib:aleph_ew}
 shows the important weight of the $\tau$ polarisation measurement
 in the Z
leptonic coupling determination.

\section{Conclusion}
From the ${\cal A}_\tau$ and ${\cal A}_e$ values obtained through the $\tau$
polarisation measurement: 
$${\cal A}_\tau=0.1451\pm 0.0059~,\hspace{1cm}
{\cal A}_e=0.1504\pm 0.0068~,$$ 
the following ratios of the effective couplings
are computed:
\begin{eqnarray*}
g_V^\tau/g_A^\tau&=&0.0729\pm 0.0030~,\\
g_V^e/g_A^e&=&0.0756\pm 0.0035~,
\end{eqnarray*}
$$\frac{g_V^\tau/g_A^\tau}{g_V^e/g_A^e}=0.964\pm0.060~.$$
 The universal value
$${\cal A}_{e\mbox{-}\tau}=0.1474\pm 0.0045~,$$ translates into a determination of the effective
weak mixing angle
\begin{equation}
 \sin^2 \theta_{\mathrm{W}}^{\mathrm{eff}} = 0.23147 \pm 0.00057~.
\end{equation}
Using both the $\tau$ polarisation and the leptonic forward-backward asymmetries
measured by ALEPH,
the value
\begin{equation}
 \sin^2 \theta_{\mathrm{W}}^{\mathrm{eff}} = 0.23130 \pm 0.00048
\end{equation}
is obtained.
\section*{Acknowledgements}
We wish to express our thanks to our colleagues from the accelerator divisions for 
the excellent performance of LEP. We also thank the engineers and support personnel at
our home institutions for contributing to the success of ALEPH. Those of us who are
guests at CERN thank the laboratory for its hospitality and support.
\newpage
\appendix
\section{Tau direction estimator}
\label{dir_estim}
\begin{figure}[ht]
  \begin{center}
    \epsfig{file=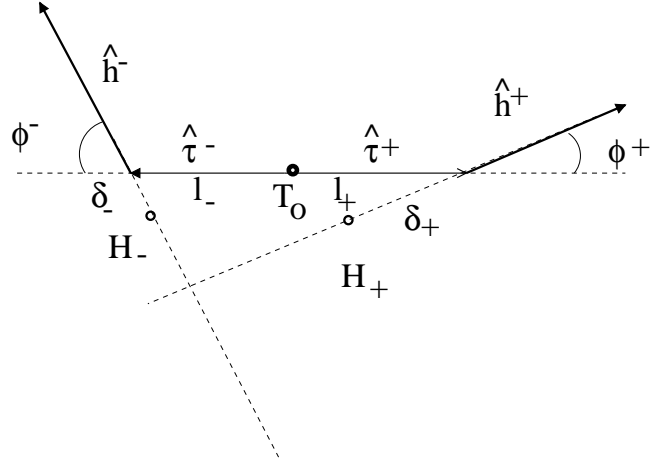,width=8.5cm}
  \end{center}
  \caption[.]{\label{fig:irena}\protect\small Decay of a $\tau$ pair.}
\end{figure}
\subsection{The ideal case}
\longpage

First consider the case
of a $\tau$ event  where both $\tau$'s decay  into a single pion 
with momentum vector measured perfectly (Fig.~\ref{fig:irena}). 
Let  $\mathrm{T}_0$ be   the interaction point and $\mathrm{H}_{\pm}$ points 
on each reconstructed track,
in the vicinity of the decay points.
The vectors $\hat{h}^{\pm}$ and $\hat{\tau}^{\pm}$ are unit vectors 
along the $\pi^\pm$ and $\tau^\pm$ momenta respectively, $l_\pm$ are  the $\tau^\pm$ decay lengths
and $\delta_\pm$  the distances of $\mathrm{H}_\pm$ to the $\tau\,^\pm$ decay points.
From $\hat{\tau}^+=-\hat{\tau}^-$ and the relation (Fig.~\ref{fig:irena})
$$\overrightarrow{\mathrm{OT}}_0+l_\pm\hat{\tau}^\pm=\overrightarrow{\mathrm{OH}}_\pm+\delta_\pm\hat{h}^\pm~,$$
where $\mathrm{O}$ is an arbitrary origin, the equation
\begin{equation}
  l\hat{\tau}=\delta_-\hat{h}^--\delta_+\hat{h}^++\overrightarrow{\mathrm{H}_+\mathrm{H}}_-
  \label{lapremiere}
\end{equation}
follows, where  
$l=l_++l_-$ and $\hat\tau=\hat\tau\,^-$.

Taking for $\mathrm{H}_\pm$ the end points of 
the vector  ${{\Vec{D}_h}}$  defined~\cite{3dip}\footnote{A similar analysis of
the ideal case is given in Ref.~\cite{kuhndir}.}
as the minimum approach vector between the two reconstructed tracks, the previous equation becomes
\begin{equation}
  l  {{\hat{\tau}}}  = l {{\Vec{A}_h}} + {{ \Vec{D}_h}}~,   
  \label{eq_dh}
\end{equation}
where the vector $\Vec{D}_h$ is orthogonal to the hadronic plane spanned by $\hat{h}^+$ and $\hat{h}^-$
and the vector $\Vec{A}_h$ lies in this plane.
\begin{figure}[t]
  \begin{center}
    \epsfig{file=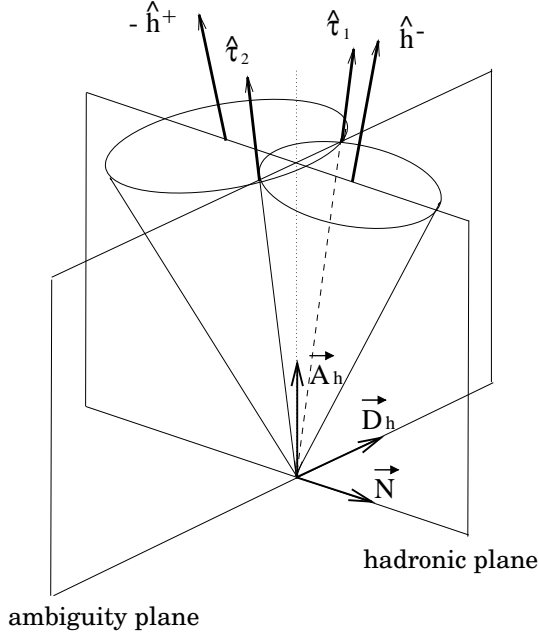,width=7.5cm}
  \end{center}
  \caption[.]{\protect\small Definition of vectors $\Vec{A}_h$, $\Vec{D}_h$, and $\Vec{N}$. }
  \label{f_fig1}
\end{figure}

The  angles
$\phi^\pm$ (Fig.~\ref{fig:irena}) are given by Eq.~\ref{eq_phi},
$$\cos\phi^\pm=\frac{2E_{\tau^\pm}E_{h^\pm}-m_\tau^2-m_{h^\pm}^2}{2p_{\tau^\pm}p_{h^\pm}}~,$$
 consequence of the $\tau$ mass constraint, where $E$, $p$, and $m$ are energy, momentum,
 and mass of $\tau$'s and hadrons ($\pi$'s).
Accordingly, the two kinematically found $\tau^-$ directions, defined by the unit
vectors $\hat{\tau}_1 $ and $\hat{\tau}_2 $,  are  the intersections of two cones 
having $-\hat{h}^+$ and $\hat{h}^-$ for axes and  $\phi^\pm$ for opening angles.
The vector $\hat{\tau}_1+\hat{\tau}_2$
lies then in  the hadronic plane and is, therefore, orthogonal to $\Vec{D}_h$ 
(Fig.~\ref{f_fig1}).
This implies
\begin{equation}
  {{ \Vec{D}_h}} \cdot {\boldmath{ \hat{\tau}_1}} = - \  {{\Vec{D}_h} } \cdot 
  {{\hat{\tau}_2}}~. 
  \label{dhdot}
\end{equation}
The vector ${{\Vec{A}_h}}$,  projection of $\hat{\tau}$ on the hadronic plane,
is  the internal bisector of $\hat{\tau}_1$ and $\hat{\tau}_2$. 
These properties, with Eq.~\ref{eq_dh}, lead to 
$$ l =   \frac{|{{\Vec{D}_h}}|}{ \sqrt{1-|{{\Vec{A}_h}}|^2}}~, $$
since $l$ is, by definition, a positive length.
The positivity of $l$ and Equations \ref{eq_dh} and \ref{dhdot} imply that the
physical $\tau$ direction is determined by the  relation
\begin{equation}
  {{ \vec{D}_h}} \cdot { \hat{\tau}_{\mathrm{true}}} \ge 0  ~.
\end{equation}

\longpage
For hemispheres with a 3-prong decay, the vector $\hat{h}^{\pm}$
is defined as equal to the sum of the three momenta and originating from the 3-track 
common vertex.

\subsection{The general case}

When  $\pi^0$'s are produced, the vector $\Vec{A}_h$, which is kinematically determined,
is still measurable, but
the exact reconstruction of ${{\Vec{D}_h}}$  
given in the previous section is not possible
and an approximation must be made.

The vectors $\hat{h}^\pm$ are now unit vectors along the sums of the hadron momenta
in each hemisphere.
Two vectors, which make with $\Vec{A}_h$ an orthogonal system, can be unambiguously measured:
$$\Vec{h}_\perp=\hat{h}^-\times\hat{h}^+~,$$
which is normal to the hadronic plane, and
$$\Vec{N}=\hat{h}_\perp\times\hat{\tau}=\hat{h}^+\cos\phi^-+\hat{h}^-\cos\phi^+~,$$
which is normal to the ambiguity plane spanned by $\hat{\tau}_1$ and $\hat{\tau}_2$.
In terms of $\Vec{h}_\perp$ and $\Vec{N}$,  the analytical expressions of $\Vec{A}_h$
and $\Vec{D}_h$ are
\begin{eqnarray*}
  \Vec{A}_h&=&\Vec{N}\times\Vec{h}_\perp\,/\,\Vec{h}_\perp^2~,\\
  \Vec{D}_h&=&\Vec{h}_\perp\,(\overrightarrow{\Delta \mathrm{H}}\cdot\Vec{h}_\perp)
\,/\,\Vec{h}_\perp^2~.
\end{eqnarray*}
The vector $\overrightarrow{\Delta \mathrm{H}}$ is equal to $\overrightarrow{\mathrm{H}_+\mathrm{H}}_-$,
where $\mathrm{H}_\pm$ are arbitrary points on the lines drawn along the vectors $\hat{h}^\pm$ from the
$\tau^\pm$ decay points, lines which coincide with the reconstructed tracks in the ideal case of 
the previous section.

\longpage
The charged tracks reconstructed in each hemisphere are used to define an approximation of $\Vec{D}_h$.
Let $\hat{c}^\pm$ be unit vectors along the momenta of the charged particles,
$\mathrm{C}_\pm$ points on the charged tracks in the neighbourhood of the decay points, 
$\Vec{c}_\perp=\hat{c}^-\times\hat{c}^+$, and 
$\overrightarrow{\Delta\mathrm{C}}=\overrightarrow{\mathrm{C}_+\mathrm{C}}_-$.
If the vectors $\hat{c}_\perp$ and $\hat{h}_\perp$ are collinear, the 
space planes parallel to the hadronic plane and containing the 
$\tau^\pm$ decay points contain also the charged tracks. The vector
$\Vec{D}_h$ is then equal to the minimum approach vector between the
charged tracks.
In the most general case,
an equation similar to Eq. \ref{eq_dh} can be written~\cite{3dip}:
\begin{equation}
l\hat{\tau}=l\Vec{A}_c+\Vec{D}_c~, 
\end{equation}
where the vector  ${{\vec{D}_c}}$, whose length is proportional 
to the minimum distance between the charged tracks,
 is collinear to $\Vec{h}_\perp$:
\begin{eqnarray*}
\Vec{A}_c&=&\Vec{N}\times\Vec{c}_\perp\,/\,(\Vec{h}_\perp\cdot\Vec{c}_\perp)~,\\
\Vec{D}_c&=&\Vec{h}_\perp\,(\overrightarrow{\Delta \mathrm{C}}\cdot\Vec{c}_\perp)
\,/\,(\Vec{h}_\perp\cdot\Vec{c}_\perp)~.
\end{eqnarray*}

As the relation~(\ref{eq_dh}) is still true, the expression 
of the $\Vec{D}_h$ vector can be written as :
\begin{equation}
 {{\Vec{D}_h}} = {{ \Vec{D}_c}} + l\, ({{ \Vec{A}_c}}  - {{ \Vec{A}_h}}) ~.
\label{eq_dh_real}
\end{equation} 

In the equation~(\ref{eq_dh_real}), the vectors 
${\Vec{D}_c}$, ${\Vec{A}_c}$, and  
${\Vec{A}_h}$ are measurable, but the decay length remains unknown. 
Taking the mean  $\tau$ decay length sum $\bar{l} = 4.6$~mm 
as a value for $l$, an approximation of $\Vec{D}_h$, 
denoted $\Vec{D}_{h_{\mathrm{eff}}}$, is obtained. 

The vector $\Vec{A}_c$ lies in the ambiguity plane and,
since both $\Vec{D}_h$ and $\Vec{D}_c$ are collinear to $\Vec{h}_\perp$, 
the projection of $l\Vec{A}_c$
on $\Vec{A}_h$ is $l\Vec{A}_h$, analytically
$ \Vec{A}_c\cdot\Vec{A}_h=\Vec{N}^2/\Vec{h}_\perp^2=\Vec{A}_h^2$. Therefore, the vector 
$\Vec{D}_{h_{\mathrm{eff}}}$ is proportional to $\Vec{D}_h$, 
irrespectively of the value of $\bar l$.
As in the ideal case,
\begin{equation}
\Vec{D}_{h_{\mathrm{eff}}}\cdot\hat{\tau}_1=-\Vec{D}_{h_{\mathrm{eff}}}\cdot\hat{\tau}_2~,
\end{equation}
and the  $\tau$ direction giving the positive value for 
${{ \Vec{D}_{h_{\mathrm{eff}}}}} \cdot {{ \hat{\tau}}} $
can be  selected.

 Because of the approximation made when $\pi^0$'s are present
and  resolution effects, this selected direction is only correct on a statistical
basis and probabilities, computed from reference distributions of 
$ \Vec{D}_{h_{\mathrm{eff}}}\cdot\hat{\tau}$, are assigned to the two solutions.  

\newpage


\begin{thebibliography}{99}
\bibliographystyle{unsrt}
\bibitem{polara}
ALEPH Collaboration, {\it Measurement of the polarization of $\tau$ 
leptons produced in                     
 $Z$ decays}, Phys. Lett. B {\bf 265} (1991) 430. 
\bibitem{polarb}
ALEPH Collaboration, {\it Measurement of the tau polarisation at the Z
  resonance}, Z. Phys. C {\bf 59} (1993) 369.
\bibitem{polar}
ALEPH Collaboration, {\it Improved tau polarisation measurement}, Z. Phys. C {\bf 69} 
(1996) 183.
\bibitem{zfit} D.~Bardin {\it et al.}, Z. Phys. C {\bf 44} (1989) 493;
Comp. Phys. Comm. {\bf 59} (1990) 303; Nucl. Phys. B {\bf 351} (1991) 1;
 Phys. Lett. B {\bf 255} (1991) 290; CERN-TH 6443-92 (May 1992) and hep-ph/9412201; 
{\it ZFITTER v.6.10 A Semi-Analytical Program for Fermion Pair Production in \hbox{e$^+$e$^-$}
  Annihilation}, DESY 99-070, June 1999.
\bibitem{optimal}
M.~Davier, L.~Duflot, F.~Le Diberder and A.~Roug\'e, Phys. Lett. B {\bf 306} 
(1993) 411.
\bibitem{3dip} ALEPH
  Collaboration, {\it Measurement of the $\tau$ lepton lifetime with 
the three-dimensional impact parameter method},
                    Z. Phys. C {\bf 74} (1997) 387.
\bibitem{michelparam}
D.E.~Groom et al. (Particle Data Group), Eur. Phys. J. C {\bf 15} (2000) 1.
\bibitem{kuhn}  J.H.~K\"uhn, E.~Mirkes, 
Z. Phys. C {\bf 56} (1992) 661;
 C {\bf 67} (1995) 364 (erratum).
\bibitem{detector}
  ALEPH Collaboration, {\it ALEPH: A detector for electron-positron
    annihilation at LEP},
           Nucl. Instrum. Methods A {\bf 294} (1990) 121.
\bibitem{perfo} ALEPH Collaboration, {\it Performance of the ALEPH
    detector at LEP},
           Nucl. Instrum. Methods A {\bf 360} (1995) 481.
\bibitem{bib:aleph_ew}
 ALEPH Collaboration, {\it Measurement of the Z Resonance Parameters
   at LEP}, 
  Eur. Phys. J. C {\bf 14} (2000) 1.
\bibitem{koral}
S.~Jadach and Z.~W\c{a}s, Comp. Phys. Comm. {\bf 66} (1991) 276.
\bibitem{unibab}
H.~Anlauf,  {\it et al.} Comp. Phys. Comm. {\bf 79} (1994) 466.
\bibitem{babamc}
M.~B\"ohm, A.~Denner, W.~Hollik, Nucl. Phys. B {\bf 304} (1998) 687;
F.A.~Berends, R.~Kleiss, W.~Hollik, Nucl. Phys. B {\bf 304} (1998) 712;
Computer program BABAMC, courtesy of R.~Kleiss.
\bibitem{phot02}
J.A.M.~Vermaseren, in {\it Proceedings of the IV International
 Workshop on Gamma Gamma Interactions}, eds. G.~Cochard and P.~Kessler
 (1980).
\bibitem{bib:4ferm} 
J.M.~Hilgart, R.~Kleiss and F.~Le Diberder,
Comp. Phys. Comm. {\bf 75} (1993) 191.
\bibitem {hbr}
ALEPH Collaboration, {\it Tau hadronic branching ratios},
 Z. Phys. C
{\bf{70}} (1996) 579. 
\bibitem{bib:etaomeg}
 ALEPH Collaboration, {\it A study of $\tau$ decays involving $\eta$ and $\omega$ mesons}
    Z. Phys. C {\bf 74} (1997) 263.
\bibitem {lbr}
ALEPH Collaboration, {\it Tau leptonic branching ratios}, Z. Phys.
C {\bf 70} (1996) 561.
\bibitem{spectrala}
ALEPH Collaboration, {\it Measurement of the Spectral Functions of
  Vector Current Hadronic Tau Decays}, Z. Phys. C. {\bf 76} (1997) 15.
\bibitem{spectralb}
ALEPH Collaboration, {\it Measurement of the Spectral Functions of
  Axial-Vector  Hadronic $\tau$ Decays and Determination of
  $\alpha_S(M^2_\tau)$}, Eur. Phys. J.  C. {\bf 4} (1998) 409.
\bibitem{brold}
ALEPH Collaboration, {\it Measurement of Tau Branching Ratios}, Z. Phys. C {\bf 54}
(1992) 211.
\bibitem{acol} R.~Alemany, N.~Rius, J.~Bernabeu, J.J.~Gomez-Cadenas, A.~Pich,
               Nucl. Phys. B {\bf{379}}, (1992) 3.
\bibitem{lund}
T.~Sj\"ostrand, Comp. Phys. Comm., {\bf 82} (1994) 74.
\bibitem{kzero}
ALEPH Collaboration, {\it $K^0$ production in one-prong $\tau$
  decays},
Phys. Lett. B {\bf 332} (1994) 219; ALEPH Collaboration, {\it One-prong
  $\tau$ decays with kaons}, Eur. Phys. J. C. {\bf 10} (1999) 1.
\bibitem{irena} I. Nikolic, Thesis, Universit\'e de Paris-Sud, Orsay,
LAL 96-27 (1996).  
\bibitem{opal}
OPAL Collaboration, {\it A Precise Measurement of the Tau Polarization and its 
Forward-Backward Asymmetry at LEP}, Z. Phys. C {\bf 72} (1996) 365.
\bibitem{l3}
L3 Collaboration, {\it Measurement of $\tau$ Polarisation at LEP}, 
Phys. Lett. B {\bf 429} (1998) 387.
\bibitem{delphi}
DELPHI Collaboration, {\it A Precise Measurement of the $\tau$ Polarisation at LEP-1},
Eur. Phys. J. C {\bf 14} (2000) 585.
\bibitem{sld}
SLD Collaboration, {\it A High-Precision Measurement of the Left-Right Z Boson
Cross-Section Asymmetry},
Phys. Rev. Lett. {\bf 84} (2000) 5945.
\bibitem{kuhndir}
J.-H. K\"uhn, Phys. Lett. B {\bf 313} (1993) 458.
\end{thebibliography}
\end{document}